Faculdade de Engenharia da Universidade do Porto

# High-level Approaches to Detect Malicious Political Activity on Twitter

Miguel Sozinho Ramalho

Masters Thesis

U.PORTO

FEUP FACULDADE DE ENGENHARIA UNIVERSIDADE DO PORTO

Mestrado Integrado em Engenharia Informática e Computação

Supervisor: André Restivo, University of Porto, PhD

Co-Supervisor: Ashton Anderson, University of Toronto, PhD Co-Supervisor:

Hugo Sereno, University of Porto, PhD

# High-level Approaches to Detect Malicious Political Activity on Twitter

Miguel Sozinho Ramalho

Mestrado Integrado em Engenharia Informática e Computação

July 21, 2020

# Abstract

Democracy, as a manifestation of the primary desires of nations, rests upon the premise that political elections allow for the citizens of these nations to express their views based on facts and truth. Naturally, it is impossible for every person's view to be clear and realistic. However, in recent years, the tools available to lengthen that gap have grown in reach and impact, far beyond any acceptable threshold. Online Social Networks (OSNs) are one such example. Indeed, these have become political weapons – Facebook and Twitter being among the most infamous cases. This problem was first prompted during 2016, in the context of the United States presidential elections, where a foreign nation was showed to have interfered – Russia. This example is not an exception.
It was the inception of a problem that has come to stay.

Our work represents another step into the detection and prevention of these ever-more present political manipulation efforts. We, therefore, start by focusing on understanding what the state-ofthe-art approaches lack – since the problem remains, this is a fair assumption. We find concerning issues within the current literature and follow a diverging path. Notably, by placing emphasis on using data features that are less susceptible to malicious manipulation and also on looking for high-level approaches that avoid a granularity level that is biased towards easy-to-spot and low impact cases.

We designed and implemented a framework – Twitter Watch– that performs structured Twitter data collection, applying it to the Portuguese Twittersphere. We investigate a data snapshot taken on May 2020, with around 5 million accounts and over 120 million tweets (this value has since increased to over 175 million). The analyzed time period stretches from August 2019 to May 2020, with a focus on the Portuguese elections of October 6th, 2019. However, the Covid-19 pandemic showed itself in our data, and we also delve into how it affected typical Twitter behavior.

We performed three main approaches: content-oriented, metadata-oriented, and network interaction-oriented. We learn that Twitter's suspension patterns are not adequate to the type of political trolling found in the Portuguese Twittersphere – identified by this work and by an independent peer - nor to fake news posting accounts. We also surmised that the different types of malicious accounts we independently gathered are very similar both in terms of content and interaction, through two distinct analysis, and are simultaneously very distinct from regular accounts.

Keywords: Twitter, Online Social Networks, Twitter API Data Collection, Political Trolling Detection
ii

# Resumo

A Democracia, enquanto manifestação dos desejos primitivos de nações, assenta sob a premissa de que as eleições políticas permitem aos seus cidadãos expressar a sua visão com base em factos e verdade. Naturalmente, é impossível que todas as pessoas tenham uma visão nítida e realista, mas os últimos anos foram marcados pelo aparecimento de ferramentas que aumentam esta discrepância, cujo alcance e impacto cresceu bem para além de níveis aceitáveis. As redes sociais online são um desses casos. De facto, estas tornaram-se armas políticas – sendo o Facebook e o Twitter os casos mais gritantes. Esta questão remonta a 2016, durante as eleições presidenciais norte-americanas, onde foi demonstrada a interferência de uma nação externa – Rússia. Este exemplo não é uma exceção, antes foi a génese de um problema que veio para ficar.

O nosso trabalho é um mais um passo no caminho da deteção e prevenção destes esforços de manipulação política cada vez mais presentes. Portanto, focamo-nos em perceber quais as limitações atuais do estado da arte – se o problema prevalece, é correto assumir que estas existem. De facto, encontrámos problemas preocupantes na literatura e, portanto, decidimos seguir um caminho divergente. Particularmente, ao colocar a ênfase no uso de atributos dos nossos dados que sejam menos suscetíveis à manipulação maliciosa e também ao investir em abordagens de alto nível que procuram evitar o nível de granularidade que é tipicamente enviesado para casos óbvios de deteção e de baixo impacto.

Concebemos e implementámos uma arquitetura – Twitter Watch– capaz de recolher dados do Twitter de forma estruturada, e aplicámo-la à *Twittersphere* portuguesa. Investigámos uma amostra capturada em maio de 2020, com cerca de 5 milhões de contas e mais de 120 milhões de tweets (atualmente este valor já ultrapassa os 175 milhões). O período analisado vai de agosto de 2019 a maio de 2020, com um especial foco nas eleições legislativas de 6 de outubro, de 2019. Contudo, a pandemia originada pelo Covid-19 manifestou-se nos dados recolhidos e também investigámos o seu impacto no comportamento anterior, na rede social.

Levámos a cabo três abordagens principais em termos de análise: orientada a conteúdo, orientada a meta-dados, e orientada a uma rede de interações. Aprendemos que os padrões de suspensão do Twitter não são adequados ao tipo de comportamento de *troll* com que nos deparámos – neste trabalho e vindo de uma colaboração com um par académico – nem com contas que partilham *fake news*. De facto, também compreendemos que os diferentes tipos de contas maliciosas que recolhemos, de forma independente, são bastante semelhantes tanto em termos de conteúdo como interação, sendo estas análises independentes, e simultaneamente muito distintas de contas regulares.

Keywords: *Twitter, Online Social Networks, Twitter API Data Collection, Political Trolling Detection*

# Acknowledgements

Two types of acknowledgments are due here.

The first, explicit, goes to my three supervisors. Professor Ashton, thank you for believing in this work from the start, when we were one Atlantic ocean apart, and for still believing it when we were one kilometer plus one pandemic apart. Professors André and Hugo, thank you for showing me that people can stay strong against a system that is oriented towards quantity and not quality, for giving me the freedom to forge my own path, oh, and for those late-night meetings too. To all three, a big thanks for your support, tips, and feedback throughout this project.

The second, implicit, is to all the people who were there for me in these past years, and to some few, way before that. Your contributions made me who I am today. Names are not written as I did not want to risk leaving anyone out. I thank those who showed me the patience, wisdom, love, and support I know I did not always deserve.

Um sargento nunca abandona os seus soldados,

Obrigado a todos,

Miguel

vi

“We all woke up, tryna tune to the daily news
Looking for confirmation, hoping election wasn’t true
All of us worried, all of us buried, and the feeling’s deep
None of us married to his proposal, make us feel cheap
Still and sad, distraught and mad, tell the neighbor about it
Bet they agree, parade the streets with your voice proudly
Time passing, things change
Reverting back to our daily programs
Stuck in our ways, lust”

Kendrick L. Duckworth *in* Lust

# Contents

# List of Figures

# List of Tables

# List of Listings

# Glossary

Docker is a platform as a service (PaaS) product that uses OS-level virtualization to deliver software in packages called containers 30, 32, 42, 43

Docker Compose is a tool for defining and running multi-container Docker applications 32, 43

Gensim is an open-source library for unsupervised topic modeling and natural language processing, using modern statistical machine learning 60, 75

GitHub Pages is a static web hosting service offered by GitHub since 2008 to GitHub users 33

hydrate In the context of APIs, hydrating means getting the complete object from an incomplete version of that object by means of an API call 39

Jupyter is an open-source project providing interactive computing notebooks and other services xiii, 29, 35, 38

MongoDB is a general purpose, NoSQL document-based, distributed database built for modern application developers and for the cloud era 30, 32, 34, 40, 41, 45

Neo4J is a native graph database, built from the ground up to leverage not only data but also data relationships 30

Pushbullet is a service that provides, among other services, a push notification API, along with web and mobile clients to receive them 34, 38

Sklearn (*a.k.a.* Scikit-learn) is a free software machine learning library for the Python programming language 60, 61

State of Emergency is a government self-declared situation that empowers the government to implement policies that it would otherwise not be able to instate 48, 49, 53

TensorFlow is an end-to-end open-source platform for Machine Learning 82

Twittersphere is the collective group of Twitter users and their interactions 2, 10, 27, 31, 40, 75, 86, 87, 90–93

# Abbreviations

# Chapter 1

# Introduction

"I've never been a good estimator of how long things are going to take."

Donald Knuth

## 1.1 Context

Twitter [1] is one of the most popular Online Social Networks (OSNs) in existence, having more than 149 million daily active users worldwide, as of Q3' 2019 [2]. Due to its widespread adoption and the social backgrounds it encompasses, it has also become a prime platform for political discussion. Simultaneously, it exists alongside a fully-featured Application Programming Interface (API) [3] that can be used, among other things, for automating account interactions with the OSN. This API allows for a multitude of useful real-time tasks, such as public risk monitoring [42], advertisement [31], customer service [40], public opinion tracking [68], and stock market prediction [68], to name a few.

Twitter and its massive adoption can be traced back to several significant moments in recent history. First, we can trace it back to the creation of the internet by Tim Berners-Lee in 1989, which was later reported in his infamous 1992 paper [9]. By 1997, the first version of OSNs appeared [47]

[1] https:/twitter.com

[2] https://investor.twitterinc.com

[3] https://developer.twitter.com/en/docs

– as first attempts at using the power of this connected new world. SixDegrees [4] and AsianAve [5] are two primordial attempts at that. Since then, many new Social Networks (SNs) have surfaced, each with its specific rules and environment, their adoption rates growing with the democratization of the personal computer. More so, in the past decades, due to the extremely high penetration smartphones have as a social commodity [6]. The combination of this ease of access and connectivity led to a level of adoption and use of OSNs that made them a constant presence in many people's lives.

## 1.2 Problem Definition

However, Twitter and other OSNs' benefits come with risks. One such risk is that the automation power it provides can be used with ill intent. Increasing the relevance of topics [49], influencing opinions [6], or even suppress the rise of insurgent movements [67] are only a few examples. On top of this, nothing prevents users from creating fake accounts, regardless of having good or bad intentions. Twitter is, therefore, an undeniably complex digital ecosystem that stands vulnerable to both small scale independent political influence as well as to orchestrated attempts at wide-scale political manipulation – some go as far as coining these phenomena as Social Cyber Security (SCS) [10] [28].

## 1.3 Motivation

With all the above considerations in mind, the present work focuses on shining a new light on the process of detecting, analyzing, and measuring the impact of malicious political interference on Twitter. First, there are still a lot of unsolved issues in current research, as we will see in the following chapters. Second, societies all over the world should have tools that allow them to better regulate and monitor political events so that ill-intended actors will not be able to keep on taking advantage of the vulnerabilities in OSNs. Therefore, there is a clear need to evolve the research lines of online opinion manipulation further. This need should be mainly focused on practical application. Due to this, aiming to build a context-free strategy to tackle this problem is detrimental. Most approaches observed nowadays are both limited to context and focus on a granularity level that reinforces their limitations, as will be further explained in Section 3.3 (p. 18).

## 1.4 Goals

This work will focus on understanding what is the current stance in regards to detecting political manipulation campaigns on Twitter, identifying limitations in the state-of-the-art approaches, pursuing a line of research that is oriented towards overcoming these limitations, and applying it to the Portuguese Twittersphere.

[4] http://sixdegrees.com/

[5] https://en.wikipedia.org/wiki/AsianAve

[6] https://en.wikipedia.org/wiki/List_of_countries_by_smartphone_penetration

## 1.5 Hypothesis

This work aims at testing the following hypothesis:

> Do different high-level approaches to Twitter data analysis, when applied to distinct types of malicious activity, a) lead to meaningful results and b) agree among themselves?

In this case, high-level will refer to approaches that avoid the account-level analysis, like building classifiers based on the features of one account, and instead focus on looking at substantially larger amounts of data pertaining to a larger set of accounts. In this sense, we look at data with an intent of uncovering patterns that are related to the behavior of the overall pool of accounts analyzed, or specific to custom account types within that pool. Although the full explanation for choosing the hypothesis is a consequence of the work presented in Chapter 3 (p. 15) and Chapter 4 (p. 23), it is advanced in this section as a means for the reader to have it in mind throughout the remaining parts of this work. For more details on the hypothesis and its validation, please refer to Section 4.3.1 (p. 25).

## 1.6 Structure

This document is divided into six other chapters. Chapter 2 (p. 5) will provide sufficient knowledge of how Twitter works and what some of the fundamental definitions used throughout this thesis. Chapter 3 (p. 15) contains a protocol description followed by the actual execution of a Systematic Literature Review (SLR) oriented towards mapping current research efforts in the fight against social manipulation for political purposes on Twitter. This chapter also works as a catalyst for understanding the limitations of current research. Chapter 4 (p. 23) builds upon this understanding and provides a collection of current approaches that are detrimental for solving the limitations identified in Chapter 3 (p. 15). It also includes a more detailed definition of the problem and the hypothesis, and how it will be validated. Chapter 5 (p. 27) describes a framework for massive structured data collection using the Twitter API. This framework is then used to generate the data analyzed in Chapter 6 (p. 47). Finally, Chapter 7 (p. 89) focuses on summarizing the rest of the work and providing a synthesis of the main ideas presented and conclusions drawn, bringing a full-circle view on how we can look and combine high-level approaches as a useful tool for political manipulation detection.

Chapter 2

# Background

> "A reliable way to make people believe in falsehoods is frequent repetition, because familiarity is not easily distinguished from truth. Authoritarian institutions and marketers have always known this fact."
>
> Daniel Kahneman, *in* Thinking, Fast and Slow

This chapter describes the main concepts required to understand the rest of the work by having a common ground of terminology. It first goes through a more in-depth description of Online Social Networks (OSNs) and Twitter's inner workings. It then delves into what strategies and goals are typically associated with OSNs manipulation, and what main categories exist to describe it.

## 2.1 Online Social Networks

A Social Network (SN) is a collection of people connected by their social relationships and interactions – these have been present ever since humankind's earliest times. With the advent of the World Wide Web [9], however, the underlying idea to SNs soon started being mapped to the online world.

OSNs are defined as "online communities among people with common interests, activities, backgrounds, or friendships" [56]. OSNs date back to the end of the 20th century with the appearance of rudimentary online communities like Six Degrees [7] and others [31]. Since then, the democratization of technology potentiated by the appearance of smart-devices like smartphones led to mass adoption of these virtual environments. OSNs like Facebook [8], Twitter, Reddit [9], Instagram [10], LinkedIn [11] and others have since came up. Figure 2.1 was adapted from [31] and shows a number of OSNs created between 1997 and 2001. These platforms give users a myriad of possibilities like establishing social presence [13], managing contacts [31] or engaging with news [40]. Their continuing success is undeniably linked with their addiction-oriented design [21] and it is not expected that they should lose their users' attention, time and investment any time soon [21].

[7] http://sixdegrees.com/

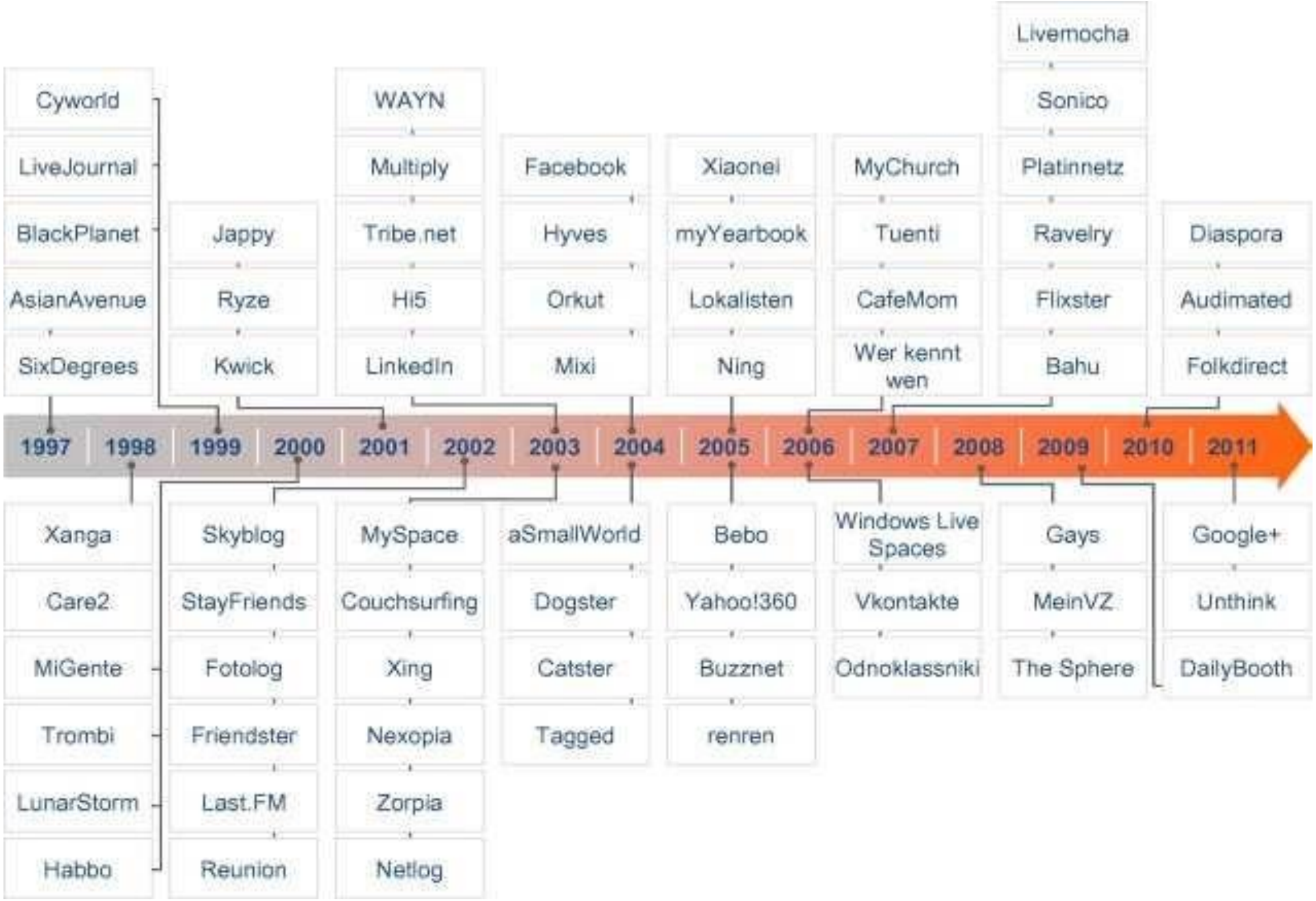

Figure 2.1: OSNs creation timeline

OSNs give people an online persona and allow for interactions with content or people, typically through shared interests. Because of this, they can be represented as graphs. The exact representation varies between different SNs. In Twitter, users can follow other users, unidirectionally. Whereas Facebook friendships have to be bidirectional so that an actual link exists between accounts. Although this is the most typical way of representing an OSN – with users being nodes and their direct connections being edges – it is also possible to build other types of graphs. One example is a graph where Twitter hashtags are the nodes, and an edge exists between two hashtags if both co-occur in a tweet, this is illustrated in Figure 2.2 (p. 7) taken from [62].

---

[8] https://www.facebook.com/

[9] https://www.reddit.com/

[10] http://instagram.com/

[11] https://www.linkedin.com/

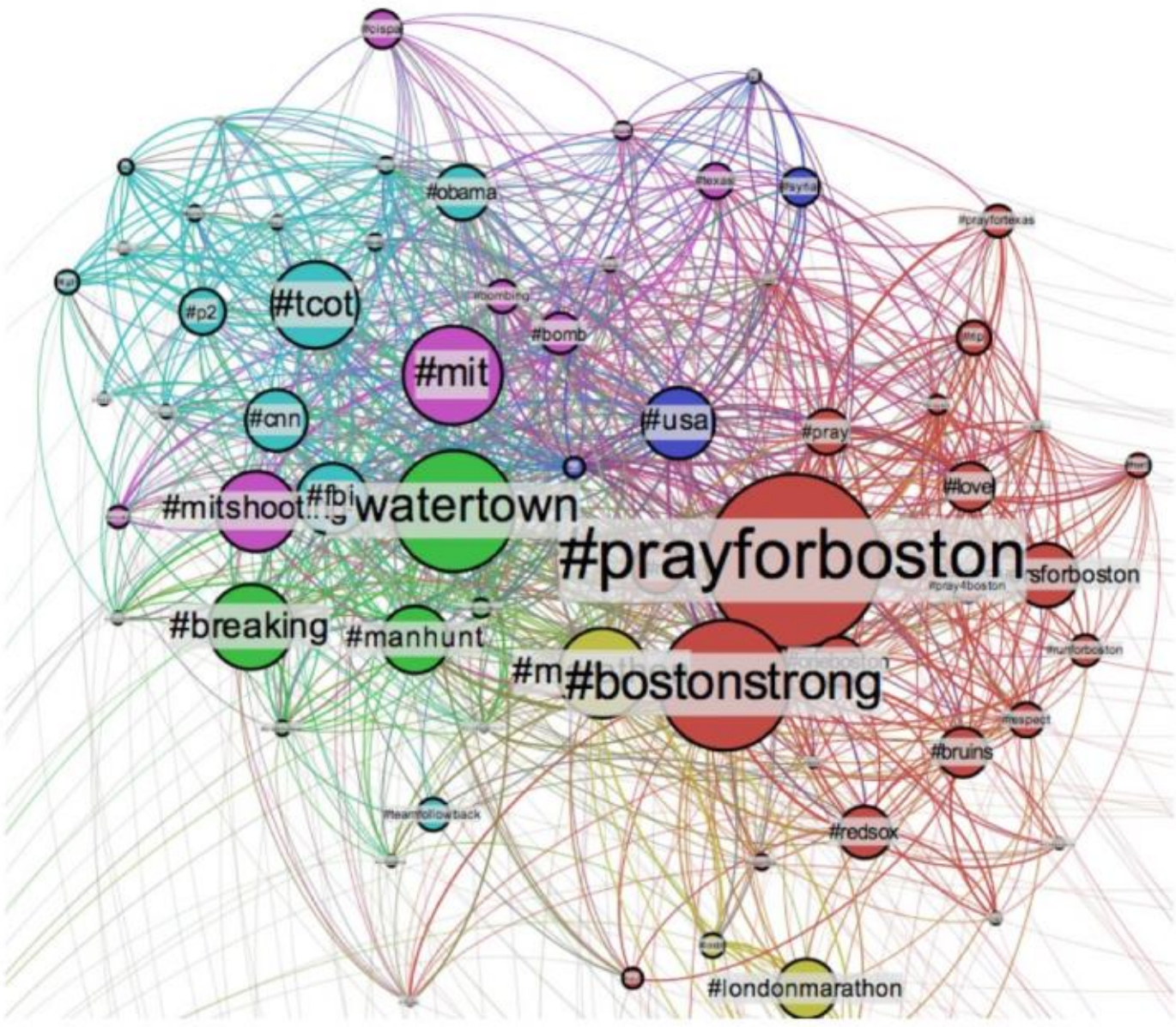

Figure 2.2: Example of a Twitter graph where nodes are hashtags and edges represent hashtag co-occurence

Besides the social relevance of OSNs, they also attract academic interest (as is the case of the current work), public interest – as they can be used for social good [27] – and malicious applications – as we will see in Section 2.3 (p. 10). Figure 2.3 is adapted from [47] and provides a quick taxonomy over OSN types and goals. Table 2.1 (p. 8) shows a comparison in terms of taxonomy and outreach graph types of the OSNs mentioned above.

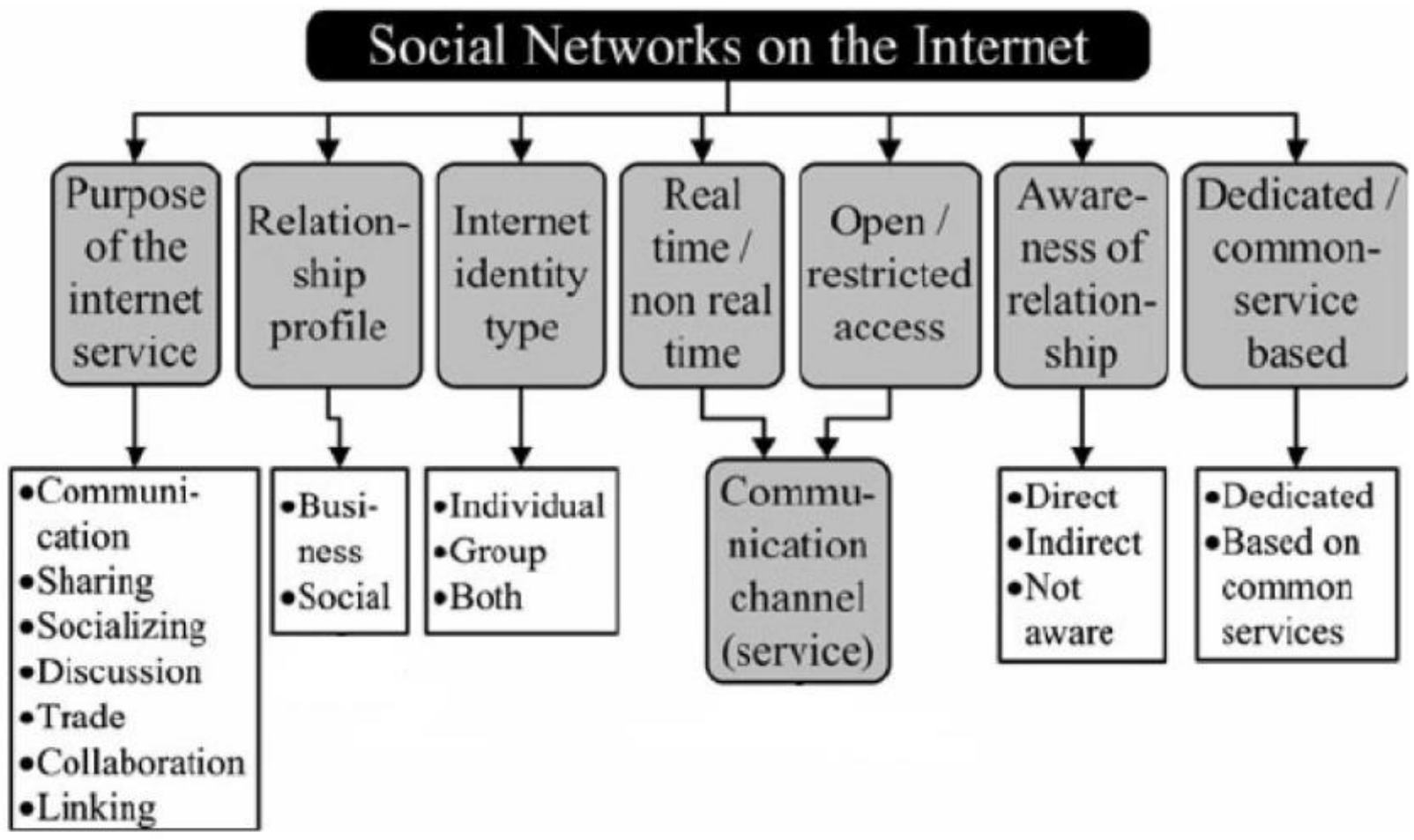

Figure 2.3: OSN taxonomies

Table 2.1: Overview of main OSNs by size and main characteristics

| | LinkedIn | Facebook | Reddit | Twitter | Instagram |
|---|---|---|---|---|---|
| Creation year | 2002 | 2004 | 2005 | 2006 | 2010 |
| Montlhy active users | 260M | 2500M | 430M | 330M | 500M |
| Purpose | S,L | C,S,SO | S,SO,D | S,SO,D | C,S,SO |
| Relationship profile | B | B,S | S | B,S | S |
| Indentity type | both | both | individual | both | both |
| User connections | U | U/D | D | D | D |

*Purpose labels*: S=sharing, L=linking, C=Communication, SO=Socializing, D=Discussion

*Relationship labels*: B=Business, S=Social

*User Connection labels*: U=undirected, D=Directed

## 2.2 Twitter – Rules of the Game

Twitter will be the OSN studied in this work. As such, more attention is dedicated to understanding its inner workings, how its data can be accessed and how behavior can be automated on it.

### 2.2.1 History

Twitter is a microblogging service [40] released on July 15th, 2006. Its initial format allowed users to create an account and post 140 character-long messages [40]. These messages are called tweets. More recently, in 2017, this limit changed. Nowadays, tweets can have up to 280 characters.
Officially, this is meant to allow people to express themselves more easily [12].

Besides writing tweets – tweeting – users can also follow other accounts, retweet, reply to tweets, like tweets, use mentions and hashtags, and search. As users start following other accounts, their timeline represents what those accounts do on the OSN. A retweet is a way of replicating what someone else has tweeted on their timeline. If the retweet includes a comment it is called a quote tweet or retweet with comment. Those who follow a user can see these retweets in their timelines. A Retweet is conceptually a tweet that links back to another tweet. Replies are similar, but their goal is to engage in discussion in the context of a tweet, leading to a discussion thread for a given tweet. Likes represent support for a tweet and will also appear on the user's timeline. Every tweet text can include mentions and hashtags. Mentions are of the form @twitter and will reference an existing account from inside the tweet. Similarly, hashtags appear as #portugal and represent topics – Twitter uses them to calculate trending topics and also to help people search for relevant content [40]. For a more detailed explanation of Twitter real-world usage, please refer to The Twitter Book [53]. Figure 2.4 (p. 9) shows the relationships among the different entities and is taken from [12] and it should be noted that the arrow symbols are merely the direction of the relationships and not Unified Modeling Language (UML)'s inheritance symbol.

[12] https://blog.twitter.com/en_us/topics/product/2017/tweetingmadeeasier.html

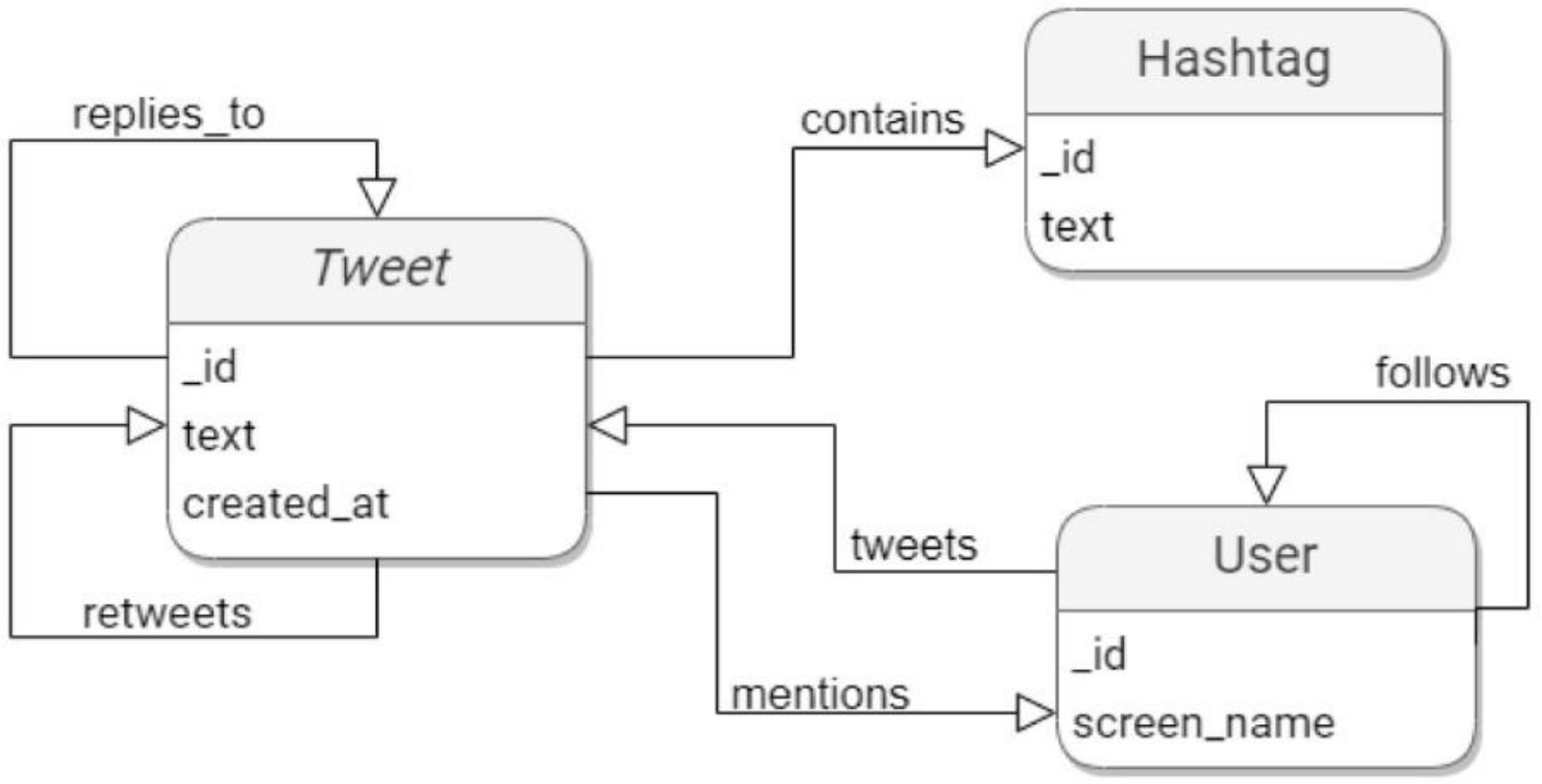

Figure 2.4: Simplified model of Twitter entities and relationships

Nowadays, Twitter is used by individual users, but also by news agencies, companies, personalities, and politicians. However, there is also another type of account: bots. These are automated accounts that can be used for giving regular weather updates like @bbcweatherbot but also for other, not as virtuous, purposes, as described in Section 2.3 (p. 10).

### 2.2.2 Application Programming Interface

Twitter has a Representational State Transfer (REST) Application Programming Interface (API) [7] for developers, researchers, and other agents. The Twitter API is complete, meaning any action that can be performed via Twitter's User Interface (UI) can also be performed through the API. Additionally, the API also contains functionalities oriented towards managing businesses, and these will be neglected here as they are irrelevant for research purposes.

In terms of data collection, the most common actions performed are searching (for users, hashtags, or just keywords), getting information for a given account (followers, friends), getting the timeline for that account, getting the likes and retweets of a given tweet and also the *retweeters* and *likers*. For a full description of the endpoints and returned results, please refer to the Docs [8].

The standard (free) version of the API's search endpoint returns only a 1% sample of recent Tweets in the past seven days. In practice, this does not limit any studies, since we can have a complete timeline of tweets from any specified user, it is only the search by term or hashtag that is restricted to the one week window and 1% sample. Another important aspect is the rating limits. These somewhat limit the amount of raw data one can fetch over time but can be accommodated by waiting for the limits to expire (typically in 15min windows), or taking advantage of the independence between API endpoints.

[7] https://developer.twitter.com/
[8] https://developer.twitter.com/en/docs

## 2.3 Manipulation Strategies

Much as a consequence of the automation freedom the Twitter API provides, wicked applications quickly rose in the Twittersphere. Although this automation is under strict regulation [13], the problem remains. This section contains more information on the types of non-allowed behavior that is typically present on Twitter, what concepts surround it, along with useful nomenclature.

### 2.3.1 Misinformation and Disinformation

Shannon's Information Theory [59] represents the earliest significant attempt at a mathematical definition of information. Despite its relevance in terms of data communication, this definition is too strict. In the context of OSNs, information needs to be interpreted in its well-known variant, or as described in [64], a "communicative construct which is produced in a social context". Indeed, in terms of SNs, it is hard to escape the concepts of information, misinformation, and disinformation.

In [36], misinformation is defined as a type of information since it has no inherent constraint on its veracity. Misinformation can be defined as simply as incomplete information [43] or as thoroughly as inaccurate, uncertain, vague, or ambiguous [36] [37].

When misinformation is produced with the intent of spreading a falsehood, it is treated as disinformation. In [22], it is argued that disinformation is not a subset of misinformation since a piece of information that is true but is still misleading is indeed disinformation and might not qualify as misinformation – disinformation needs only be misleading, according to those authors. In this work, however, we separate the concept of disinformation and misinformation by its deceptive intent or lack thereof, respectively.

Table 2.2, adapted from [36], contains a comparison of these three concepts. The label Y/N should be interpreted as context and time-dependent.

Table 2.2: Comparing Information, Misinformation and Disinformation [36]

| | Information | Misinformation | Disinformation |
|---|---|---|---|
| True | Y | Y/N | Y/N |
| Complete | Y/N | Y/N | Y/N |
| Current | Y | Y/N | Y/N |
| Informative | Y | Y | Y |
| Deceptive | N | N | Y |

Y=Yes, N=No, Y/N=Yes and No, depending on context and time

[13] https://help.twitter.com/en/rules-and-policies/twitter-automation

### 2.3.2 Fake News

Nowadays, the concept of misinformation and disinformation is closely related to that of fake news. However, a clarification of what is understood by "news" is in order.

For clarity reasons, we shall adopt the view of news as a byproduct of journalism [63] that is, in its common format, an account of recent, exciting, and meaningful events [38]. Consequentially, fake news is the intersection between news and either misinformation or disinformation. In a recent survey [66] of fake news typology, the authors identified the following types of fake news: news satire, news parody, fabrication, manipulation, advertising, and propaganda. Additionally, more pragmatic definitions focus on verifiably false [2] viral posts made on social media that imitate legitimate news reports [63].

### 2.3.3 Trolls vs Bots

A clarification is required given the lack of agreement in the literature as to what is a troll, what is a bot, and what lies between them.

Some authors define trolls as human-managed accounts that have a negative online behavior [30] whose goal is to attract as many responses as possible [20]. Simultaneously, some authors define bots as automated agents with similar goals [26]. Others refer to the term social bots [67] as an indistinct version of either trolls or bots. Furthermore, some of these accounts that post malicious content do so in a hybrid manner by combining automated and possibly centralized command with human-produced content and insights [10] [39]. These techniques aim at staying under the radar since they dilute the most apparent bot activity giveaways.

Since this nomenclature is not universal, we feel we need to clarify how we use it. In this paper, we use the term bot to refer to an automated account; troll to refer to potentially manually managed accounts, even though in practice it is possible to be fooled by automated accounts that resemble human behavior, so this is the one term that can fluctuate in the automation spectrum; malicious account as referring to either trolls or bots; automated account, automated agent, social bot, to refer to bots. So, if we observe a high level of automation, we consider that account to be a bot, otherwise we will use the term troll. Additionally, not all bots are malicious, but all trolls are malicious. With this compromise, we can focus on the effects of malicious behavior rather than on specifying what originated it. Besides, it also lifts the dangerous assumption [28] that these accounts are not managed by a human today and a script tomorrow.

### 2.3.4 Manipulation Nomenclature

Trolling has accurately been described as a Social Cyber Security (SCS) problem [10]. Following that perspective, the authors in [27] have compared trolling with hacking. There is a target – OSNs. There are vulnerabilities – the operating principles of OSNs. There are exploits – that take advantage of these vulnerabilities.

Simultaneously, there is also a distinction between black hat, gray hat, and white hat trolls – as is the case with computer hackers [27]. Black hat trolls are driven by personal gain and can elude ethical issues. Grey hat trolls have the opposite behavior, typically to push altruistic or anti black hat troll agendas. White hat focus on identifying both previous types and report them to appropriate entities so that they can be removed. In this work, our primary focus is that of identifying black hat (malicious) trolls. However, gray hat trolling activity may have comparable properties and can thus also be labeled as trolling by detection techniques.

Another convention defined for this work is that of using the terms manipulation campaign, opinion manipulation campaign, manipulation efforts, malicious campaign, bot campaign, and social bot campaign interchangeably. Although it might not be fully coherent with the black/gray hat distinction in the previous paragraph, it will simplify common terminology without impacting the experimental work.

In addition to the previous section, but with less focus on defining recurrent terms and more on providing an overview of the trolling phenomenon, we present Table 2.3 (p. 13) taken from [27]. This table comprehensively consolidates the multitude of techniques and approaches that trolls can use and is a good reference point to take into account when classifying opinion manipulation campaigns in OSNs.

Table 2.3: List of Trolling Techniques taken from [27]

| Technique Name | Technique Description |
|---|---|
| Source Hacking, “Journobaiting” | Planting false claims or posing as sources to dupe mainstream media, especially in the wake of a crisis event. |
| Keyword Squatting | Associating a rarely used keyword or search term, especially one that becomes suddenly popular in the wake of a crisis event, with disinformation or propaganda. |
| Denial of Service | Overwhelming a public space with information or data designed to promote exhaustion and disaffection, or generally frustrate sensemaking efforts. |
| Sockpuppetry | The creation and management of social media accounts designed to appear as authentic individuals, typically using “data craft” [1]. |
| Persona Management, Botherding | The co-ordination of multiple sockpuppet accounts or their algorithmic automation as a botnet |
| Ironic Hedging / Bait and Switch | Using hateful or extreme language and imagery in a way that creates plausible deniability about intent and empowers messagers and some interpreters to downplay sincerity and seriousness. |
| Political Jujitsu | Soliciting attack from an opponent to elicit sympathy from political allies, ground victimization narratives, facilitate recruitment, and justify counterattack. |
| Controlling the Opposition | Using sockpuppet accounts to pose as representatives of an oppositional group |
| Astroturfing | Using sockpuppet accounts to create the illusion of a bottom-up social movement or social outcry. |
| Wedge Driving | Inserting narratives designed to create divisive infighting among social groups. Often part of an overarching “divide and conquer” strategy. |
| Memetic Trojan Horses | The popularization of seemingly banal content that opens the Overton Window [8] by prompting commentary from mainstream journalists |
| Deep Fakes | Altering photographs and videos to change the original message, in a way that is difficult to detect. |
| Concern Trolling | Disingenuously expressing concern about an issue in order to derail discussion and damper consensus. Posing as a concerned ally or objective third party in order to make criticisms more palatable. |
| Brigading / Dogpiling | Bombarding a targeted individual or space with *ad hominem* attacks from multiple accounts |
| Conspiracy Seeding | Spreading “breadcrumbs” on social media and anonymous forums to nudge participants towards conspiracist thinking |
| Algorithmic Gaming | Exploiting the functioning of an algorithm or related databases to elicit a result not intended by its designers. |

## 2.4 Summary

In this chapter, we have established common ground in terms of background knowledge required to understand the following chapters better. This knowledge includes an understanding of what OSNs are, what are the main OSNs in existence, how Twitter and its API work, known trolling practices and relevant trolling nomenclature.

# Chapter 3

# Systematic Literature Review (SLR)

"Sometimes you will hear leaders say, 'I'm the only person who can hold this nation together.' If that's true then that leader has truly failed to build their nation."

Tim Marshall
*in* Prisioners of Geography

The present chapter describes the protocol and results of a Systematic Literature Review (SLR) conducted with the intent of gaining in-depth knowledge of the types of approaches used nowadays to detect malicious content and its spread within Twitter.

## 3.1 Systematic Literature Review Protocol

This section contains a formal definition of the process used during the SLR. It intends to ensure both the soundness of the procedure as well as its replicability. SpLuRT [14], a TypeScript command-line tool, was chosen to organize the search and filtering stages.

First, we raised three Survey Research Questions (SRQs). Secondly, we constructed a search query aimed at finding academic papers that can answer these questions along with inclusion and exclusion criteria. Then, on November 23rd, 2019, we performed the search on the Scopus [15] and DBLP [16] databases, resulting in a total of 2,787 papers. This initial set was filtered as described in
Figure 3.2 (p. 17), and the final paper count obtained was 26. Each of these papers was analyzed and summarized, although we only include aggregated considerations of their content, namely in Section 3.3 (p. 18). Finally, we propose answers to the SRQs and included detrimental takeaways that will guide the rest of the work in this thesis.

[14] https://github.com/arestivo/splurt

[15] https://www.scopus.com/

[16] https://dblp.org/

### 3.1.1 Survey Research Questions (SRQs)

In the context of uncovering, classifying, and measuring the impact of malicious political disinformation and misinformation efforts on Twitter:

SRQ1 What data types and metrics are currently extracted?
SRQ2 What techniques are being used to process the different data types?
SRQ3 To what end are the analyses conducted?

### 3.1.2 Search Methodology

The search was guided by the following exclusion and inclusion criteria.

#### Exclusion Criteria

Exclude papers that are:

- not written in English;
- missing Digital Object Identifier (DOI);
- marked as Informal Publications (I.P.);• marked as Trade Journals (TJ);
- surveys.

#### Inclusion criteria

Include papers that:

- are at least as recent as 2016;
- have a title related to the SRQs;
- have an abstract related to the SRQs;• are cited according to the following rules:
    - ≥ 10 if from 2016;
    - ≥ 5 if from 2017;
    - ≥ 3 if from 2018;
    - ≥ 0 if from 2019.

### 3.1.3 Search Query

Below is the search query constructed to look for papers that helped answer the SRQs:
3.2 Search

```
twitter AND (politics OR political OR election) AND (bot OR troll
OR
↳agents OR actors OR ``fake news'' OR misinformation OR
↳disinformation OR ``information operation'')
```

## 3.2 Search

Figure 3.1 presents a diagram of the filtering steps and includes information on how many papers were excluded in each step.

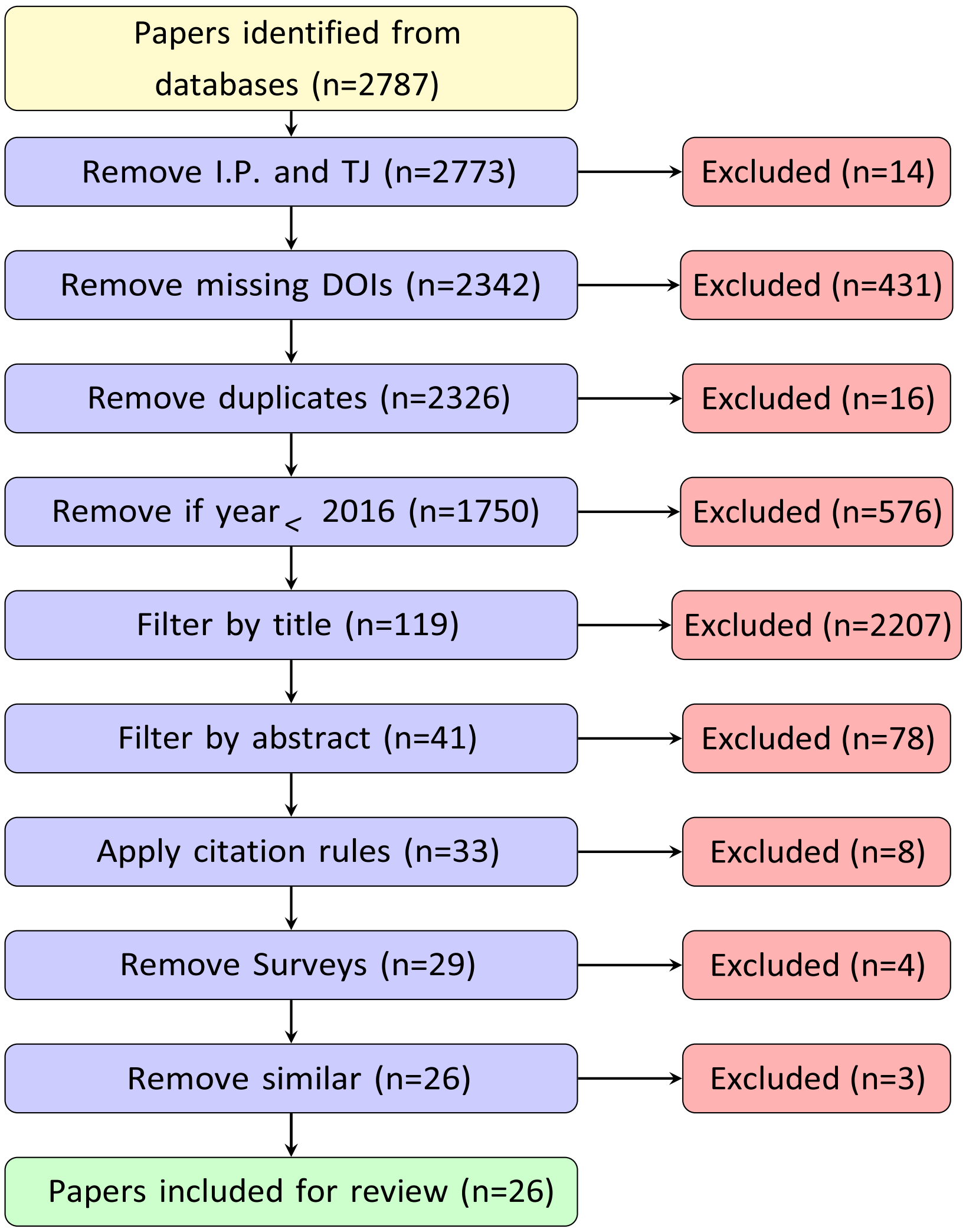

Figure 3.1: Diagram of the exclusion and inclusion criteria application for the SLR

In total, 26 papers were identified for review. The next section contains an aggregated analysis from all of them.

## 3.3 Discussion

This survey constituted an enlightening research effort. It served as a structured way of answering the SRQs, to identify trends, new approaches, and systematic flaws in the research. This discussion focuses on clarifying each of the above points.

### 3.3.1 SRQ1 – What data types and metrics are currently extracted?

Although the surveyed papers share some common nomenclature in terms of the types of features used, there is no one major classification system. As such, we are proposing a standard set of feature types that cover the main sources of features. It should be noted that a large subset of features can be calculated and engineered, and thus we focus on the goals of the features used, rather than on their original nature. The proposed feature types are:

Account Includes features that can be read directly from the account metadata returned by the Application Programming Interface (API) like creation date, username, description # (number of) lists, and verified status.

Content Includes features derived from tweet content, like #words, vocabulary, sentiment, semantic representation, topics, URLs used, #hashtags, #mentions, and writing style.

Network Includes features that describe accounts in terms of presence in the network like #followers, #followees, #triangles it belongs to, centrality, social interactions, network presence, adapted H-index [32], and Pagerank [48].

Activity Includes behavior metrics like frequency of posts, active hours, active days of the week, and #posts per day.

Others This class serves to include custom metrics like political orientation labels, or labels taken from other tools.

Table 3.1 (p. 19) contains an aggregated view of the feature classes used by each paper that handles these features, regardless of the goal.

3.3 Discussion

Table 3.1: Comparison of feature types used per paper (excluding 3 papers with no feature use reported)

| Paper | Account | Content | Network | Activity | Others |
|---|---|---|---|---|---|
| [24] | | • | • | • | |
| [26] | • | | • | • | |
| [46] | | • | • | • | |
| [58] | | | • | | |
| [7] | • | | | | • |
| [60] | | | • | | |
| [10] | • | • | • | • | |
| [29] | | • | | | |
| [57] | | | • | | • |
| [57] | | | • | | • |
| [65] | | • | • | | |
| [39] | | • | | | |
| [61] | | | • | | |
| [14] | | • | | | |
| [34] | • | • | • | • | |
| [4] | | • | | | |
| [18] | • | • | | • | • |
| [25] | | • | • | | |
| [50] | • | • | • | • | • |
| [17] | • | | | • | |
| [54] | | | | • | • |
| [66] | | • | • | | |
| [23] | | • | • | | |
| [69] | • | • | • | • | • |
| Total count | 8 | 15 | 15 | 10 | 6 |

Although it makes little statistical sense to assume this sample is representative of the whole set of papers, we can report a few curious observations. First, even though account data is straightforward to obtain and use, it is not as used as content or network. Second, only four papers try to use all of the data types (excluding others); this might be a good indicator that the literature does not focus on using extensive amounts and sources of data.

### 3.3.2 SRQ2 – What techniques are being used to process the different data types?

We have identified several groups of techniques. These groups are agnostic to the goals of the studies. For instance, in [50], the authors used Decision Trees (a technique for Machine Learning (ML) classification or regression tasks) as a stepping stone to measure feature relevance. One of these groups – specialized tools – is dedicated to external tools and not techniques.

The identified groups are ML classifiers, data representation, community detection, sampling techniques, and specialized tools. Below is a list of each technique per group followed by the papers in the study that use them. This list excludes libraries and frameworks that are not relevant to the context of the study. Each group of techniques is sorted by decreasing number of papers that use them.

### Machine Learning Classifiers

- (7) Neural Networks [58] [39] [14] [34] [18] [17] [23]
- (4) Random Forest [24] [10] [4] [17]
- (4) Support Vector Machine (SVM) [14] [34] [4] [18]
- (3) AdaBoost [46] [39] [4]
- (3) Decision Trees [4] [18] [50]
- (3) Naive Bayes [24] [4] [18]
- (2) Logistic Regression [39] [66]
- (2) Long short-term memory Neural Networks (LSTMs) [39] [54]
- (2) Convolutional Neural Networks [14] [54]
- (1) K-Nearest Neighbors [4]
- (1) SGD classifier [39]
- (1) Sequential Minimal Optimization [24]
- (1) Logit Boost [17]
- (1) Gradient Boosting [34]

### Data Representation

- (5) Word2Vec/Dov2Vec [14] [25] [54] [66] [69]
- (4) Latent Dirichlet Allocation (LDA) [46] [4] [23] [69]
- (3) TF-IDF [14] [54] [66]
- (2) GloVe [39] [66]
- (1) BERT [14]
- (1) Relief Algorithm [18]

### Community Detection

- (4) Louvain clustering [65] [61] [25] [23]
- (2) Label Propagation (LPA) [58] [25]
- (1) Link Communities [25]
- (1) Oslom [25]

## 3.3 Discussion

### Sampling Techniques

- (3) SMOTE [39] [34] [17]
- (1) Edited Nearest Neighbours (ENN) [39]
- (1) TOMEK [39]

Specialized Tools

- (4) DeBot [26] [57] [50] [54]
- (3) Botometer/Botornot [7] [28] [50]
- (2) BotWalk [26] [50]
- (1) Sentimetrix [58]
- (1) Sentistregnth [4]
- (1) ORA Network Tool [65]

All in all, we have the ML techniques group as being more prevalent than the others. This over-representation is explained by the fact that the majority of papers filtered focus on creating bot detection systems. It is also interesting to see that embedding-based techniques like Word2Vec, GloVe, and BERT are standard for semantic representation of text, but not for other types of representations.

### 3.3.3 SRQ3 – To what end are the analyses conducted?

We have found that the main research lines are: performing data-oriented *ad-hoc* analyses for a given context, developing bot (or similar) classifiers. Others focus on using Social Network Analysis (SNA)-inspired approaches (like community detection), or on creating consensual taxonomies for future research.

### 3.3.4 Conclusion

The most valuable takeaway from this study is identifying a systematic fallacy in the literature. Most approaches either lack the eyesight to acknowledge the well-defined scope of their data and models, or lack the will to test it. The fallacy is (wrongly) assuming that bot behavior is not evolving and adaptable to detection systems when designing the very systems that aim to thwart bots. Some authors report dataset-specific conclusions like “bots generally retweet more often” [26] or “social spambots and fake followers are more inclined to post messages with a positive sentimental fingerprint” [4] as eternal postulates. These are taken as absolute truths when they are perfectly susceptible to adaptive strategies taken by bot creators. Some authors recognize this limitation [69] [10]. In [28], the authors start by asking “whether the detection of single bot accounts is a necessary condition for identifying malicious, strategic attacks on public opinion, or if it is more productive to concentrate on detecting strategies?” and then follow through with suggesting a shift in the current approaches to bot detection. It is precisely this conclusion that we take from this SLR process.

Furthermore, we believe that high-level approaches are detrimental for developing robust approaches to detect political interference on Twitter. The most promising ones focus on capturing the evolution of the Online Social Network (OSN) rather than on pinpointing automated accounts, these include the high-level *ad-hoc* analyses but also the community detection efforts such as the ones in [25] [23]. As such, the remaining of this work, will focus on testing different high-level approaches that focus on leveraging data features that are not susceptible to adaptive strategies, or at least at using the ones that seem less so.

Chapter 4

# Problem Statement

"Power is okay, and stupidity is usually harmless.
Power and stupidity together are dangerous."

Patrick Rothfuss
*in* The Name of the Wind

## 4.1 Assumptions

Now that we have a good understanding of the context, the main approaches found in the literature, and their limitations, we can focus on defining the problem we will tackle and how to do so. Firstly, it is good to restate that approaches which leverage bot detection fall short of scalability and are highly susceptible to adaptive strategies. Secondly, we conclude that the problem of opinion manipulation on Twitter should be tackled from a high-level perspective, by taking a *bird's eye view* of the system – leading us to Assumption 1:

Assumption 1. *High-level approaches (to uncover political manipulation on Twitter) are less prone to irrelevant patterns and conclusions than fine-grain ones.*

Thirdly, we have seen that many insights on how to detect bots, taken from previous studies, are quite clearly only relevant for a small amount of time. Such is due to the level of effortlessness that exists for bot creators to overcome them. Examples include the inconsistency of daily schedule activity with that of humans, lack of personal data, or even ones that require some more commitment like account age. Taking this into account, we establish Assumption 2:

Assumption 2. *The best way to design scalable and durable methods (of manipulation detection on Twitter) is to use data features that are harder for malicious orchestrators to manipulate.*

The insights from Chapter 3 (p. 15) lead to the conclusion that two things are hard to forge and manipulate. The first one is the content of what is said – if the goal is to change someone's mind on a topic that topic needs to be covered. The second one is influence, presence or interaction – if the goal is to reach many users, or a particular group of users, it becomes unavoidable to establish a strong presence and reach in the network. Therefore we shall focus on using data that contains latent information on these two inescapable facets of manipulation in Online Social Networks (OSNs).

All in all, the problem we investigate is assumed to require an approach that is both high-level and as hard to manipulate as possible. With these two pillars, we now have a properly defined foundation for understanding the origin of the hypothesis stated in Section 1.5 (p. 3):

> Do different high-level approaches to Twitter data analysis, when applied to distinct types of malicious activity, a) lead to meaningful results and b) agree among themselves?

To answer it, we will conduct three different exploration efforts: one content-oriented (*cf.* Section 6.3.1, p. 59), one metadata-oriented (*cf.* Section 6.3.2, p. 74), and one structure-oriented (*cf.* Section 6.3.3, p. 80). These are our high-level approaches that focus on using hard to manipulate data, with metadata representing the weakest of the three in terms of manipulability. Although we have already described the meaning of high-level in this context (*cf.* Section 1.5, p. 3), we have not introduced the meaning of distinct types of malicious activity. These distinct types correspond to groups of accounts that are considered malicious for distinct reasons. Section 6.2 (p. 52) contains a detailed description of each of these and how they are obtained.

### 4.1.1Preliminary Study

Before we can answer our hypothesis, however, we need to understand each high-level approach individually. This is where we introduce the need for a preliminary study. This study consists in understanding what type of information we can extract from each approach, and how meaningful it is, essentially answering the first point of our hypothesis. Only then will we be able to ascertain the second point – the existence of agreement between the results of each approach. To achieve this goal, we will delve into understanding the information gathered, for each of the three approaches, in their respective sections; and then briefly recall this in their individual summary, in: Section 6.3.1 (p. 59), Section 6.3.2 (p. 74), and Section 6.3.3 (p. 80) corresponding to content-oriented, metadata-oriented, and structure-oriented, respectively.

## 4.2 Hypothesis Plausibility

At this point, it becomes clear that the hypothesis stems naturally from the analyzed literature. It requires existing techniques like semantic content representation, embeddings, clustering, and other forms of representation. These techniques can easily be combined and tweaked to the specific context at hand. The existence of previously mentioned similar works and also parallel research efforts further support that this is neither a lost cause nor a finished one. Thus we argue that the presented hypothesis is plausible.

## 4.3 Related Work

Although no single approach found in the literature attempts to validate our hypothesis, many recent works touch subtasks that we shall have to go through in order to validate it.

In terms of uncovering topics from textual content, one of the most standard approaches is Latent Dirichlet Allocation (LDA) [11] – a statistical model that was significantly used (*cf.* Section 3.3.2, p. 20) and even referred to in the Systematic Literature Review (SLR) process. More recent works have been using word embeddings to uncover topics in streaming data [16] and even as a stepping stone for topic inference and correlation [35].

The authors of [55] propose an algorithm that is particularly relevant for the present case, namely due to its ability to combine content and structure, allowing us to measure the aforementioned duality of content and network influence. [23] shows another attempt at combining content and network information, especially interesting as they also use LDA for topic detection, but introduce a temporal notion into their model. Other approaches focus on performing community detection on custom-built graphs that already embed content and topology [61] [25]; this allows for more traditional algorithms to be used. Other alternatives include deep-learning-inspired Generative Adversarial Networks (GANs) [52] that can generate overlapping communities where each vertex has a strength that describes its level of membership towards a given community. In terms of evaluation, it is also relevant to consider the work of [25] that proposes a new metric to evaluate the Internal Semantic Similarity (ISS) of communities.

Along our work, we will complement each of our approaches' choice with references to other related work that, for being related only to particular tasks and decisions, is found to be more beneficial when located at the moment it becomes relevant.

### 4.3.1 Hypothesis Validation

In order to validate the proposed hypothesis, we will use validation at intermediary steps in the process of developing Machine Learning (ML) models. We will gather Twitter data and enrich it with labeled examples obtained from different sources, in order to ensure our results are not susceptible to biased data. Most of our exploratory work will focus on comparing different highlevel strategies, and their results will reflect their potential viability in validating our hypothesis. In this sense, it is hard to establish one single metric that will reveal whether our exploration processes Problem Statement works in different contexts. We find it relevant, however, to establish them nonetheless. Especially in the hopes that future work in our research line will continue revealing the feasibility of high-level approaches that focus on content and interaction. One final aspect of our validation process is that we are collaborating with a peer working on a parallel problem – creating a troll classifier for the Portuguese context – and our results will be compared to theirs in an attempt to mitigate the difficulty in non-abstract validation methods.

Even with respect to other works that focus on validating bot detection models, it remains hard to validate them, knowing it is commonly accepted they are validated against incomplete example sets. In fact, it is known [29] [46] that the accounts that Twitter suspends and blocks are found with a conservative mindset – their goal is to have high precision and not high recall.

Overall, we will compare the results of our independent exploration efforts to ascertain if their results make sense when interpreted together. We will also rely on multiple sources of labeled data – one of which a peer working in a similar issue – to mitigate dangers of biased data. Finally, we will perform most of our data collection and then data exploration with a focus on ensuring that both data quality and the results reached are statistically significant and free from typical mistakes and biases found in similar works.

## 4.4Research Methodology

The main goal of this work is to push the current state-of-the-art of detecting malicious political content on Twitter further, ideally by producing a visual tool – or a first draft at one – that can put the developed solution in the hands of the public, the media, and Twitter itself.

In order to do so, we shall experiment with state-of-the-art tools for each of the sub-tasks in our work: data collection, data representation, temporal representation of data, topic detection, community detection, and visualization of the evolution of these topic-aware communities over time.

## 4.5Summary

This chapter has drawn a line on where state-of-the-art research is, relating to political opinion manipulation on Twitter, its issues, and potential that more theoretical lines of research contain. We have stated three assumptions that will support the investigation of the hypothesis. We have argued that this is a plausible hypothesis and how we shall validate it, even with the known constraints.

# Chapter 5

# Twitter Watch

---

---

“Humans were always far better at inventing tools than using them wisely.”

Yuval Noah Harari
*in* 21 Lessons for the 21st Century

The first requirement for analyzing Twitter data is, well, having data. Although it is not uncommon for researchers to reuse already existing datasets for the validation of new approaches [14] [23], this option was not viable to us, as the most recent dataset focusing on the Portuguese Twitter context that we could find was from 2016. At the time of starting this work, we wanted to focus on the 2019 Portuguese legislative elections, occurring on October the 6th, 2019. Since no data was available for immediate use, we had to collect it. This section reveals how we achieved this goal and also how and why we developed a new Twitter data collection framework – Twitter Watch.

## 5.1 Tool Gap

As specified in Section 2.2.2 (p. 9), Twitter has an Application Programming Interface (API) that can be used for a multitude of purposes, we will focus solely on data collection for research purposes. Since our goal is to focus on a specific context – the Portuguese Twittersphere – and on a particular political event – the 2019 legislative elections – we needed to find a tool capable of extracting a relevant dataset to conduct our study. We did not find such a tool. Indeed, we listed the requirements that such a tool should fulfill in order to generate a relevant dataset of both users and tweets:

Data Collection Requirements 5.1: Initial requirements

1. Capture a specified period of time;
2. Start from a list of relevant Twitter accounts and dynamically find new potentially relevant ones;
3. Restrict the collected data to the Portuguese content as much as possible;
4. Detect suspension of accounts as they occur;
5. Properly explore the API limits, as that is a potential bottleneck;
6. Save the data in a way that facilitates the subsequent analysis;

7. Ensure a coherent structure and avoid redundancy;
8. (Optionally) Allow regular backup of dataset snapshots;
9. (Optionally) Notify failures in real-time;
10. (Optionally) Ensure a logging mechanism to follow both progress and failures during the collection process;
11. (Optionally) Provide a visual interface that facilitates monitoring the collection process;
12. (Optionally) Be adaptable to different collection goals through an easy configuration;
13. (Optionally) Allow flexibility in how the data collection evolves.

With the above requirements in mind, we looked for tools that would either satisfy them or be flexible enough to accommodate them through easy changes.

On one side, we have commercial tools like Hootsuite [17], Sysomos [18], or Brandwatch [19] that are both commercial and abstract the access to the data, but focus on using the search endpoints by looking at hashtags or search terms, as explained in Section 2.2.2 (p. 9) this means only a seven-day window of past data is available. Although this work had some initial efforts of data collection surrounding the election period (*cf.* Section 5.2, p. 29), the usage of such approaches focusing on endpoints and a very narrow window for that collection discouraged the use of both that endpoint as the sole source of data and of tools that relied heavily on it. These observations mean these tools fail many of the mandatory requirements, the most limiting being 1, 4, 6.

On the other side, we have open-source tools like Socialbus [20], TCAT [21], sfm-twitter-harvester [22], or Twitter-ratings [7] that are more oriented towards research data collection. These tools, however, are limited. Socialbus and TCAT are configured to filter tweets by users or by topics/keywords, but these need to be specified beforehand, and any dynamic evolution is relying on restarting the system, therefore not meeting requirement 2. Sfm-twitter-harvester can be interacted with as either a Representational State Transfer (REST) or streaming API, this means it is more flexible but lacks the persistence desired when building a dataset, it can actually be seen as an abstraction layer above the API, and fails to meet requirements like 2, 6. These tools are still found to be too generic and don't add a lot of value for the current use case when compared to the API wrappers available like Tweepy [23] and python-twitter [24].

[17] https://hootsuite.com/

[18] https://sysomos.com/

[19] https://www.brandwatch.com/

[20] https://github.com/LIAAD/socialbus

[21] https://github.com/digitalmethodsinitiative/dmi-tcat

[22] https://github.com/gwu-libraries/sfm-twitter-harvester

[7] https://github.com/sidooms/Twitter-ratings

This initial desire to find a suitable tool was unmet, and the gap remained open. Before actually implementing a solution that would close it by fulfilling the above requirements, we hit some metaphorical walls that are described in the next section as a reference point for anyone interested in achieving a similar effort.

## 5.2 Failures

Initial approaches were *ad-hoc* and, unsurprisingly, faulty. Even before starting to collect data, we focused on understanding Twitter API, its different response formats and objects [25], endpoints [26], rate limits [27] and response codes [28]. In the end, we also developed a simple open-source scraper [29]that generates two JavaScript Object Notation (JSON) files containing error and response codes that can be used to better understand the interactions with the API.

Then, having chosen python-twitter [15] as the API wrapper to use for the collection process, we identified all the accounts from the Portuguese political context that fell into one of the following categories:

- Political party account;
- President of a political party.

The final number of accounts found was 21, and this process was conducted in October 2019.

Starting from this seed of accounts we designed a single page Jupyter notebook that underwent the following phases of:

1. (A) Get updated profile information on the seed accounts;
2. (B) Get all the followers of (A);
3. (C) Get all the followees of (A);

---

[23] https://github.com/tweepy/tweepy

[24] https://github.com/bear/python-twitter

[25] https://developer.twitter.com/en/docs/tweets/data-dictionary/overview/ intro-to-tweet-json

[26] https://developer.twitter.com/en/docs/tweets/search/api-reference

[27] https://developer.twitter.com/en/docs/basics/rate-limits

[28] https://developer.twitter.com/en/docs/basics/response-codes

[29] https://github.com/msramalho/twitter-response-codes [15]https://github.com/bear/python-twitter

4. (D) Get the 25 most recent tweets from (A);
5. (E) Get the retweeters of (D);
6. (F) Get the retweets of (D);
7. (G) Get the 25 most recent tweets from (B);
8. (H) Get the retweeters of (G);

9. (I) Get the retweets of (G);
10. (J) Get the 25 most recent tweets from (C);
11. (K) Get 10 tweets for every user in the database that did not have collected tweets;
12. (L) Calculate the 10,000 most common hashtags;
13. (M) Use the search API to get 200 tweets for each hashtag in (L).

All the data was saved to a MongoDB instance, which has the out-of-the-box benefit of ensuring no duplicate documents exist, combining this with Twitter object model's `_id` field, means that the redundancy requirement (7 in the requirements list) was easily achieved. We executed this script in a time window that encapsulated the October 6th elections.

Although this dataset is created coherently, there are a few subtle inherent limitations to the way the collection steps are designed. Firstly, by not having the full tweet history, we cannot conduct any analysis on how Twitter's overall usage varied through time (on the Portuguese context). The same goes for analysis of each user's usage patterns, and other analysis that require complete temporal data. Then, there is a hidden assumption that malicious activity will necessarily be within accounts that are either followers or followees of the seed accounts, or of the retweeters identified in (E) and (H). Then, (L) and (M) enrich and add variety, but their benefit is not exploited since no new accounts are expanded from those collected tweets. Also, and in line with a limitation common to all the open-source tools mentioned above, they did not allow for requirement 4 to be met, as the suspension of accounts was not easy to monitor or record. This approach was far from ideal to what we required.

As the initial research focus was on combining textual content with structural information, we also went down another insidious path – attempting to save followers and followees of every account in the MongoDB instance. This proved hard due to quickly reaching MongoDB's maximum document size of 16MB [30]. Working around this was not advisable [17]. A few options were considered, but we ended up going for something outside our comfort zone – Neo4J – a graph database designed specifically to save relationships between entities.

After setting up a Neo4J Docker instance, we started using the old collection process with a few code changes that would ensure the follower/followee relationships would be saved. We had some success, as is visible in Figure 5.1 (p. 31), we were able to capture accounts, and their follow relationships. However, this proved to be yet another dead end when we observed that the database-writes became the bottleneck of the process, and not Twitter's API, no tuning or batch writing solved this problem.

[30] https://docs.mongodb.com/manual/reference/limits/#bson-documents

[17] https://jira.mongodb.org/browse/SERVER-23482

5.3 Architecture

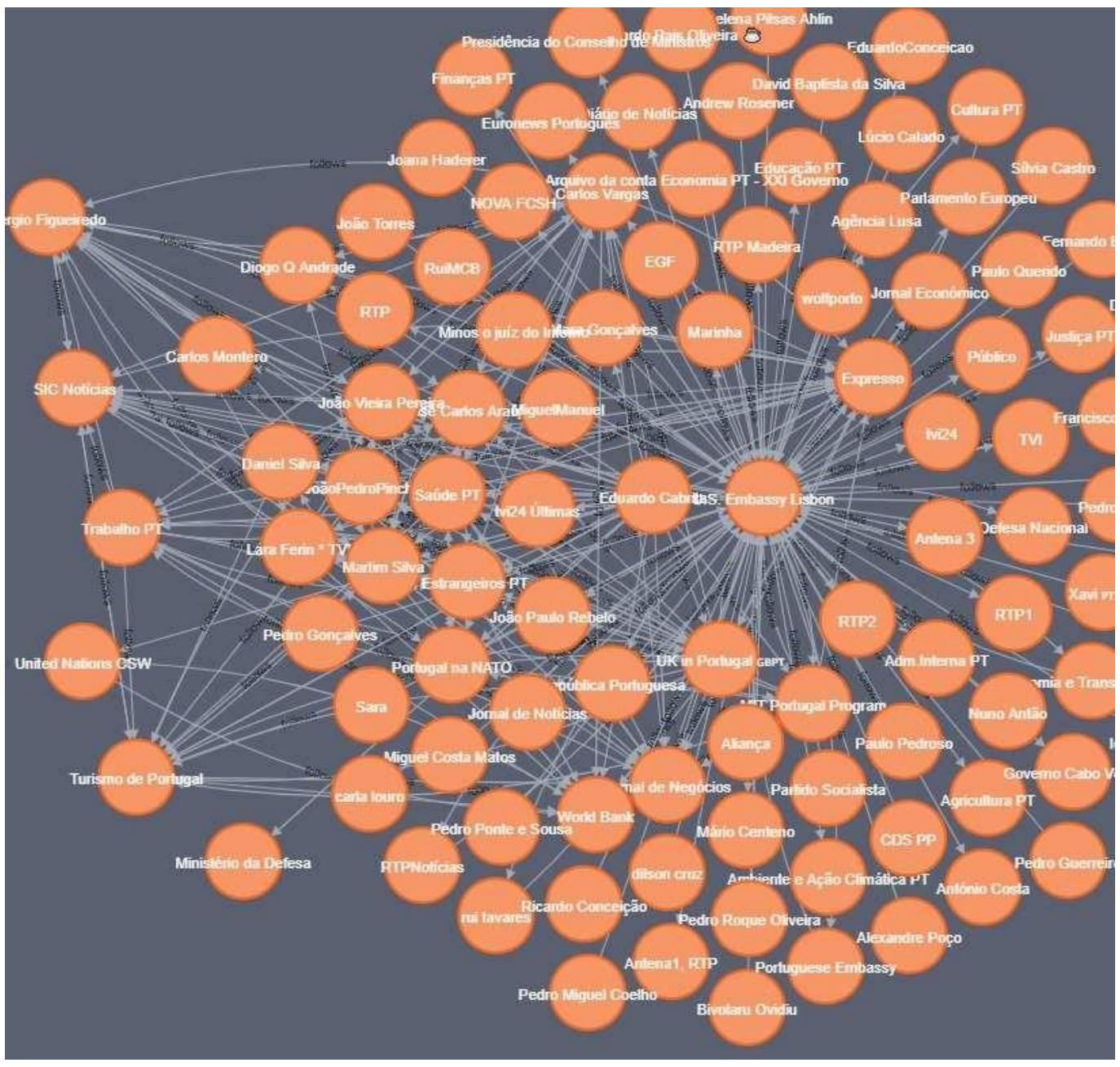

Figure 5.1: Example visualization of Twitter "follow relationships" saved in Neo4j

After having spent a significant amount of time on struggles in the collection process, we took a step back and decided to focus on designing a more deliberate system, even if at a greater time cost, which could answer the requirements above, as well as ensure two new ones:

Data Collection Requirements 5.2: Additional requirements

1. Separate watched from non-watched accounts, the first type consisting of accounts with content posted in Portuguese and ideally within Portugal's Twittersphere;
2. For every account that was marked as being watched, all their tweet history should be collected.

This new system was dubbed Twitter Watch. The next section introduces the architecture designed to answer Requirements 5.1 and 5.2.

## 5.3 Architecture

Twitter Watch's high-level architecture can be split into User Interface (UI), backend, and external services. Figure 5.2 (p. 32) contains a visual representation of this architecture, its components, and their interaction.

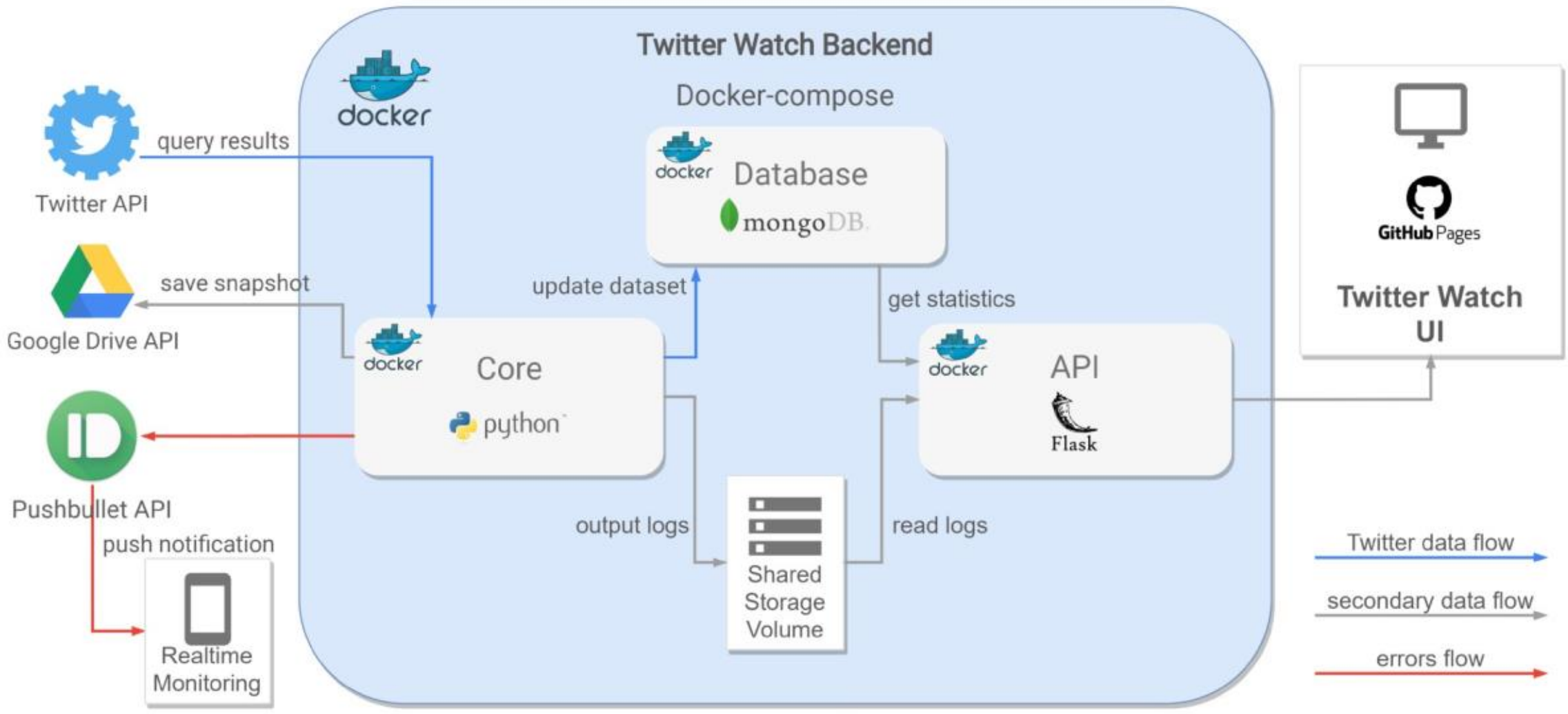

Figure 5.2: Twitter Watch Architecture

### 5.3.1 Backend

The main functionality of Twitter Watch is its backend. This backend is organized in Docker containers. Docker Compose is used to orchestrate these containers and their interactions. The architecture requires three containers, as described below:

#### Database container

This is a standalone MongoDB instance where all the data collected is saved. This container is exposed to the outside of the Docker Compose network mainly so it can be interacted with from the outside even while data is being collected.

#### Core container

The core container contains all the logic behind the collection process, and that will be further detailed in Section 5.4 (p. 34). This container runs a Python-ready environment and includes the vast majority of code developed for the backend.

#### API container

This container is designed to provide information on the data collection process as requested by the UI. It shares a Docker storage volume with the core container so that it can serve the execution logs generated during the collection process. It can also interact with the database container in order to query statistics on the collected data, such as the number of accounts or tweets saved. Flask [18] is the chosen Python framework for implementing this service.

[18] https://flask.palletsprojects.com/en/1.1.x/

### 5.3.2 User Interface (UI)

The UI is a pragmatic effort to streamline the management of the data collection process through an easier and faster diagnosis of the system's state. The UI itself is agnostic to the which Twitter Watch backend it is connecting to and accepts as input the Internet Protocol (IP) address of any Twitter Watch API instance. The most updated version of the UI in production is available at msramalho.github.io/twitter-watch. The UI is hosted on GitHub Pages [31], it was developed with Nuxt.js [20] and Vuetify [32]. In terms of content, the current version has two main pages: statistics and logs.

The statistics page, as seen in Figure 5.3, contains plots of how the number of users and tweets recorded in the database evolve over time, as well as a plot of the database size evolution through time. Additionally, it contains some overall statistics like the current number of users and tweets, and can easily be expanded to accommodate more information.

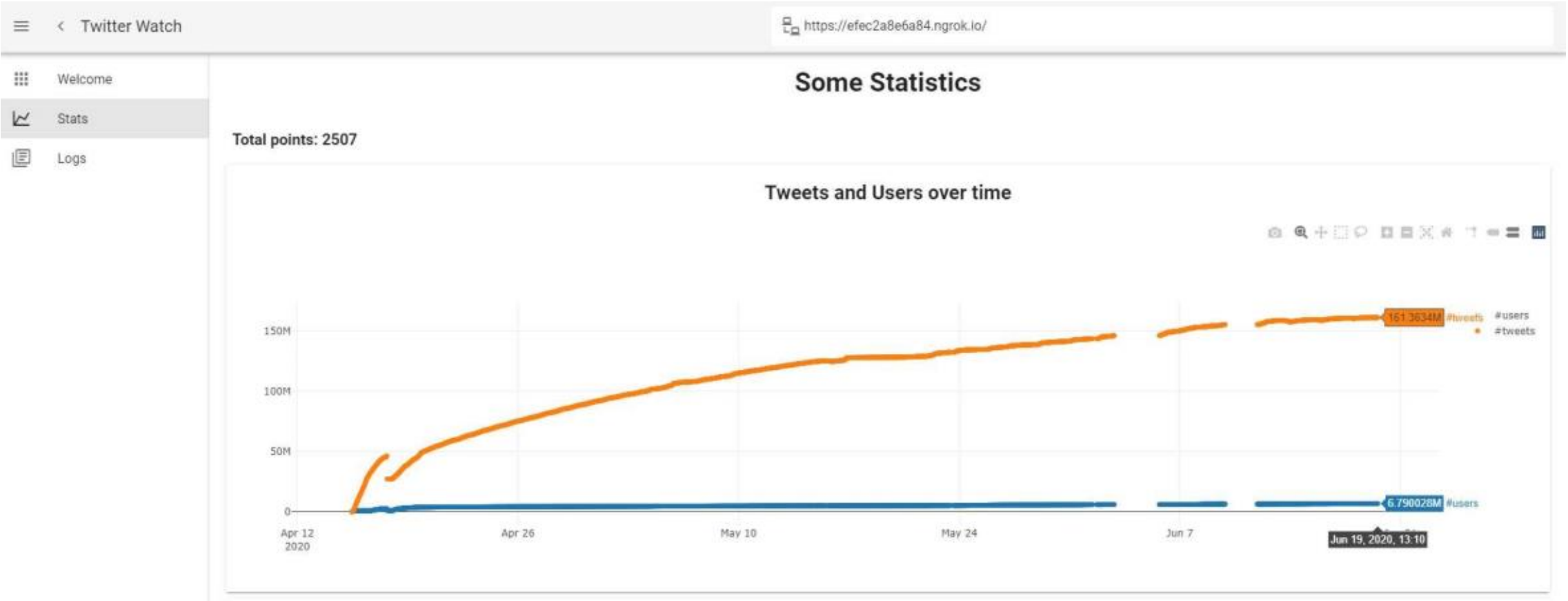

Figure 5.3: Twitter Watch UI statistics page

The logs page, as seen in Figure 5.4 (p. 34), has two main sections. On the left, there is a panel to explore all the different scripts that are run by the backend, as specified in Section 5.4 (p. 34). Currently running scripts are marked with a green dot that, when hovered, displays information on how long they have been executing. Furthermore, expanding any of the scripts reveals a list of the individual logs for the date and time their execution started. When one of these logs is clicked by the user, its output is fetched and displayed on the right panel (*cf.* Figure 5.4, p. 34).

[31] https://pages.github.com/ [20]https://nuxtjs.org/
[32] https://vuetifyjs.com/en/

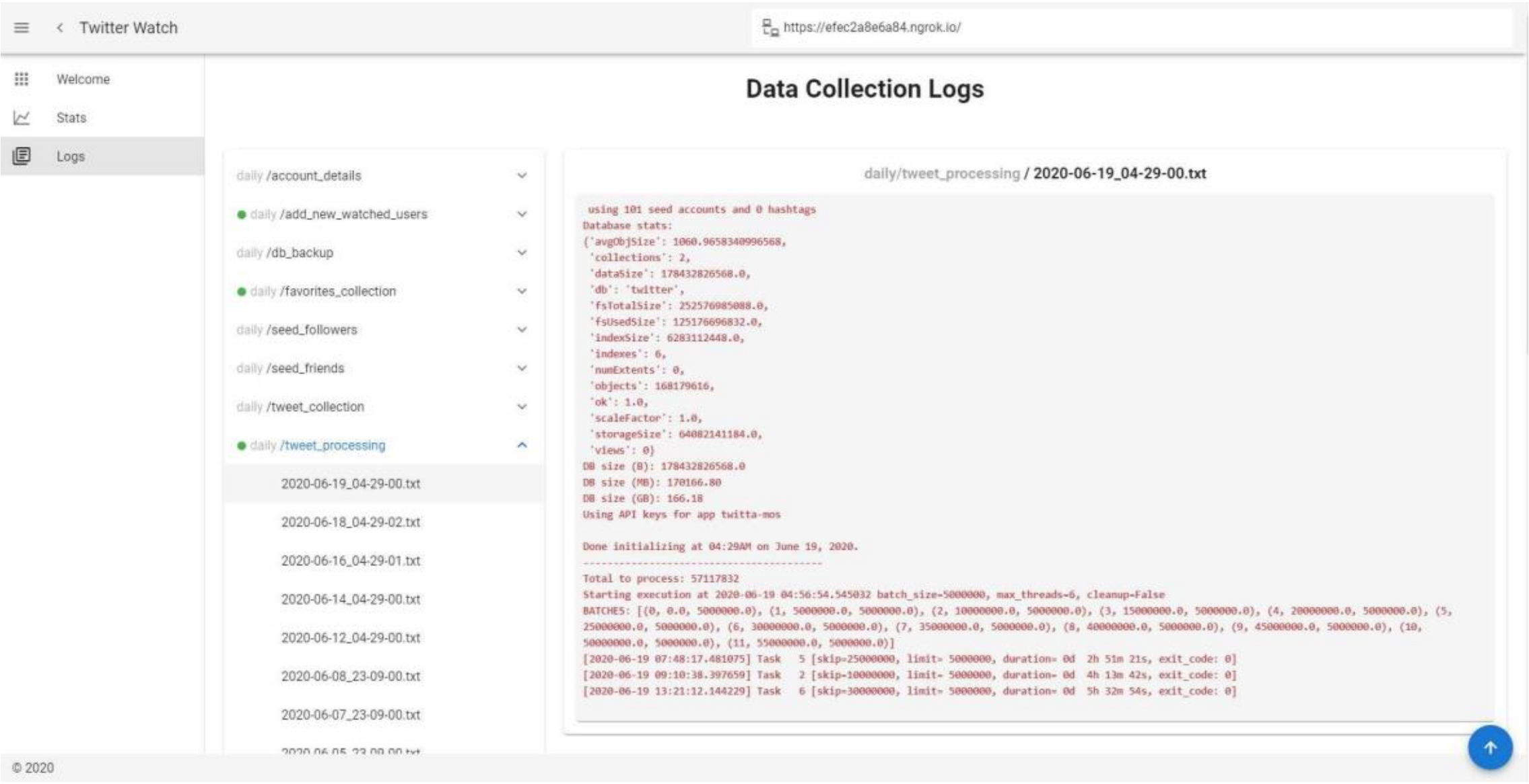

Figure 5.4: Twitter Watch UI logs page

### 5.3.3 External Services

Twitter Watch interacts with external services each with a particular goal. The most obvious one is the Twitter API, a quintessential part of Twitter Watch, where all the data comes from. On top of that, Twitter Watch uses Google Drive API [33] for storing regular snapshots of the database in case there is a problem with the deployment. In practice, these snapshots are used for local data analysis as they constitute an easy way to download the compacted data and rebuild the database in any local deployment of MongoDB. Finally, out of necessity, we found that it is useful to have a way to receive notifications about unexpected collection errors. To achieve this, we used Pushbullet [23], a free push notification service that has a mobile client that can receive real-time push notifications. Push notifications were mostly used to detect errors in the most fragile processes like interaction with the Twitter API, building the database compressed snapshot, or uploading it to Google Drive. The logic behind this push notifications mechanism is isolated in the core container code and can easily be invoked anywhere else in the application, where any future user of Twitter Watch needs.

## 5.4 Implementation

Twitter Watch's implementation rests on a scheduling system that is capable of launching parallel processes, each with its logic, and the combination of individual tasks interacts indirectly, as these get their input from the database and write their output on it too – a holistic system. Although the system is designed to be both flexible and customizable, the out-of-the-box version is already capable of meeting all the desired requirements. Part of the flexibility comes from a configuration file used to dictate how the dataset should evolve over time, this configuration file is further explained in Section 5.4.1 (p. 36). The rest of the flexibility stems from the creation of a semantic folder structure along with the ability to add new features without having to change any core code. This last point is achieved by isolating all the collection process logic into Jupyter notebooks, which are automatically interpreted according to their location in the aforementioned semantic folder structure.

[33] https://developers.google.com/drive/ [23] https://www.pushbullet.com/

Figure 5.6 (p. 36) contains a high-level view of the backbone of Twitter Watch. It shows four different execution steps: launch, setup, run-once, and scheduled tasks. Each task is written in its own Jupyter notebook. The key to differentiating them is their location in the folder structure, namely the one visible in Figure 5.5. Each folder in the `collection` folder has a specific execution routine that maps directly into the implementation (*cf.* Figure 5.6, p. 36), as is described below.

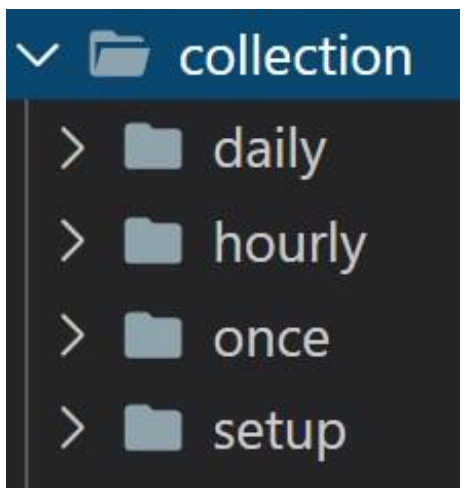

Figure 5.5: Jupyter notebooks semantic folder structure

Initially, when the application is launched, there is a single line of execution. First, the configuration file is parsed and validated; then, the output folder structure is created if it does not exist; lastly, the Jupyter notebooks, where the collection logic is written, are converted into Python (*.py*) script files, as these can then be easily executed in separate processes.

Afterward, the now converted Python script files inside the `setup` folder are executed. In our implementation, this step has tasks related to inserting an initial set of accounts into the database for subsequent exploration (these seed accounts come from the configuration file), and executing database migrations (these can be anything from creating indexes on the database to restructuring the database and can vary through time). These setup tasks are executed as parallel threads. Once all the setup tasks have completed, both the run-once tasks and the scheduled tasks launcher loop are executed in parallel.

The run-once tasks differ from setup tasks by being able to co-exist with the recurrent tasks launched by the scheduled launcher. These tasks are often tasks that only need to be executed once, or that will be running non-stop until the application is manually closed. One-time data migrations fall into this category. In our case, we had a task running non-stop that simply logs the number of documents in each database collection at a custom interval.

Finally, the scheduled launcher is the piece of the puzzle responsible for parsing the filenames inside the daily and hourly tasks and running them at the specified time. Note that other schedules can be added in the future, like weekly executions, with little effort due to the abstraction mechanisms implemented. Note also that the filenames of the files within the `daily` and `hourly` folders are used to infer the exact desired execution time. For instance, the filename

`daily/03_00_tweet_collection.py` is instructing the launcher that it should be run precisely three hours and zero minutes after the application is launched, on a daily basis. This simple hack proved quite useful in practice, as it facilitates easy re-arrangement of the order in which scripts should be executed, even by playing with overlapping or distancing scripts based on their resource usage.

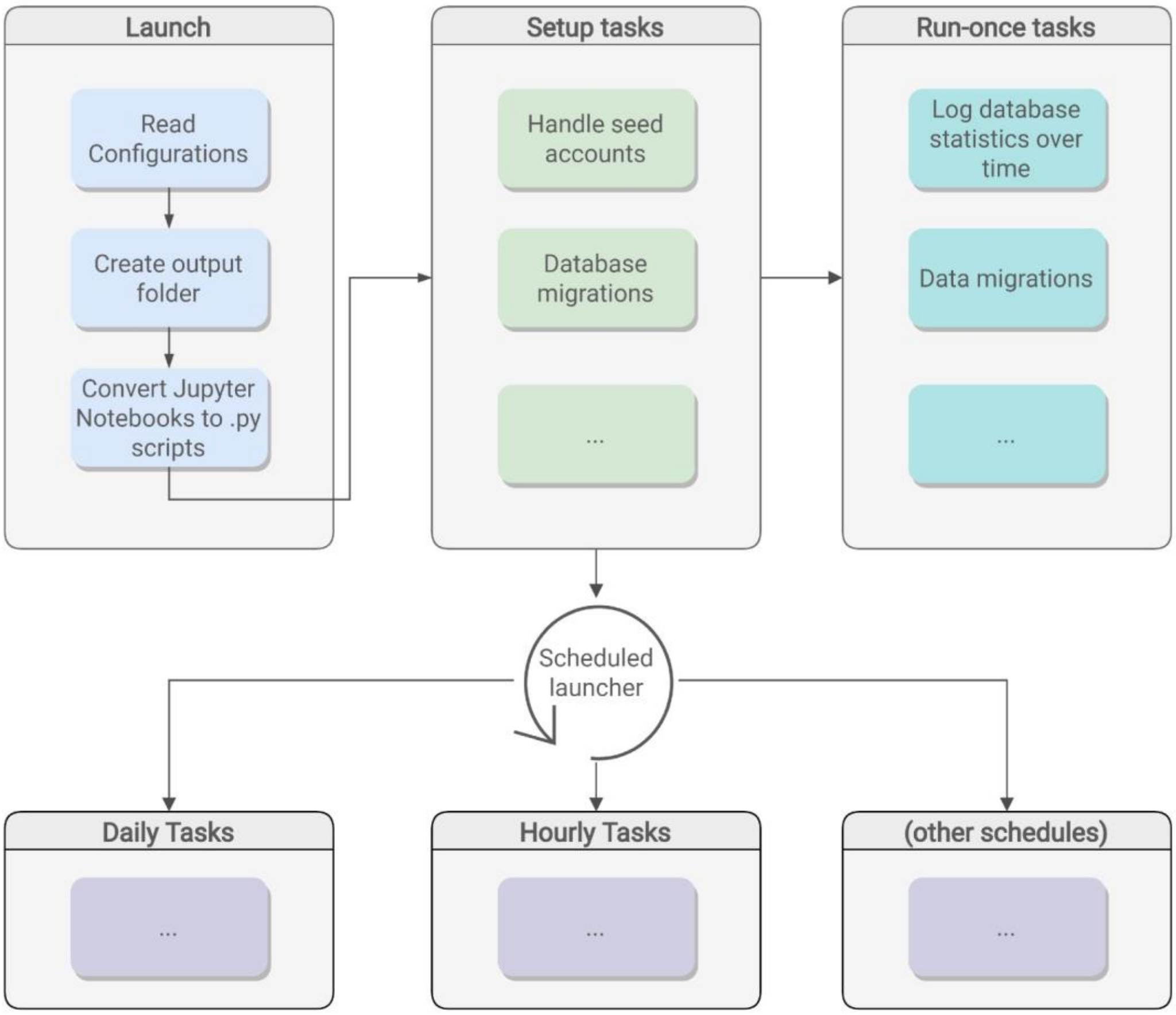

Figure 5.6: Twitter Watch Implementation

### 5.4.1 Configuration File

The configuration file works as a contract of settings that are unanimously used by all tasks, it is written in JSON and is automatically loaded into a global variable that becomes accessible to each process with a single import statement. Code Listing 1 (p. 37) contains a simplified example of the version used for the duration of this project.

```
{
"seed": {
"usernames": []
},
"collection": {
"limits": {
"max_watched_users": 100000000,
"max_daily_increase": 25000,
"max_daily_increase_ratio": 0.1,
"min_appearances_before_watched": 10
},
"ignore_tweet_media": false,
"oldest_tweet": "Sun Aug 1 00:00:00 +0000 2019",
"newest_tweet": "Sun Aug 1 00:00:00 +0000 2030",
"search_languages": ["pt", "und"],
"max_threads": 8,
"min_tweets_before_restricting_by_language": 10
},
"mongodb": {
"address": "mongodb://USERNAME:PASSWORD@mongo:27017/",
"database": "twitter",
"drive_api_backup_enabled": true
},
"notifications": {
"pushbullet_token": "API TOKEN"
},
"database_stats_file": "out/db_logs.csv",
"seconds_between_db_stats_log": 10,
"api_keys": "TWITTER API KEYS FILE"
}
```

Listing 1: Simplified example of the JSON configuration file

Below is an explanation of the most relevant fields in the configuration file:

- The `seed` field contains usernames of the seed accounts that serve as a starting point for the collection process;
- The `collection` field specifies what the behavior of the collection process should be like, for instance:
    - `limits` is a set of restrictions on the total size of the dataset and how it can grow in size on each day;
    - `oldest_tweet` and `newest_tweet` restrict the time span during which Tweets are to be collected;

- `search_languages` is a list of language codes [34] used to restrict tweet collection on accounts with more than `min_tweets_before_restricting_by_language` tweets, while not having a majority of them in one of the provided languages;
  - `max_threads` is a global restriction on the number of threads each process uses when is parallelized through the class described in Section 5.5 (p. 41).
- The `mongodb` field and its inner fields indicate the database access credentials, the default database name to use, and whether to perform this database backup to Google Drive (*cf.* Section 5.3.3, p. 34) or not;
- The `notifications` field is used to provide the API credentials for Pushbullet (*cf.* Section 5.3.3, p. 34);
- `database_stats_file` and `seconds_between_db_stats_log` specify the location and period of the database statistics collection process that is always running in the background;
- `api_keys` is simply the filepath of a JSON file where a list of our Twitter API credentials are stored.

This configuration file can quickly grow to meet the developer's end-goals since adding a field in the JSON file will make it immediately available in any of the Jupyter notebooks.

### 5.4.2 Collection Logic

With the above knowledge, we can now delve into explaining how the different collection tasks work. A first remark has to do with making the most out of Twitter API keys, namely by isolating each used endpoint in its task, since the rate limits apply at the endpoint level. Knowing this, it should also be noted that the functionality was designed to be as modular as possible. The next subsections reveal the way the current implementation extracts Twitter data by going into detail on the most relevant implemented Jupyter notebooks and respective tasks.

#### Seed followers (daily)

This task runs daily and iterates over all the seed accounts specified in the configuration file updating the database with any new followers of those accounts by querying the GET followers/ids endpoint. All these accounts are marked as watched accounts.

#### Seed friends (daily)

This task is in all aspects similar to the previous one, differing only by focusing on friends (*a.k.a.* followees) instead of followers, through the use of the GET friends/ids endpoint.

---

[34] https://developer.twitter.com/en/docs/developer-utilities/supported-languages/

api-reference/get-help-languages

### Account details (daily)

This task runs daily and iterates over all the accounts without a screen name property. This happens because accounts are added to the database by other tasks whenever a new account is found, but they typically contain a single _id property and not the complete profile information. It uses the GET users/lookup endpoint to hydrate the account objects. Appendix A (p. 105) contains an example of an object returned by this endpoint.

### Tweet collection (daily)

This task runs daily and iterates over all the accounts that are marked as watched, or accounts that have not yet been excluded due to the restriction imposed by the parameter search_languages (*cf.* Section 5.4.1, p. 36)). The logic behind filtering out accounts is focused on producing results, *i.e.* accounts and tweets that are relevant for the subsequent study's goals. In this case, the focus is on restricting by proximity to the seed accounts and also to a set of specified languages. So, all the aforementioned accounts are then iterated and all their tweets are collected between the oldest_tweet and newest_tweet values, using the GET statuses/user_timeline endpoint, up to a limit of 3,200 tweets imposed by Twitter (*cf.* documentation [35]). In fact, this task is optimized as every time an account's tweets are collected, the _id of the last collected tweet is saved to the database and use in future calls as the since_id param. This is in accordance with the official optimization guidelines to minimize redundant API calls [36]. Finally, this task also updates the most_common_language of each account by pre-processing its collected tweets, this property is used to restrict further iterations of the task from processing users whose most_common_language is not in search_languages, note that most_common_language is only set if an account has at least min_tweets_before_restricting_by_language tweets.

### Favorites collection (daily)

This task is similar in behavior to the previous one, with two differences. First, it collects liked tweets instead of posted ones, which is achieved through the GET favorites/list endpoint. Second, it restricts the iterated accounts to watched users only.

### Tweet processing (daily)

[35] https://developer.twitter.com/en/docs/tweets/timelines/api-reference/ get-statuses-user_timeline

[36] https://developer.twitter.com/en/docs/tweets/timelines/guides/ working-with-timelines

This task runs daily. It does not rely on making API calls. It iterates over all the unprocessed tweets in the database, so that each tweet is only processed once. For each tweet it isolates the ids of all the accounts that are related to that tweet:

- all mentioned accounts;
- the author of the tweet;
- the author of the original tweet, if this is a retweet;• the author of the original tweet, if this is a quoted tweet;
- the author of the original tweet, if this is a reply tweet.

Each of these account ids is inserted in the database if it does not exist yet and the appearances counter for each account is incremented by one. This counter is used as a minimum threshold to start including the respective account into the watched users set; this behavior is configured by the `min_appearances_before_watched` field in the configuration file. Larger values of that parameter will lead to a slower yet potentially more relevant expansion of the watched users set.

### Seed tweet processing (daily)

This task is similar to the previous one. However, it is focused only on tweets by the seed accounts defined in the configuration file and the accounts identified are directly marked as watched, due to their proximity to the seed account and expected relevance to the dataset.

### Add new watched users (daily)

The goal of this task is to iterate all users that are yet to be marked as watched or non-watched and insert the ones that meet the minimum number of appearances count as defined by the configuration field `min_appearances_before_watched`. Furthermore, the configuration field `limits` and its inner values will restrict the maximum number of newly watched users per day and, *in extremis*, will prevent the addition of any more watched users if the `max_watched_users` value has been reached. These configurations are meant to coerce the system to evolve more slowly to avoid resource overload, when that is necessary.

### Hashtag tweet collection (hourly)

This task is executed every hour. First, it collects all the tweets in the database from the past 24h originating from the seed accounts. Then, the hashtags of those tweets are gathered and merged, and are then used to perform a language-restricted search (according to `search_languages`), for each unique hashtag. We use the Standard search API, noting that this is a unique endpoint where the results are restricted to a 1% sample of all tweets in the Twittersphere of the past seven days. The end-goal of this task is to introduce some relevant variety to the collection process since these tweets will later be processed and influence the expansion process.

### Google Drive backup (daily)

This task has been mainly explained in Section 5.3.3 (p. 34) and it takes care of calling the `mongodump` command [37] from MongoDB and then uploading the resulting database snapshot to Google Drive.

[37] https://docs.mongodb.com/manual/reference/program/mongodump/

### Note on suspensions

It is important to highlight that on every API call that operates on account ids, namely looking up followers, friends, account information, tweet timeline, among others, Twitter Watch has a mechanism that wraps the errors returned by the API and, if a received error is related to the given account having been suspended (deletions, and private accounts are also recorded, for that matter), the database is updated to register this occurrence and its timestamp. This is useful for enriching the dataset with suspensions information.

### Note on long-running tasks

The current version of Twitter Watch ensures that any scheduled task set to start its execution at a given moment is only launched if its previously launched instance (1 hour before for hourly tasks, and 24 hours before for daily tasks) has finished, in order to avoid excessive resource consumption.

## 5.5 Parallelism

After having Twitter Watch collecting data for a while, we noticed that the size of the database was large, and some of the tasks, such as the daily tweet collection or tweet processing, were taking more than the ideal 24 hours threshold. Inspired by the Map-Reduce algorithm [19], we developed our own implementation of a parallel processing mechanism that each task could benefit from. This mechanism requires only isolating the logic code in each notebook to a `def task(skip, limit):` method that performs the same query on the MongoDB database but appends `.skip(skip).limit(limit)` filters to the query. Once this change is ready, all a developer has to do is invoke the `.run()` method on the `DynamicParallelism(..., batch_size, max_threads)` class. The `batch_size` and `max_threads` properties can be used to adapt the behavior of the parallel execution both in terms of the size of each batch of database documents to be processed as well as on the number of threads to use, respectively. The number of threads will default to the configuration field `max_threads`. A `.reduce()` method exists to merge all the outputs, but in practice, it has only been used for data analysis tasks and not actually in the collection process. In the end, this effort managed to significantly reduce the long-running tasks' length to under the desired 24h limit.

## 5.6 Results

First of all, the framework resulting from this effort was able to meet both the mandatory as well as the optional requirements specified in Requirements 5.1 and 5.2.

Second, the development of Twitter Watch was incremental, since such a complex and byzantine system was impossible to effectively tune in advance while assuring few wasted resources. This led us to give priority to things as they came up. In any case, we believe it served its purpose for our use case, and has led us to be hopeful about its potential to be adapted or used as-is for other research or industry efforts. In terms of results, we can look at its overall success as a consequence of both data quantity and quality it managed to collect. Although this section will focus mostly on quantity, the exploration Chapter 6 (p. 47) will inherently depend on the data quality, and we will let the results described there echo this notion quality. With that in mind, this section reports quantity results for the system where it is deployed, and the configurations used.

### Deployment server

The current deployment server has an x86_64 architecture running Ubuntu 18.04.3 LTS; having 6 Central Processing Units (CPUs), with 8GB of Random Access Memory (RAM), 236GB of Solid State Drive (SSD) disk space, and running Docker version 19.03.11. The initial setup included only 2 CPUs and 4GB of RAM; this was considered a bit limited, especially when interacting with the machine through Secure Shell (SSH) while the collection process was under execution. The final setup proved to be sufficient, but no doubt that increasing processing power and available RAM could significantly speed up some operations such as database read/writes that use indexes, which can require a lot of memory and also benefit from available processing power. In fact, special care was taken to spread out the tasks that could compete heavily for the same resources, but this was done through trial and error.

### Configurations

Although our initial approach described in Section 5.2 (p. 29) used 21 accounts as seed for the data collection process, we decided to enlarge it by taking into account all the accounts that fall under one of the categories in the following list that expands the original one:

- Political party account;
- Government ministry account;
- President of a political party;
- Minister in office;
- Secretary of state in office;
- Parliament deputy.

The final number of accounts found is 101. These accounts match with the `seed.usernames` parameter in the configuration file. This seed identification process was conducted in February 2020. Other configurations match the values presented in Code Listing 1 (p. 37) excluding, of course, the placeholder values which do not influence the logic of the collection process, like the database credentials or the API tokens.

Output

At the time of writing this report, the system is still up, and will remain so for some time. For a period of little over two months, from April 15th 2020 to June 20th 2020, we have collected almost seven million accounts (over 6,890,000) and more than one hundred and sixty million tweets (over 163,280,000), corresponding to more than 170GB of data. Luckily, compressing this data with `mongodump` leads to a file with around 30GB. Figure 5.7 shows the evolution of the number of collected accounts and tweets over this time period. Four colored areas are highlighted, from left to right:

- The first area, "docker-compose bug", corresponds to a period where the system was operational but a line in the Docker Compose configuration had been commented with the consequence of not exporting the collected data outside the Docker container, once this was fixed, the system automatically recognized the database in the real file system and started the process again, where it had left off;
- The second, "parallel processing introduced", marks the approximate time of deployment of the parallel processing mechanism described in Section 5.5 (p. 41). It is visible that before this moment, the data collection had almost flattened and quickly regained a steady increase;
- The last two areas, "system down", correspond to two different periods during which the system was offline – this was actually due to human error.

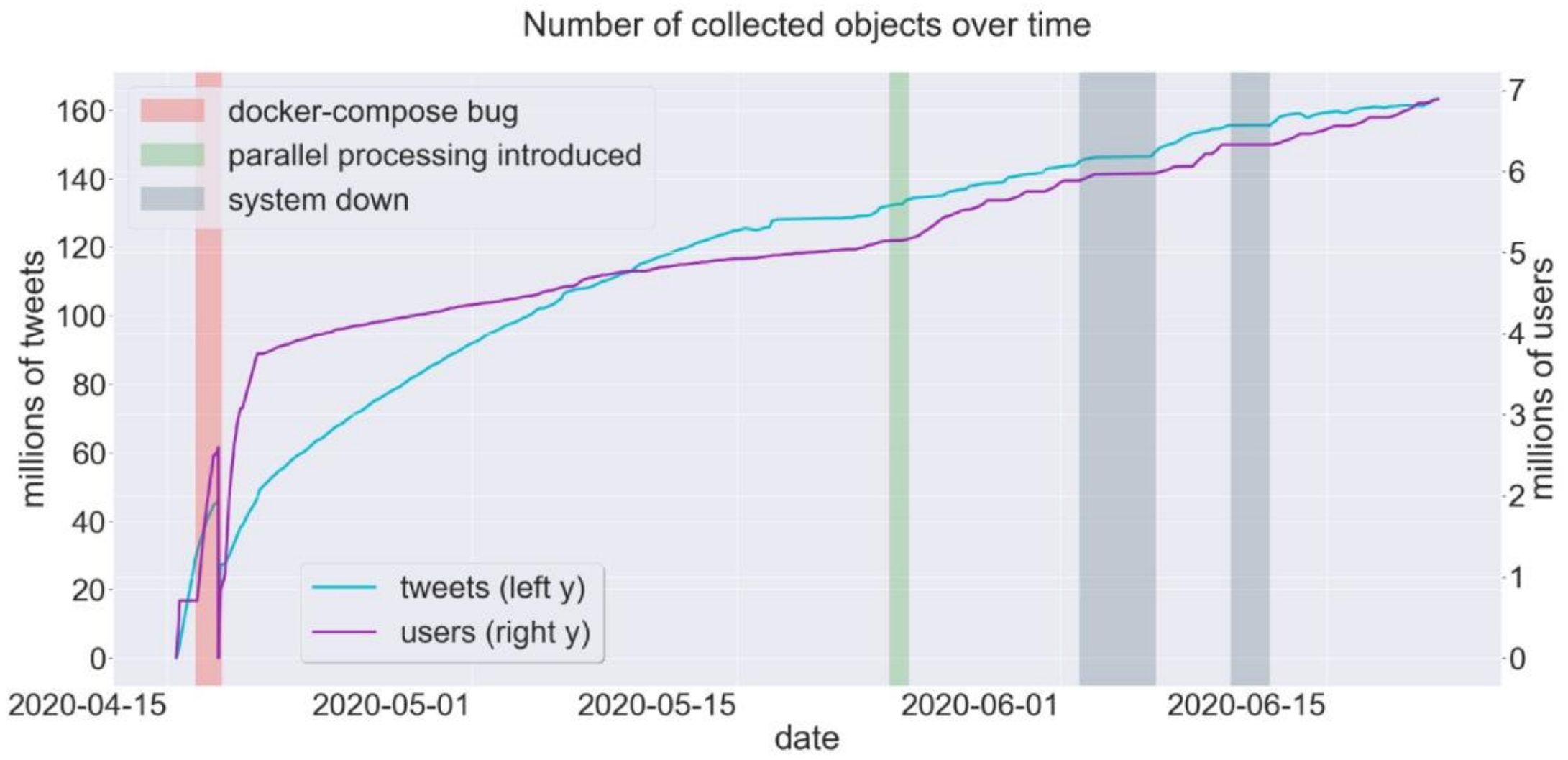

Figure 5.7: Twitter Watch total accounts and tweets collected over two months and five days (double y-scaled)

This figure also reflects a predicted phenomenon, when the system was designed: a fast initial growth, followed by a flattening of the curve.

The initial growth is mostly associated with a quick identification of the accounts that interact very intensely with the seed accounts, with the remark that their daily growth is limited by the inner fields in the `limits` configuration field. A similar effect happens with the number of tweets, which grows very rapidly. This growth is easily explained by the fact that the first time an account's tweets are collected, they are collected for the entire period from `oldest_tweet` until the moment of collection. In contrast, the following iterations for those accounts will only yield the most recent tweets, since the day of the previous collection.

The flattening of the curve, although related, at the same time, to the bottleneck that was fixed by introducing parallel processing, was not surprising. Since the collection process focuses on quickly including all the accounts that are close to the seed accounts (and restricted by the `search_languages` field too), and then it will only increase when new accounts are detected in the tweet processing task. These tweets come from accounts that interact with the accounts already in the database but also from the hashtag tweet collection task. Hypothetically, we could have defined a threshold for the total amount of collected accounts, and the system would keep on collecting tweets for that fixed number of accounts – this could be a strategy to fight an explosion of the number of accounts.

### Experienced limitations

A final note on Figure 5.7 (p. 43) to explain some slight decreases in both curves. For our use case, we defined a few run-once tasks that would remove some accounts and their tweets from the database, and that explains some slight decreases in the curve. This is due to the first limitation of Twitter Watch, the fact that it relies highly on the language of the tweets to find relevant content. This is undesirable because, for instance, the English language is widely used, and it is hard to capture realities in English speaking countries without incurring the risk of having a lot of noise in the dataset since it will eventually expand to capture tweets in the same language but from other countries. For our disgruntlement, Portuguese is the official language of Portugal and several other countries, most relevant in practice, of Brazil. Actually, the official language is Brazilian Portuguese, but Twitter marks both types as `"pt"`. Having seen a non-negligible number of Brazilian Portuguese tweets in our data, we devised a task focused on identifying accounts with a location in Brazil and removing them. This was partially effective, but not entirely as many accounts do not specify their location and, therefore, remain in the dataset. Despite this, we were able to perform data exploration tasks (*cf.* Chapter 6, p. 47) in a way where it was easy to check if Brazilian accounts were influencing a given result, namely by looking at random tweet samples and manually looking for Brazilian Portuguese content, or at least, Brazilian-related content. Other Portuguese speaking countries hardly ever came up throughout the collection and following exploration processes.

## 5.7 Summary

The above sections have laid out the current standpoint of Twitter Watch, and how it is already in a version that allows for massive and structured data collection that satisfies the desired requirements. However, we envision several improvements in terms of usability and the collection process.

In terms of usability, we have thought of adding control functionality to Twitter Watch's interface like the ability to launch or kill a given task, or even edit the configurations file on the fly.

In terms of the collection process, we can highlight that the system's scalability is limited to a vertical growth of resources, and we believe that larger collection processes might benefit from a horizontal approach, for instance, by using MongoDB's sharding mechanisms [38] to benefit from a multi-cluster system.

In terms of already verified limitations, the language limitation could be mitigated by adding a filter for the watched accounts based on the location of the account. In fact, if the location is not explicitly provided, one can even rely on existing research to infer an account's location from its posted content [44].

Finally, we believe that Twitter Watch's approach can be adapted to other data collection processes and APIs.

Overall, we have taken a top-down approach to explain the inner workings of Twitter Watch. However, we believe that anyone using it from scratch will still need to spend a couple of days getting familiar with its overall structure if they intend to develop custom tasks. Even so, we have seen it working and are satisfied with the obtained results that enabled our analyses that would otherwise be limited to, and based on, an incoherent and incomplete dataset.

[38] https://docs.mongodb.com/manual/sharding/

# Chapter 6

# Exploration

> "You'll never stumble upon the unexpected if you stick only to the familiar."
>
> Ed Catmull
> *in* Creativity, Inc.

This chapter, although only a piece of the entire developed work, represents our efforts of actually investigating the potential of using high-level methods for the detection of malicious behavior on Twitter. We start by describing the dataset we explored, a snapshot taken from Twitter Watch. As it is hardly avoidable, we also highlight several meaningful changes on Twitter behavior due to the Covid-19 [15] pandemic. Then, we report how we faced the challenge of finding labeled data and its importance. Finally, we get to explore the dataset with three complementary lines of focus: content, metadata and structure, from where we draw our conclusions and strive to open new research-worthy questions for the future.

## 6.1 Dataset Description

The dataset studied in this chapter is a snapshot of Twitter Watch taken on May 18th, 2020. However, we trimmed the snapshot to exclude tweets collected after May 15th since these had only been partially collected. Indeed, this type of limitation on the collection process was later fixed, as is reported in Section 5.5 (p. 41).

In terms of aggregated data-points, the dataset contains:

- 4,989,221 total accounts, 1,079,379 (21.6%) are watched accounts as defined in Section 5.4.2 (p. 38);
- 118,816,838 total tweets subdivided into:
    - 45,466,054 retweets (38.3%);
    - 33,078,055 original tweets (27.8%);
    - 31,449,458 replies (26.5%);
    - 8,823,271 quoted tweets (7.4%).

Figure 6.1 contains an overview of how the total daily number of the above types of tweets evolves between August 1st 2019 and May 15th 2020. Two periods are highlighted.

The first stretches from September 6th to November 6th 2019, this period was chosen as the focus of the content analysis reported in Section 6.3.1 (p. 59), and corresponds to a two-month window centered around the legislative elections of October 6th, 2019. The analyses presented in Section 6.3.2 (p. 74) and Section 6.3.3 (p. 80) focus on the entire time period since their results benefit from larger amounts of data and are also time-agnostic.

The second period corresponds to a State of Emergency imposed in Portugal due to the Covid-19 [15] pandemic, from March 19th to May 2nd, 2020. Although this period was characterized by imposed social isolation, it is worth mentioning that the weeks prior to the enforced isolation were characterized by voluntary self-isolation led by a significant part of the population.

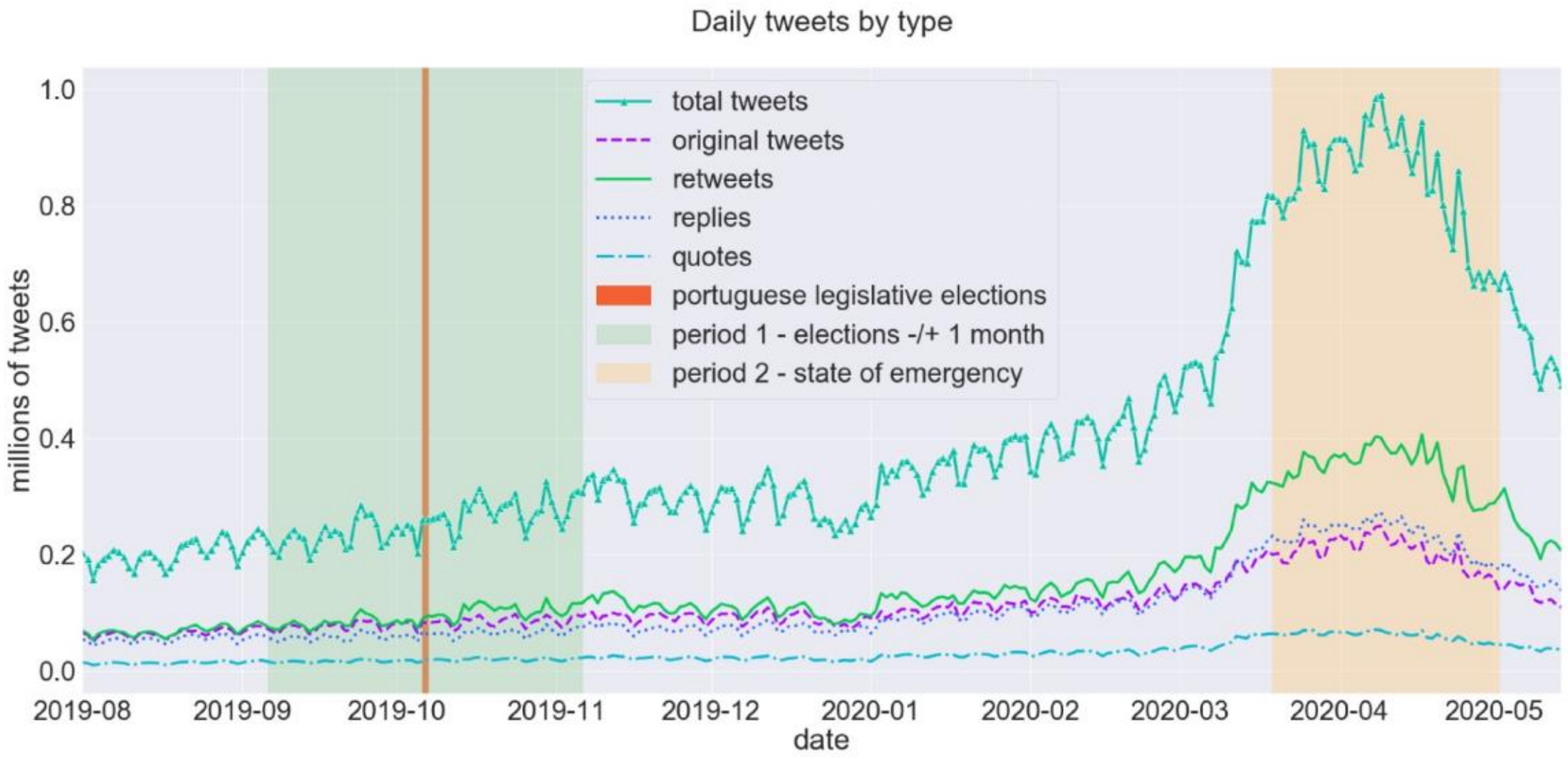

Figure 6.1: Dataset daily tweets evolution (from August 11st 2019 to May 15th 2020)

It becomes evident that Covid-19 and the social isolation period reflected in terms of user activity as there is a distinct spike in the total tweet activity that peaks at around three to four times the levels observed prior to the pandemic. Coincidence or not, it is interesting to entertain the idea (perhaps further investigate it in another research effort) that the State of Emergency period is conspicuously focused around the large spike in activity; perhaps because the increase in Twitter activity is correlated, and therefore a good indicator, to the general concern felt by the citizens. Looking at how this activity decreases and its decrease is followed by the end of the State of Emergency. Nevertheless, this decrease could also be explained by an increasing softening of the isolation measures, or even simply be indicative of people's boredom.

Another observation, less prone to guesswork but still subtle, is that the week prior to new year's eve is associated with a noticeable decrease in overall activity. Other observations include a behavior of retweets as generally being the most occurring tweet type, and that it is also the type that has a larger relative increase during the observed Covid-19 spike.

By looking at the total tweets curve, a type of seasonality is also present, even if only under normal circumstances, that is, before the Covid-19 impact becomes visible. Another visualization of the same data, Figure 6.2, highlights this seasonality pattern by separating weekdays and weekends.

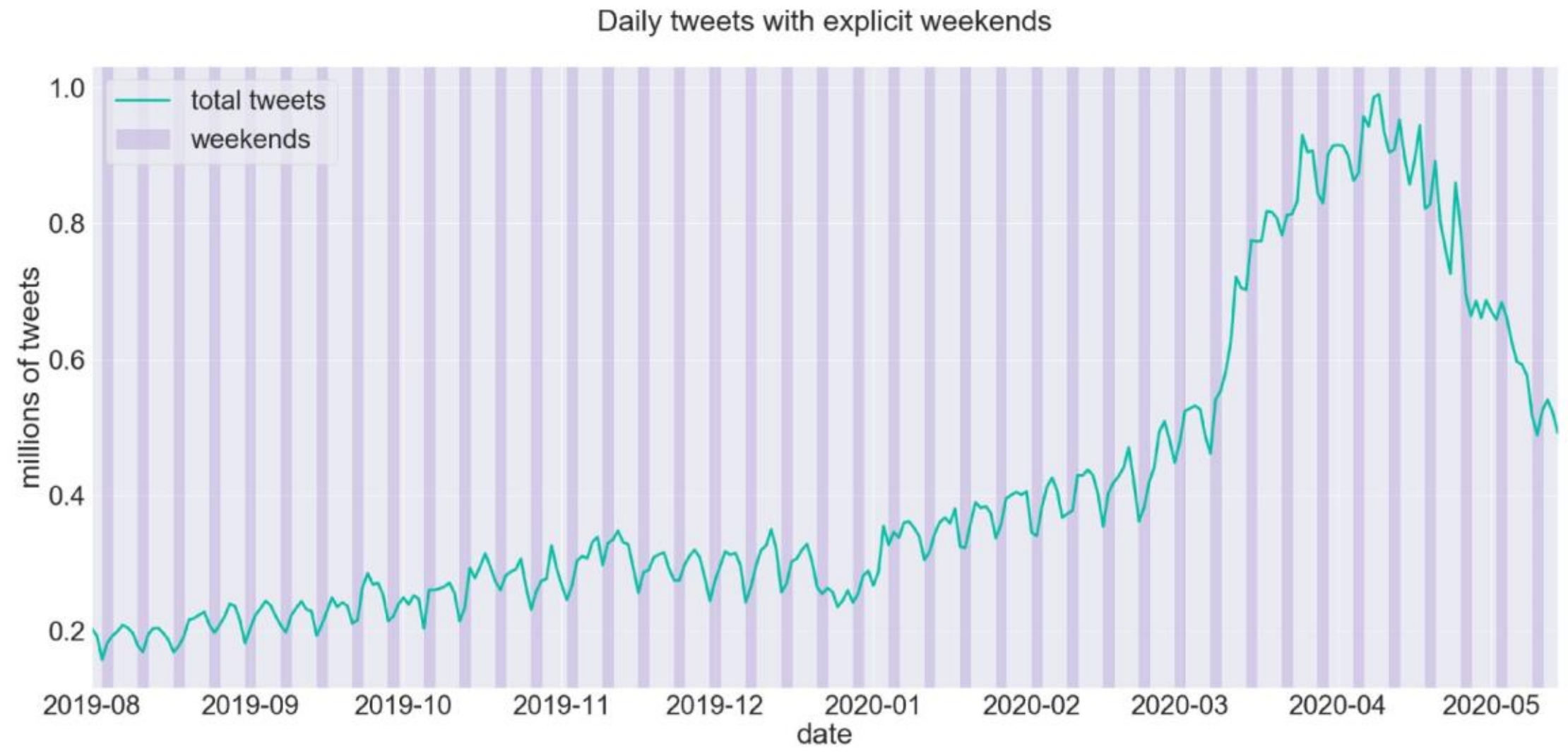

Figure 6.2: Dataset daily tweets with explicit weekends (from August 1st 2019 to May 15th 2020)

By pure observation, weekends seem to vastly correspond to the decrease in a weekly seasonality observed before Covid-19's impact. We took the liberty of calculating the average values distributed per day of the week, displayed on Table 6.1 (p. 50) in ascending order.

Table 6.1: Average total tweets per day of the week between August 8th 2019 and February 1st 2020 ( in hundreds of thousands)

| Day of the week | Average total tweets (100k) |
|---|---|
| Saturday | 2.45 |
| Sunday | 2.54 |
| Monday | 2.76 |
| Friday | 2.79 |
| Tuesday | 2.88 |
| Thursday | 2.89 |
| Wednesday | 2.90 |

Indeed, Saturday and Sunday are the days with the two lowest averages. Figure 6.3 (p. 51) displays the discrepancy between weekends and weekdays. This is confirmed by Figure 6.4 (p. 51), which shows a value of 1 if $pvalue < 0.05$ for Welsh's test, or 0 otherwise, under the null hypothesis that each pair of days of the week have the same average of total daily tweets. The conclusion is that, during this period, we have weekends consistently distinct from all other weekdays, except for the Sunday-Friday and Sunday-Monday pairs, where this is not a statistically significant claim. Such a pattern was unexpected since weekdays typically mean less free time to engage in social activities, this was not the case, suggesting that weekends might represent, for some people, time to disengage from social media. Another explanation could be that the increase in weekdays activity is related to media accounts being themselves more active and generating more novel content for commentary. In any case, an initial temptation could be to use this knowledge as yet another feature in the task of troll detection, but going back to the warnings issued in Section 3.3 (p. 18) – *any human behavior that can be automated becomes irrelevant the moment it is discovered*.

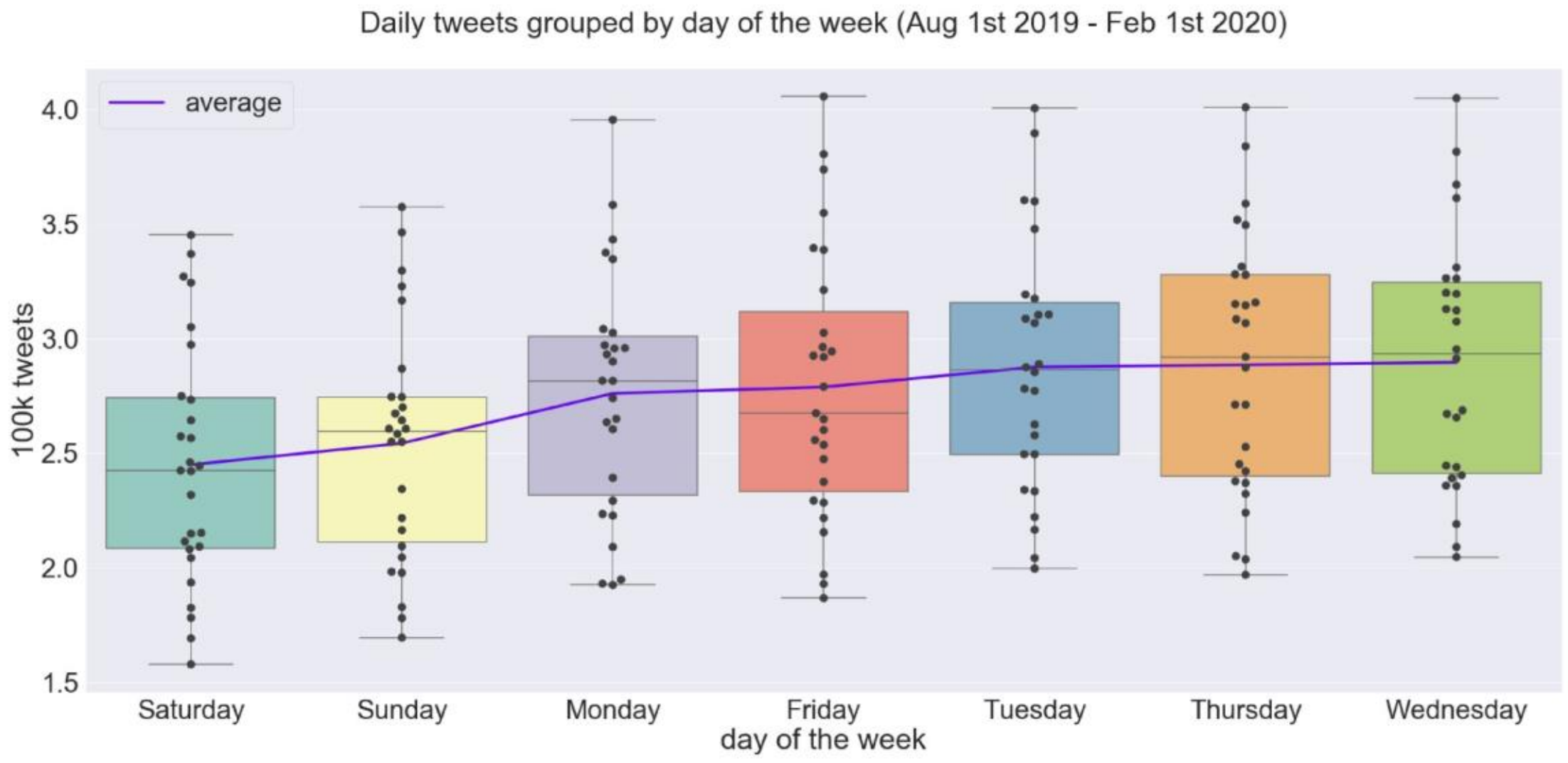

Figure 6.3: Boxplot and average daily tweets grouped by day of the week (from August 1st 2019 to February 1st 2020)

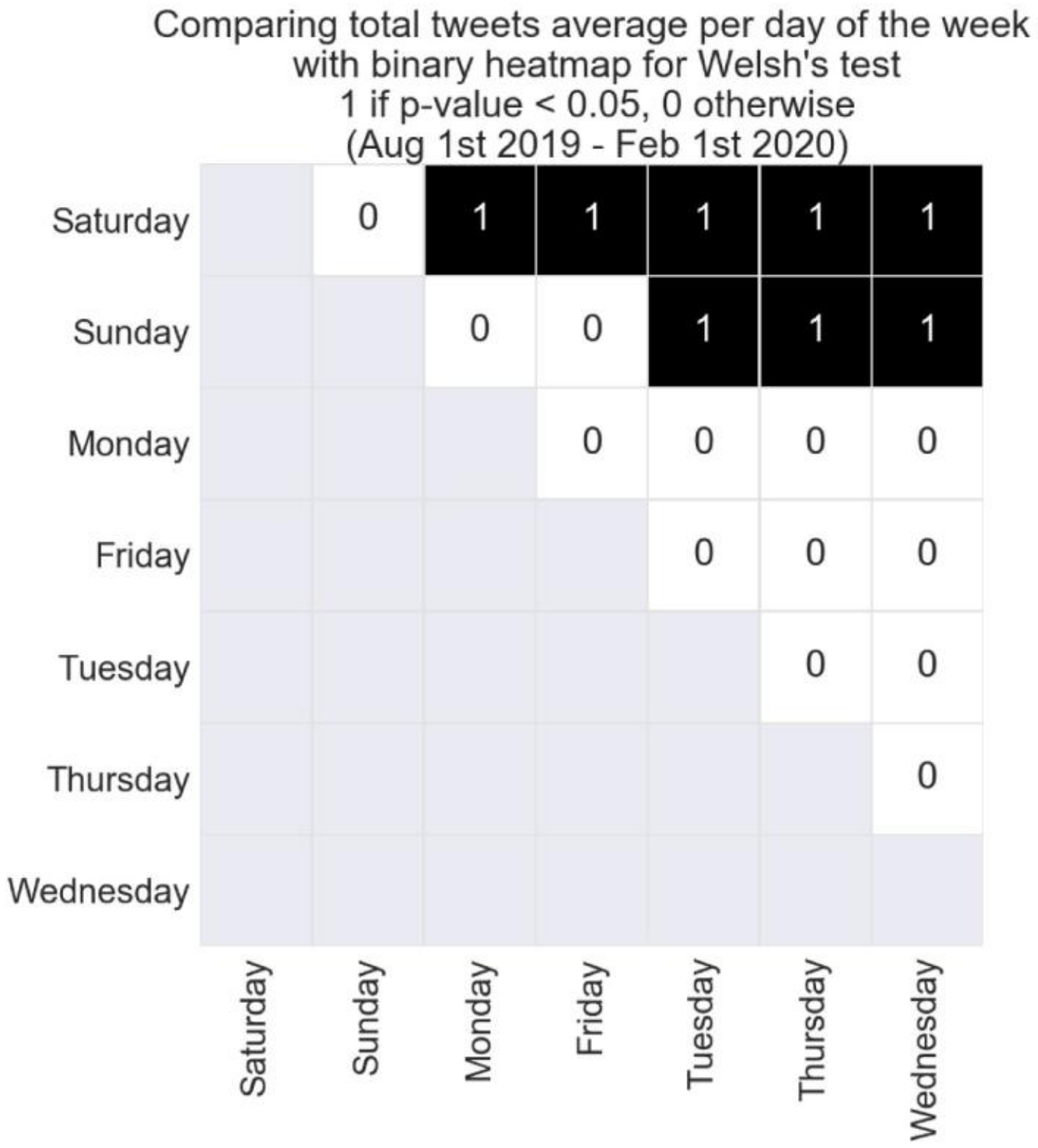

Figure 6.4: Welsh's test comparison between averages of total tweet activity for different days of the week (from August 1st 2019 to February 1st 2020)

However, when we present the same visual plot for the Covid-19 time period, in Figure 6.5 (p. 52), this discrepancy is no longer evident. In fact, $pvalue \geq 0.05$ for all days of the week combinations meaning there is no statistically significant difference between daily traffic in this time period, most likely as a consequence of Covid-19 and its societal impacts.

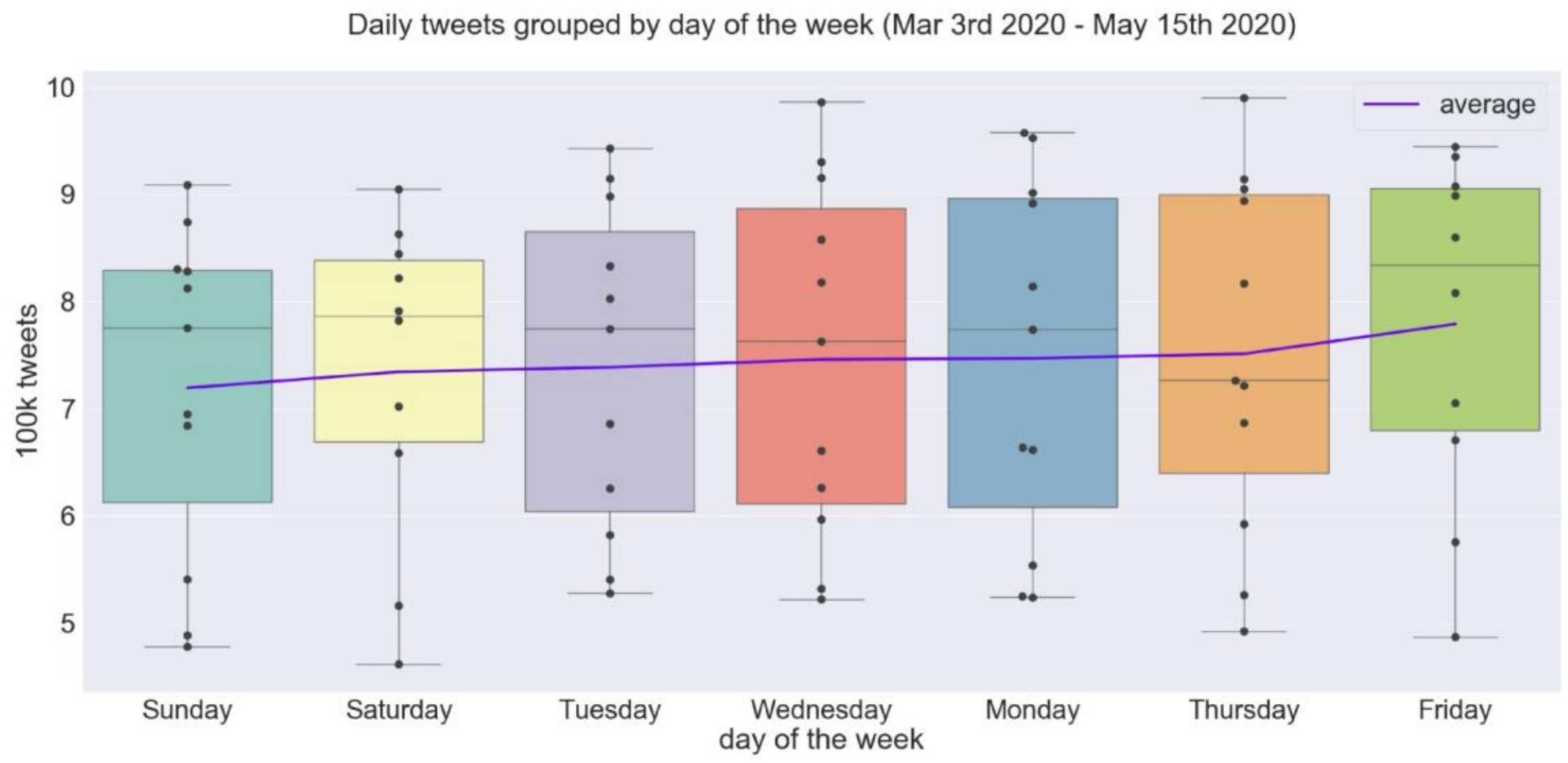

Figure 6.5: Boxplot and average daily tweets grouped by day of the week (from March 3rd 2020 to May 15th 2020)

## 6.2 Gathering Labeled Data

One crucial step in the analysis we want to develop is being able to investigate different types of accounts, namely the so-called malicious accounts. In this section, we explain how we collected labeled accounts from different sources. These labeled accounts differ in essence but represent what is here called malicious behavior. It is also essential that the accounts for which we have a guarantee of having their entire tweet history, for the period under study, are the watched accounts, to a total of 1,079,379, other accounts can have a large number of their tweets collected, but no guarantee exists on that. Interestingly enough, as will become explicit in the following sections, a significant amount of these labeled accounts were actually in the watched set. The exception being the suspended accounts described in Section 6.2.1, as these suspensions are tracked over the full set of accounts regardless of being watched or not.

### 6.2.1 Suspended Accounts From Twitter Watch

One of the advantages of Twitter Watch over other collection solutions, is its ability to identify account suspensions within a small time window of this suspension happening, typically, at most one day after it happens (*cf.* Section 5.4.2, p. 38). Twitter suspension policy is an umbrella for a variety of bad behavior on the platform [1], namely:

[1] https://help.twitter.com/en/managing-your-account/suspended-twitter-accounts

- Spam – the majority of suspensions which are related to spammy or fake usage of Twitter. The spam category encompasses cases where Twitter's rules are broken [2];
- Account security at risk – for cases of compromised accounts or credentials;
- Abusive tweets or behavior – related to threatening users or impersonating other accounts.

Due to a bug in the suspension detection feature, during the first part of the collection process, we cannot take advantage of the suspension date prior to the middle of April, since a significant amount of suspended accounts was detected at the same time, potentially much time after they had actually been suspended. Although the dataset snapshot used ends on May 15$^{th}$ 2020, we enriched the set of suspended accounts with the updated values taken on June 1$^{st}$ 2020, an addition in hindsight. However, we believe it has more benefits than potential disadvantages, especially considering that Twitter's suspensions are typically associated with a time delay. With this, we implicitly embrace the assumption that an account that is to be suspended in a timespan of less than half a year is already worth including in the suspended accounts set – mostly for believing that spam behavior is likely already present in that timespan.

In the end, we identified 94,447 suspended accounts, 10,478 (11.1%) are in the watched accounts set. Figure 6.6 contains a visual evolution of suspensions daily detection for suspended accounts between April 18$^{th}$ 2020 and June 1$^{st}$ 2020. In theory, this plot approximates the real suspension rate – notice that it corresponds to when Twitter Watch detects the suspensions not the actual day of the suspension – for instance if an account is suspended on a given day $d$ but only enters the total accounts collection on day $d+7$ its suspension day is recorded as at least $d+7$. Even with this limitation, the plot allows us to infer that the Covid-19-related State of Emergency is probably associated with a larger suspension rate. Indeed, there seems to be a time window of a couple of days after the State of Emergency ends that is followed by a decrease in suspension rates.

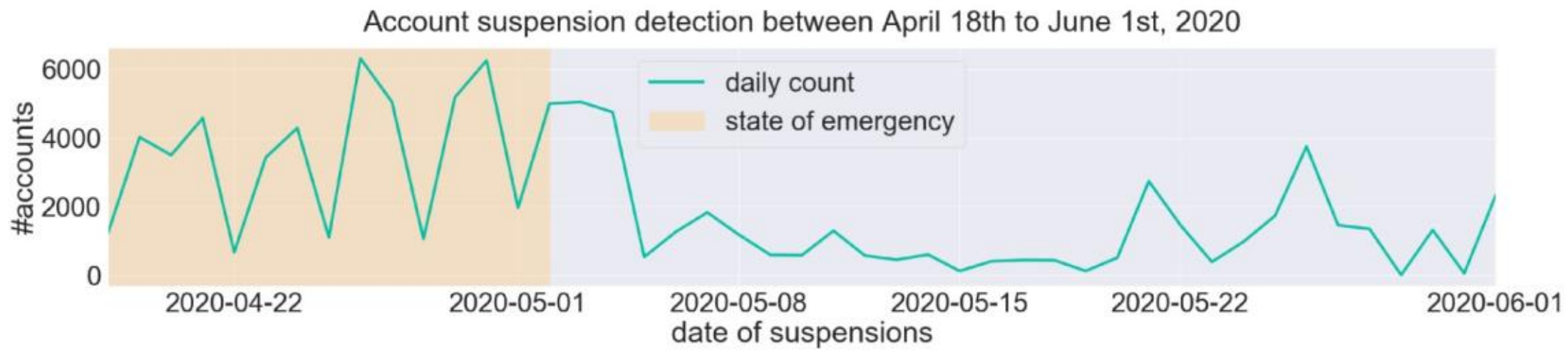

Figure 6.6: Daily suspensions detection rate from April 18$^{th}$ 2020 to June 1$^{st}$ 2020

Although Twitter Watch also collects accounts that are marked as not found, and that some of these can correspond to Twitter actively removing them, it is impossible to tell whether they are not found because the user deleted the account or because Twitter removed it. As such, we are not considering any not-found account as a blocked account.

[2] https://help.twitter.com/en/rules-and-policies/twitter-rules

### 6.2.2 Peer-shared Trolls

Concomitantly, to the present research effort, another *MSc* student at the University of Porto, Tiago Lacerda, is developing a thesis that focuses on creating a classification model to identify Portuguese trolls [41]. In a mutual collaboration effort, we have decided to share information and results. In our case, we have received a list of 287 accounts labeled as trolls, of which 278 (96.9%) were already in our watched users set. These troll accounts are obtained in two ways. The first, manual labeling, this approach consists of manually assigning a score to accounts and consider the ones above a score threshold as trolls; this method yielded 237 trolls. The remaining 50 accounts are obtained by a model trained using the manually labeled trolls. For readability reasons, whenever this particular type of labeled accounts is used in the rest of the document, it will be referred to as `peer`.

### 6.2.3 Fake News sharing

With the established knowledge that fake news is a real problem in today's society and Online Social Networks (OSNs) in particular. Especially, as fake news work as a catalyzer for political manipulation, namely in the fabrication, manipulation and propaganda categories (*cf.* Section 2.3.2, p. 11),
we considered that capturing their dispersion would be relevant to our analysis.

Appendix B (p. 107) contains a set of initial fake news websites, totaling 32, found through manual investigation, along with the four sources where they were taken from. We tried a new approach consisting of using those websites to uncover other fake news websites, as is detailed further in this section.

Using the `expanded_url` associated with shared Uniform Resource Locators (URLs) in Twitter's tweet object, we can iterate the millions of tweets in our dataset to look for tweets, and their posters, that share links to fake news websites – in any tweet form: tweets, retweets, replies, and quotes. One unfortunate limitation is that some of the identified fake news pages are Facebook pages. We have decided to ignore these cases, as Facebook's URLs are not easily mapped from the accounts page, for instance, a URL may be posted by a given page and have no identification back to that page depending on how it is viewed, and consequently shared. Because of this, 11 URLs were excluded, meaning the 32 websites were reduced to 21.

After processing our data looking for the remaining fake news tweets, we were able to identify the most predominant of these websites. Figure 6.7 (p. 55) shows the different predominance for each of the websites that were found to have been tweeted. We found that, out of this pool of websites, a small amount of them have a very high presence namely lusopt.eu [39] (530), noticiasviriato.pt [40] (501), noticiasdem3rda.com [41] (390), and bombeiros24.pt [6] (259) explain 88% of the total shares. In total, 1,898 shares were found associated with 715 different accounts.

[39] `https://www.lusopt.eu/`

[40] `https://www.noticiasviriato.pt/`

[41] `https://noticiasdem3rda.com/`

[6] `https://www.bombeiros24.pt/`

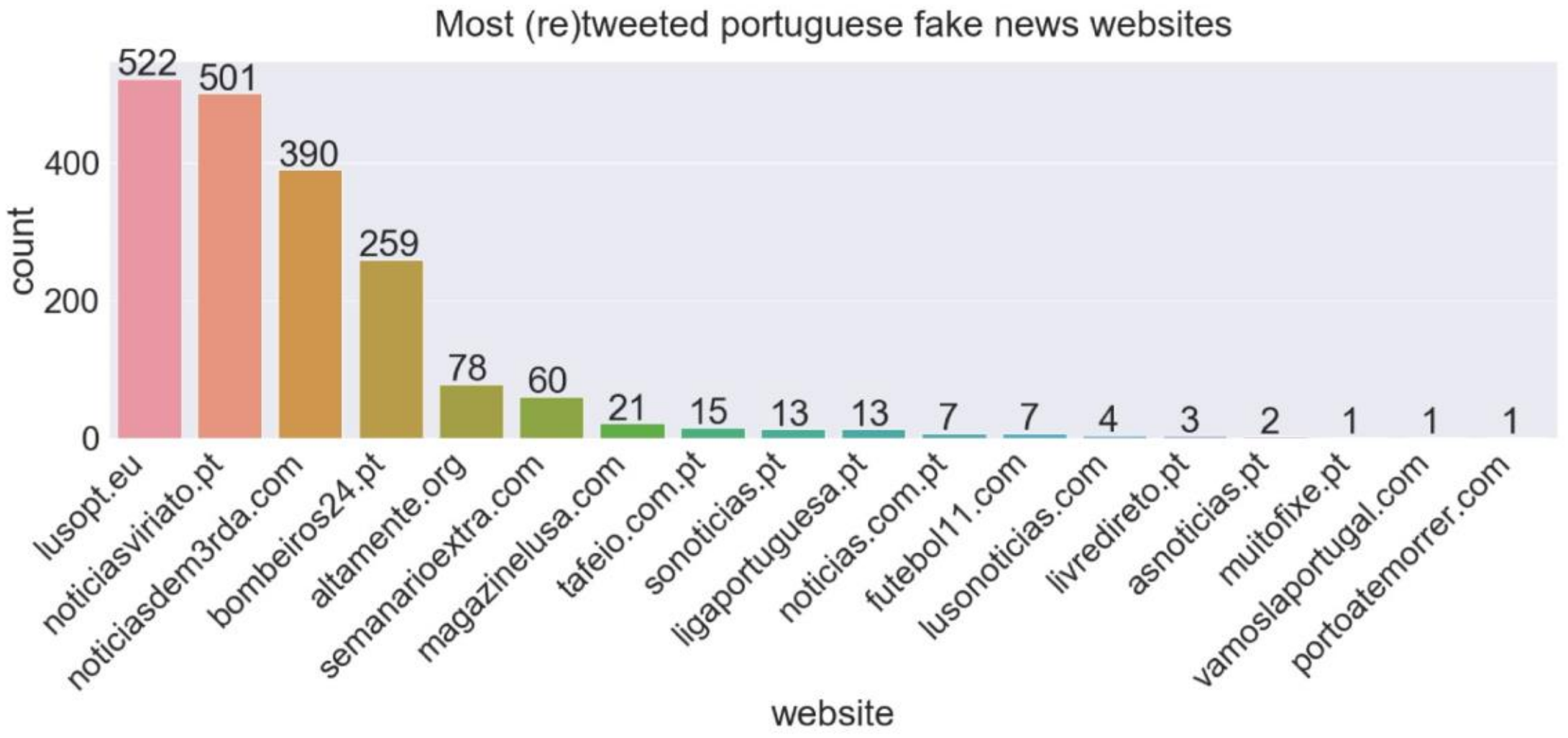

Figure 6.7: Histogram with fake news websites frequency of occurrence

Figure 6.8 contains the account handles (*a.k.a.* screen names) and the total number of fake news tweets they have posted. It should be noted that, in the first six accounts, three are facade accounts for fake news websites (@JornalQ, @NoticiasViriato, and @frasesdem3rda), the other three are accounts without clear allegiance (@casadoscaes, @DavidMagarreiro, @maragitado). Indeed, @JornalQ is the one account with the largest number of fake news URLs shared. At the time of this report is written, the account @DavidMagarreiro has disappeared, although we have information in our database that it was made private on June 16$^{th}$. After inspecting its recorded tweets (no longer accessible on Twitter), it became clear that this was a spam account since it has countless tweets equal or very similar to Code Listing 2 (p. 56).

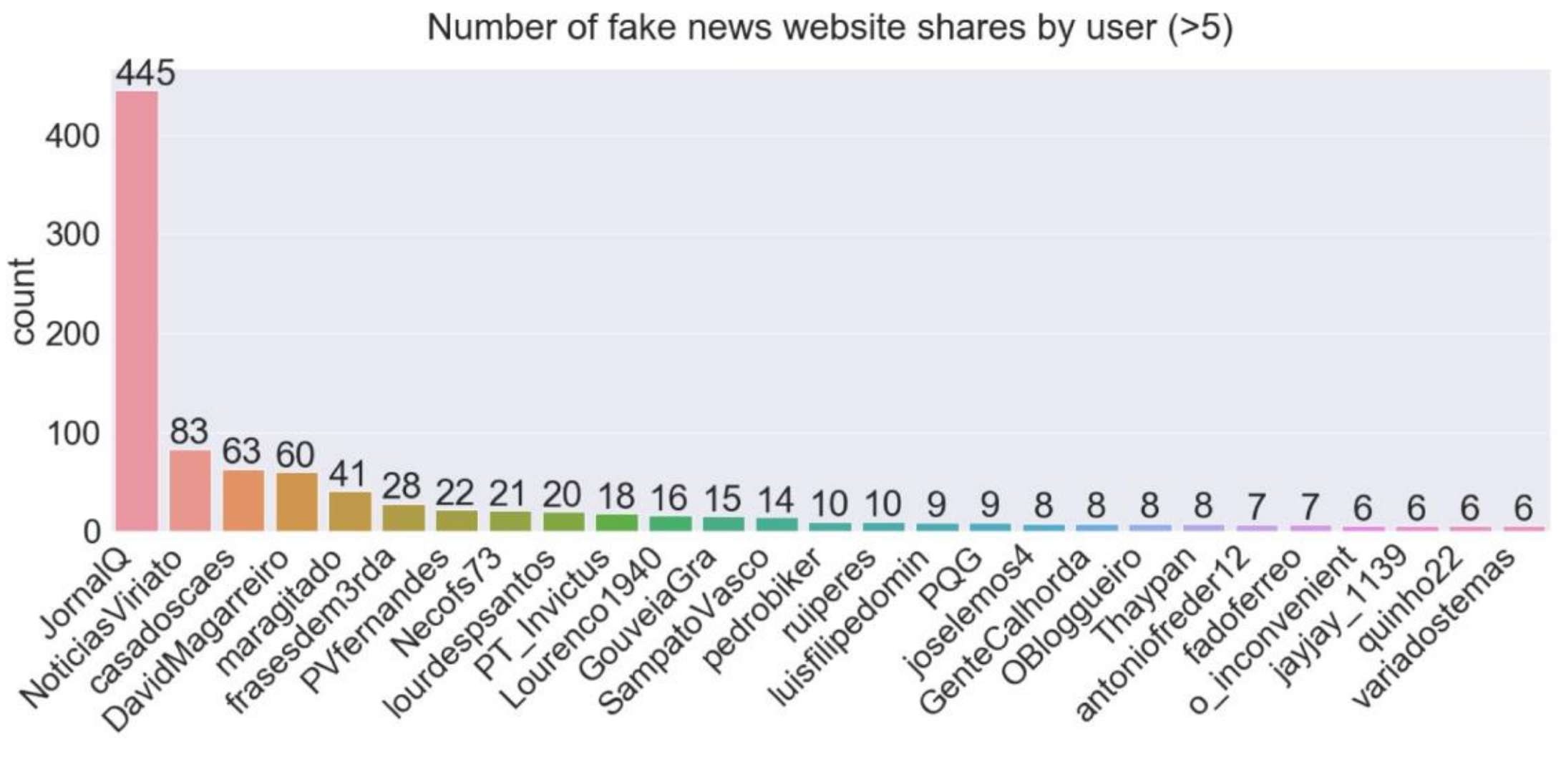

Figure 6.8: Histogram with fake news posters' total number of tweets, with a cut-off for accounts with less than five shares

```
{
'_id': 1241507379742638080,
'full_text': '#CoronavirusFrance #coronaviruswuhan
    #CoronaVirusFacts #coronavirus #coronavirusjapan
    #coronavirusdeath #COVID-19 #coverup #coronavirusaustralia
    #coronavirusoutbreak #virus #China #Chine #NEWS #coronavirusu
    #wuhan #COVD19 #COVID_19 #COVID19 #COVID2019 #COVID19PT
    #PORTUGAL https://t.co/yQtq9qIDRy'
}
```

Listing 2: Example tweet by @DavidMagarreiro

Having had success in identifying fake news posters, we decided to use the existing data in search of other fake news websites. To achieve this, we looked at the 200 most shared websites from the top 50 fake news posting accounts. From these 200, we excluded the safe websites (youtube.com [42] and instagram.com [43] are two such examples, but many were also from Portuguese newspapers). Figure 6.9 (p. 57) shows the top posted websites by those accounts that were not immediately discarded for being well known, ordered by decreasing number of shares, we have also added a focus on distinguishing those websites that were part of the initial list of fake news websites. Here, we identified, using our knowledge and through manual verification of the existence

[42] https://www.youtube.com/

[43] https://www.instagram.com/

of at least one fake news piece:

- 18 new Portuguese fake news websites, tuga.press [9] is one such example, the remaining can be perused in Appendix C (p. 109);
- 5 clickbait, with a clear focus on reposting actual news, like 4gnews.pt [10];
- 6 very biased news, either politically (like portugalglorioso.blogspot.com [11]) or religiously (like pt.aleteia.org [12]), but without clear evidence of fake news;
- many others, including foreign websites that were ignored, websites with unclear intentions, websites that were not available, and websites considered safe.

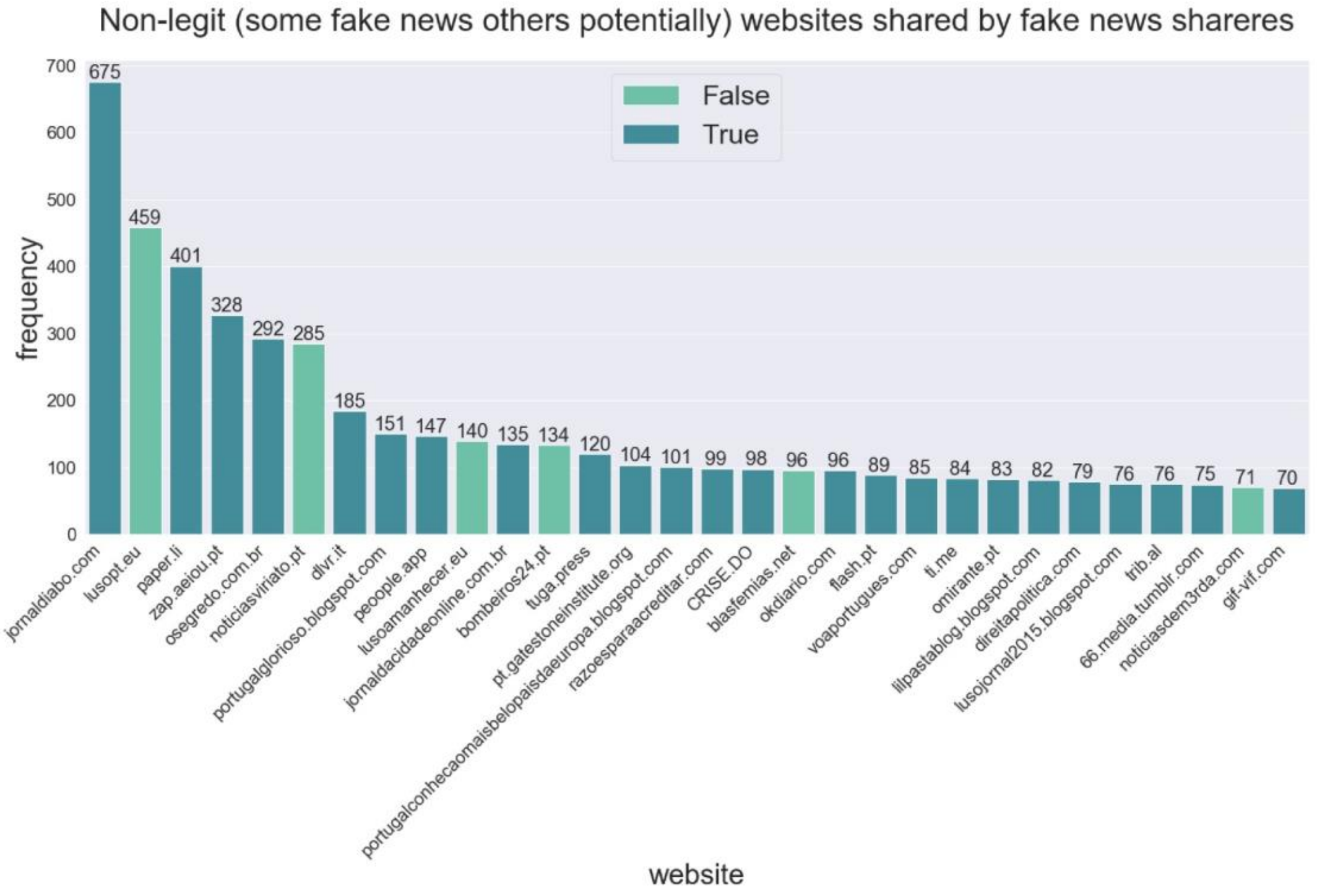

Figure 6.9: Histogram with new potential fake news websites

In total, 2,765 accounts were found by looking at the totality of all 39 (21 previous + 18 new) fake news websites. Of those, 74 of are in our peer's list, 1 has been suspended, and 2,633 (95.2%) are watched users, this final percentage was rather interesting for increasing confidence in the collection strategy since only about 21.5% of all accounts are watched.

Of the new 18 fake news websites considered, one is of a peculiarity that requires explanation: inimigo.publico.pt [13]. It is, in fact, a satire news website which precedes the general adoption of the term fake news. Under expectable conditions, it would not be included in the fake news set.

[9] https://www.tuga.press/
[10] https://4gnews.pt/
[11] https://portugalglorioso.blogspot.com/
[12] https://pt.aleteia.org/
[13] https://inimigo.publico.pt/

However, we noticed that when this website is shared on Twitter, it looks very much like a fake news article, as each article’s title gets the majority of screen space. Indeed, we calculated the conditional probability of one of those accounts publishing news from this site (*inimigo*) knowing that they published one of the others (*others*):

$$P(inimigo|others) = \frac{P(inimigo \cap others)}{P(others)} = \frac{\frac{178}{2765}}{\frac{2147}{2765}} = \frac{0.064}{0.776} = 0.082$$

This value is similar to those of other websites, whose average is 0.065. For both these reasons, we decided to consider it a fake news website as well. Even if we recognize that an analysis where it is excluded would still be acceptable.

### 6.2.4 Labeling Process Summary

All in all, the work reported in this section led to the identification of three types of different accounts. However, later in this report, in Section 6.3.1 (p. 59), we further identify 17 more troll accounts. The process through which that happens belongs in that section but we include those labeled accounts in Table 6.2 under the name *LDA found* for completion purposes.

Table 6.2: Types of labeled accounts and their presence in the watched accounts set

| Name | Total | Watched | % in Watched |
|---|---|---|---|
| Suspended | 94,447 | 10,478 | 11.1 |
| Peer | 287 | 278 | 96.9 |
| Fake news posters | 2,765 | 2,633 | 95.2 |
| LDA found | 17 | 17 | 100 |

## 6.3 High-level Exploration

Having a well-described dataset along with a varied amount of labeled accounts, we now delve into investigating the type of approaches that can be used to perform high-level detection of malicious political behavior. The following work is exploratory in nature and could have taken many different contours. That being said, we focused on three main research lines: analyzing posted content, posted metadata, and interaction structure. With that in mind, we first present a Latent Dirichlet Allocation (LDA)-based content-oriented approach in Section 6.3.1 (p. 59) followed, in Section 6.3.2 (p. 74), by a briefer hashtag exploration, and finally an interaction-oriented approach with a focus on modeling communities of accounts by clustering an embedded network, in Section 6.3.3 (p. 80).
These choices point back to the main types of approaches used in the literature, as demonstrated in Section 3.3.1 (p. 18).

### 6.3.1 Content-oriented

LDA is an unsupervised Machine Learning (ML) technique with typical applications in topic modeling. Topic modeling stands out as a very common approach to analyze Twitter content, especially when trying to find malicious activity starting from topic analysis [4] [23] [69] . Because of this, and its continued success as a state-of-the-art technique (*cf.* Section 4.3, p. 25), we now focus on using it on our Twitter data.

#### 6.3.1.1 Document Pooling

Literature shows that LDA typically works better when trained on larger documents than the typical tweet size. In other words, LDA produces worse results for microblogging [45]. Naturally, different approaches for constructing the documents fed into LDA have surfaced, and we felt the need to avoid the vanilla approach to pooling documents – using tweets as documents. [3] identifies four main ways of building documents suitable for LDA with Twitter data, namely:

Tweet pooling each tweet is one document – vanilla approach;

User pooling each document is composed of the whole history of a user – better than tweet pooling but limited to uses with small time-windows;

Hashtag pooling all tweets with the same hashtag compose a document - tweets with multiple hashtags appear more than once, tweets without hashtags are either discarded or included as tweet pooling documents;

Conversation pooling a document is a tweet, along with the cascade (upwards or downwards) of answers, replies, and comments associated with it - yielded the best results in [3].

In [45] this list is expanded to include:

Burst-score pooling involves running a burst detection algorithm to detect trending terms, followed by grouping tweets with trending topics (typically hashtags), according to a burst-score;

Temporal pooling consists of pooling all the documents within a short time frame, especially when unexpected major events, confined in time, happen.

The last two methods, by their exceptional nature, were not found to be a good fit for our use case. So, after some *ad-hoc* testing of the expectable document sizes for the previous approaches, we settled down for a combination of conversation pooling, hashtag pooling, and tweet pooling for the tweets that were not used by any of the two previous methods. However, we decided to introduce time as a variable to determine the included documents. This was achieved by performing pooling on a daily basis. Therefore our pooling approach consists of gathering documents for each day, for instance, for a given day $d$ in the analyzed period, and then we perform:

Conversation pooling

By gathering all the documents in day $d$ that are either replies or quotes (*a.k.a.* commented retweets as retweets add no new text content) and traversing their conversation tree backward.

This traversing process stops when the root tweet is found or when the next parent tweet is older than $d - 2$ (two days old). We tested values ranging from 1 through 5 with little impact on the number of documents, 2 also yields a proper combination of recency and comprehensiveness since conversation trees on Twitter are not expected to last for many days, due to its design being oriented towards interaction with the most recent content;

Hashtag pooling

By considering only original tweets (no retweets), posted in day $d$ and in Portuguese `"pt"`. These tweets are clustered by hashtag usage. If a tweet has $h$ hashtags, its text belongs to $h$ documents;

Tweet pooling

By considering the tweets from the previous step – hashtag pooling – that had no hashtags.

Following the pooling process for a given day, we clean the text of the documents by:

- Removing URLs, numbers, mentions, hashtags, punctuation, and emojis;
- Excluding stop words in Portuguese, English, Spanish, Italian, and French;
- lowercasing all text;
- Stemming words.

#### 6.3.1.2 Model training

We tested two libraries for the LDA model training: Sklearn [44] and Gensim [45]. Both these libraries implement an online (*a.k.a.* incremental) version of LDA, as originally proposed in [33]. This implementation was chosen for three reasons: it provides a faster training speed than the original implementation; it can be trained on each day's data and then evaluated – we used this during our initial experiments for comparing libraries; and, being online, can easily be extended to be updated in real-time – an advantage that we do not explore but consider for future work. Practice showed that Gensim was about ten times slower than Sklearn for the same task, so we rely on Sklearn's implementation for the rest of this report.

We focused on training an LDA model for the elections time period as defined in Section 6.1 (p. 47), ranging from September 6th to November 6th, 2019. The online version of the LDA training algorithm has a different set of hyperparameters from the original implementation, we focused on tuning the following (in Sklearn nomenclature):

**`doc_topic_prior`** prior of document topic distribution; **`topic_word_prior`** prior of topic word distribution; **`batch_size`** number of documents to use in each iteration; **`learn_decay`** controls the learning rate update; **`learning_offset`** (positive) value that downweights early iterations in online learning.

[44] https://scikit-learn.org/stable/modules/generated/sklearn.decomposition.LatentDirichletAllocation.html

[45] https://radimrehurek.com/gensim/models/ldamodel.html

In early experiments, we tried different values for the number of topics, one parameter that was not later tuned since we wanted a number of topics that was small enough to be manually analyzed and large enough to capture the diversity of the content on Twitter. The sweet spot we defined was of 64 topics, the value used for the remaining of this work.

Using a grid search implementation provided by Sklearn [16], we tested different hyperparameter values. Since this approach uses cross-validation and we are trying to feed the same order to the model as the actual time order, we used a *TimeSeriesSplit* [17] implementation designed precisely for these cases. The number of folds used for cross-validation is three, since the training process for these models is quite slow. We used the default maximizing score function, log-likelihood. Figure 6.10 and Figure 6.11 (p. 62) show how the different parameter values reflect on the score function. It should be noted that these figures contain a confidence interval since there are three points for each combination of parameters, one per fold in the cross-validation, and the confidence interval is actually the minimum, maximum and middle values, so it is a 100% confidence level interval. It can be used to observe how variable the results can be for the same set of parameters but by varying the one parameter omitted in each figure, `topic_word_prior` and `doc_topic_prior`, respectively.

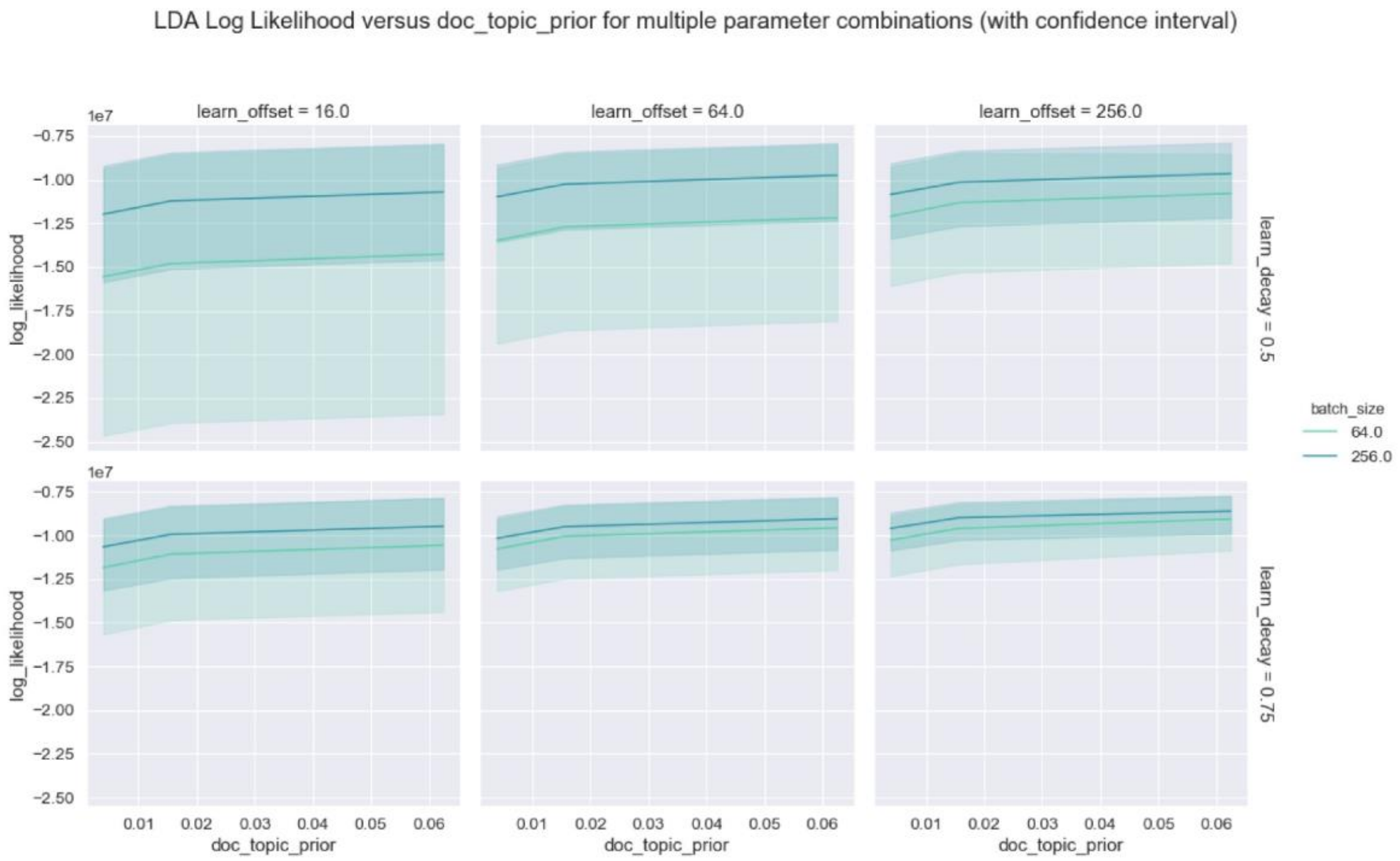

Figure 6.10: Grid search hyperparameter tuning for LDA, excluding `topic_word_prior`

[16] https://scikit-learn.org/stable/modules/generated/sklearn.model_selection.GridSearchCV.html

[17] https://scikit-learn.org/stable/modules/generated/sklearn.model_selection.TimeSeriesSplit.html

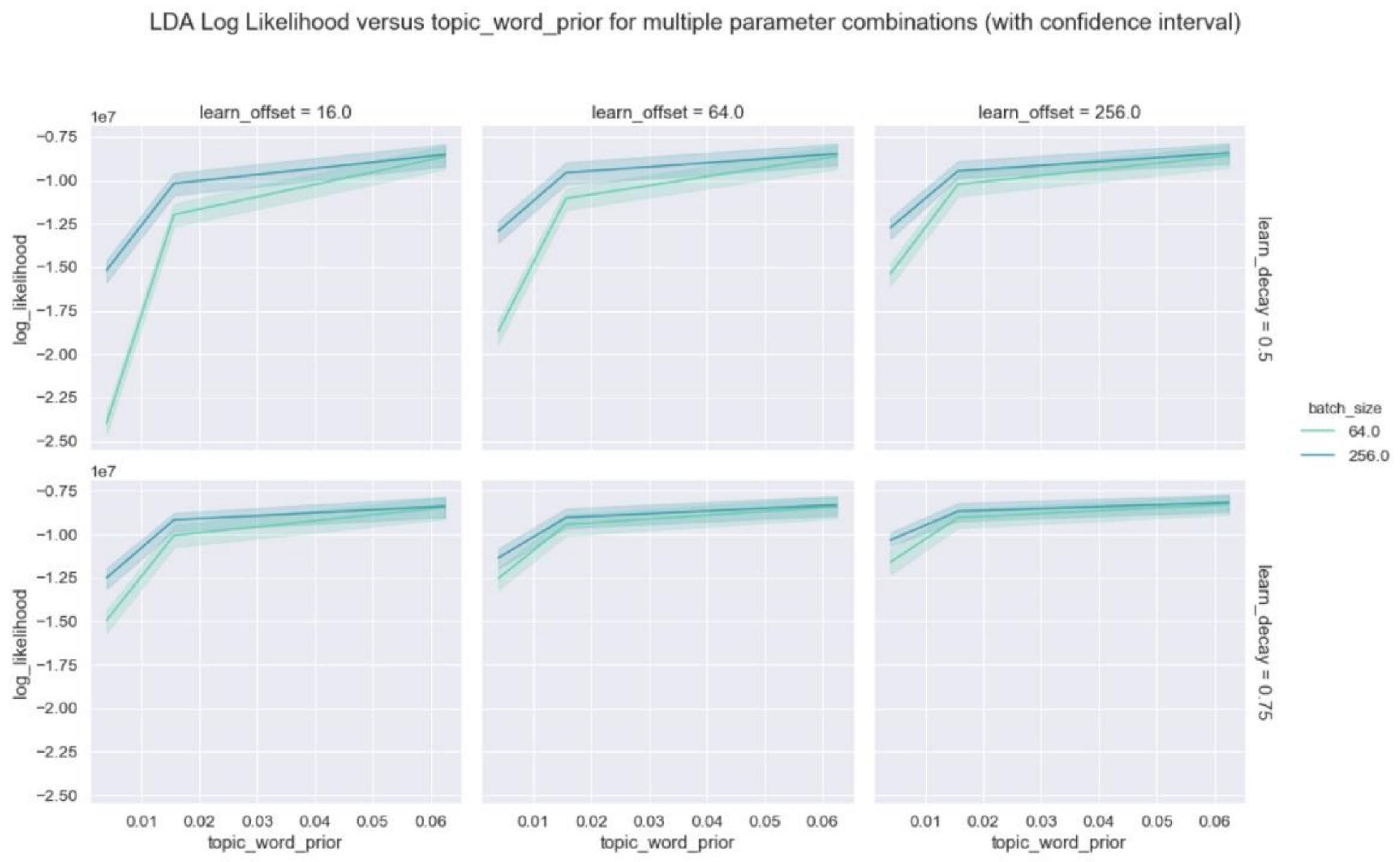

Figure 6.11: Grid search hyperparameter tuning for LDA, excluding `doc_topic_prior`

This tuning process revealed that increasing `learning_offset` or `learning_decay` ensures a faster convergence. Also, lower values of `topic_word_prior` or lower values of `doc_topic_prior` lead to worse results. We did not pursue the grid search process for larger values of `learn_offset` because the values of log likelihood are essentially stable for the best `topic_word_prior`-`doc_topic_prior` pairs, regardless of `learn_offset`. The same goes for `learn_decay`, in which case values of 1 would yield worse results.

On top of this empirical validation, we also made sure that the output of the model would make sense. To achieve this, we looked at the top 20 words representative of each topic and asserted that most of these related to relevant topics. Commonly, not all topics capture a perceivable topic since some will tend to capture stop words specific to the dataset. In any case, the next section contains a more comprehensive reflection on this matter, since we focus on manually describing each topic.

Still within the model training process, we ended up with the following parameters for the best model:

```
{
  'batch_size': 256,
  'doc_topic_prior': 0.0625,
  'learning_decay': 0.75,
  'learning_offset': 256,
  'topic_word_prior': 0.0625
}
```

Figure 6.12 (p. 63) shows how the best model – henceforth the LDA model or topic model –

describes the information on each day of the period it was trained in. This figure uses the measure of perplexity that, as its name hints, is a value that shows how “surprised” the model is by seeing a given piece of data. The lower this value is, the more confidently we can say the model explains the information it sees.

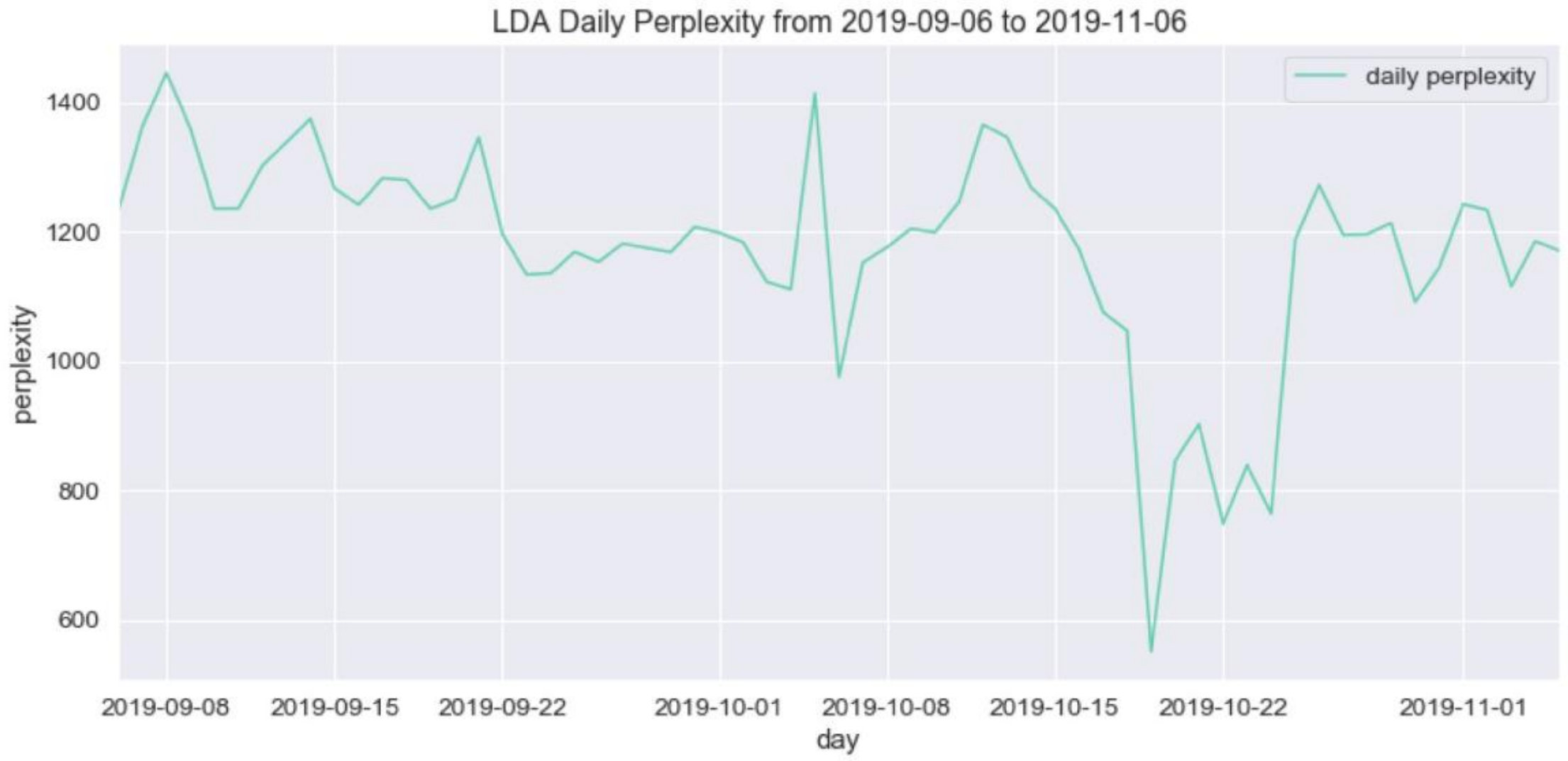

Figure 6.12: Daily perplexity values for the best trained model and the training documents, from September 6$^{th}$ to November 6$^{th}$, 2019

This plot shows two exciting phenomena. First, we see a spiked increase followed by a spiked decrease corresponding to October 5$^{th}$ and 6$^{th}$, the day before the elections and the elections day. In Portugal, the day prior to any election event is considered a reflection day and political discourse is discouraged. In theory, that would result in the discussion of different topics, making the discussions less monothematic than in previous days. The spike seen on that day in the perplexity value would be in agreement with this behavior. The elections day itself corresponds to a significant decrease in the perplexity value, perhaps because that day ends up being more monothematic in return, again the theory is aligned with the observed results. Second, there is another abrupt decrease in the perplexity values starting on October the 19$^{th}$. In respect to this phenomenon, we have investigated news reports of those days and we found this day was marked by the independence protests in Catalonia that stretched for some time afterward. Appendix D (p. 112) shows the covers of two national newspapers for that day. Although this analysis is susceptible to confirmation bias, we find it useful to forward at least one plausible explanation for the observed evidence.

#### 6.3.1.3 Topic Labeling

Following our model training we now focus on explicitly interpreting the learned topics. To this end, we used pyLDAvis [18] – an interactive visualization tool. Figure 6.13 (p. 64) shows how this interface is and exemplifies the analysis of a political topic. Of the 25 top words, on the right side of the image, we can identify 14 as clearly political, like the names of political parties (*ps,psd,cds,bloco,pcp*), political terms referring to elections, voting, and political views.

[18] https://github.com/bmabey/pyLDAvis

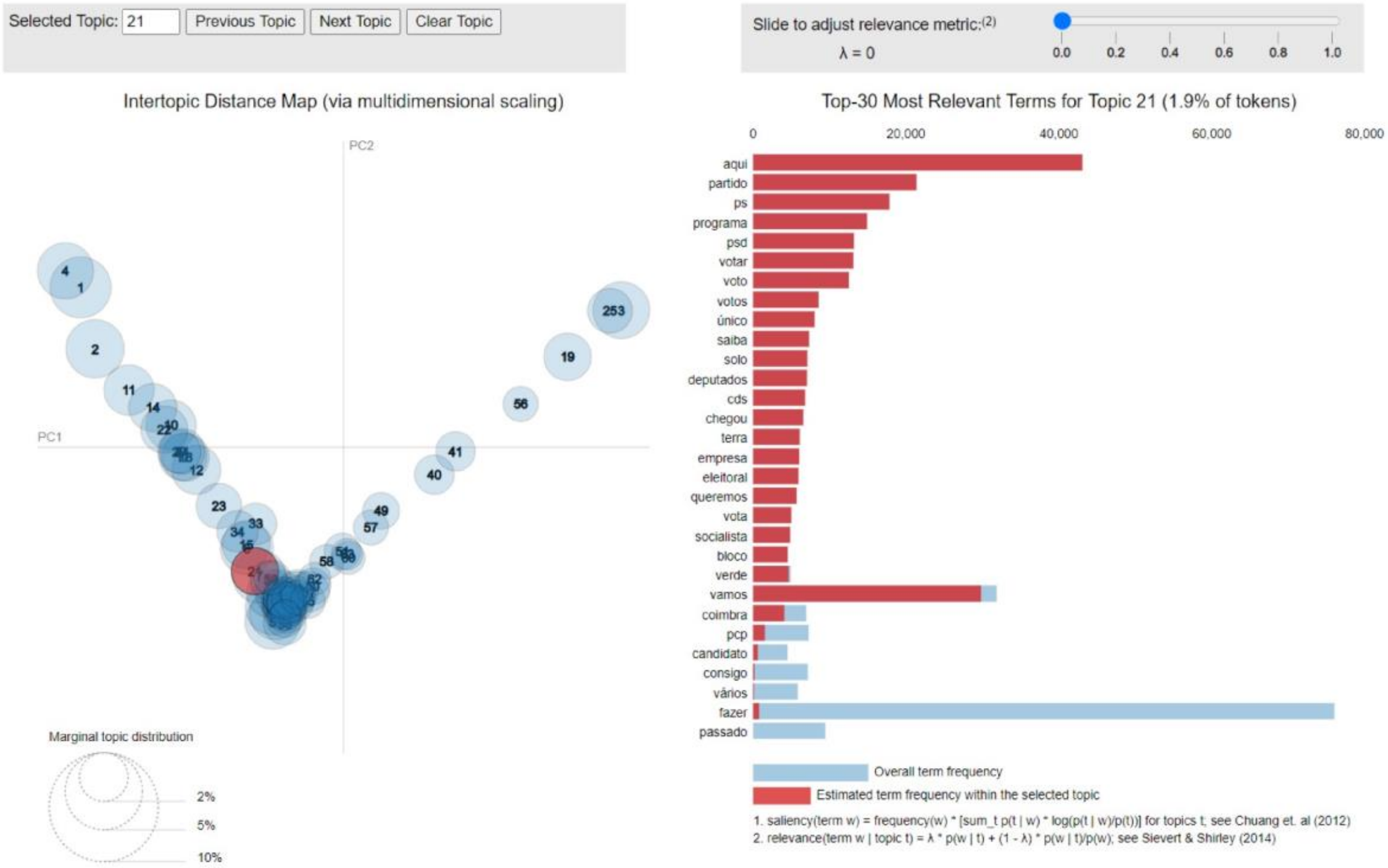

Figure 6.13: PyLDAviz print-screen of the visualization of a political topic

An expert was chosen to label each topic's category and give it a title. This expert happens to be the first author of this report, who is considered to be capable of understanding Portuguese and international terms and then group them in topics. The subsequent analysis of the 64 topics yielded:

- 10 *politics* like the one exemplified in Figure 6.13;
- 8 *international politics* – political topics in the United States of America, China, United Kingdom (Brexit), Spain and Catalonia (independence struggle), as well as in Brazil;
- 15 *junk* – captured mostly stop words or words in which no specific topic was discernible;
- 31 *miscellaneous* – every other discernible topic.

#### 6.3.1.4 Topic Analysis

With labels assigned to each topic, it is now possible to take advantage of that fact in order to compare different types of topics and their evolution through time. Figure 6.14 (p. 65) does precisely this. In it, we can see each day's average topic distribution over all the available documents. This figure also contains a highlighted three-day range around the elections of October 6$^{th}$. By immediate analysis, we can observe that a few topics are much more present than others, as is the case of topic 63 (the last one). We can also observe some spikes at different moments in time.

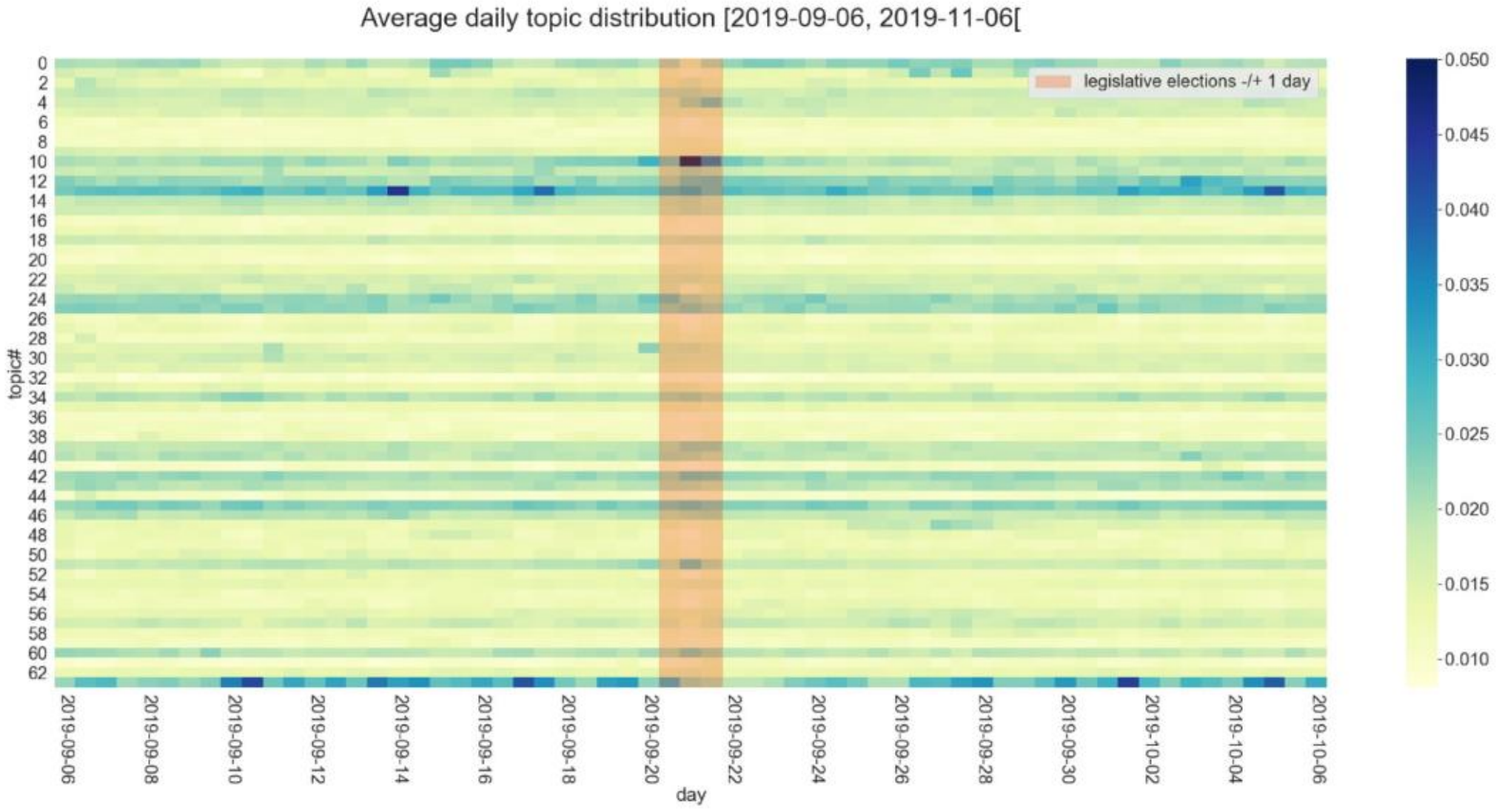

Figure 6.14: Heatmap with average topic attention from September 6th to November 6th, 2019

By matching topic descriptions with their index, we can conclude that the last topic is related to Portuguese football. If a more detailed look is given to the spikes in this topic, we can actually see that they mostly correspond to games between one of the three central Portuguese teams (SLBenfica, FCPorto, and Sporting). Another exciting aspect is how some topics show spikes either on the elections day or in the preceding days, for instance, topics at indexes 11, 51, and 61. These topics had actually been previously inserted into the *politics* set of topics. By isolating only the ten political topics, this becomes even more evident (*cf.* Figure 6.15), a good indicator that the model captures topics as we intended it to, and that the manual labeling process is apparently in agreement with reality.

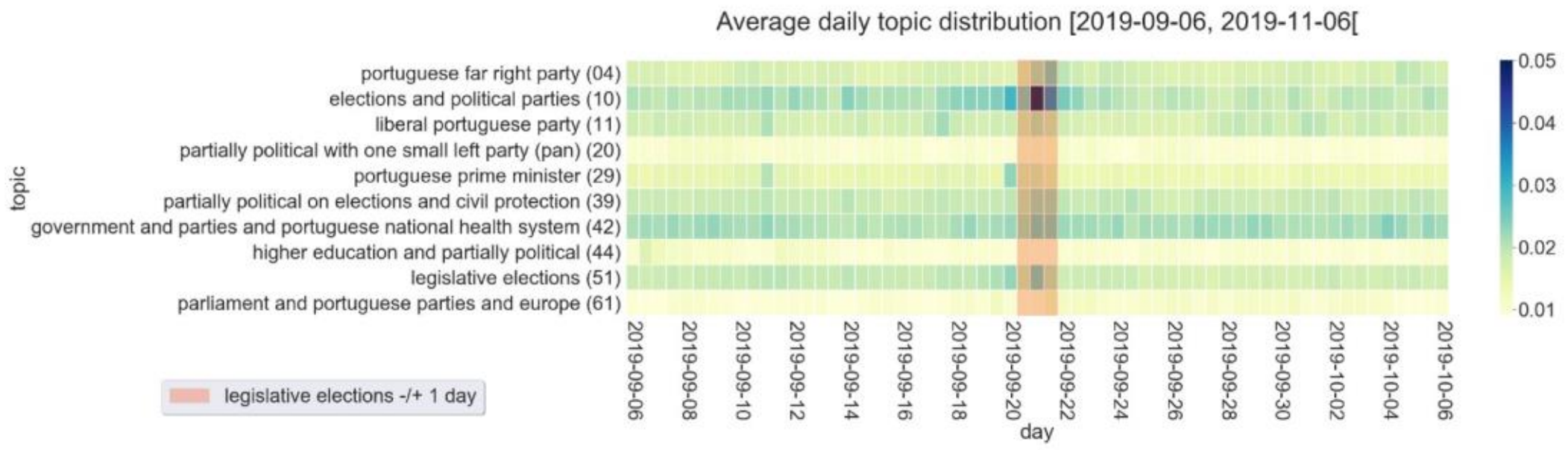

Figure 6.15: Heatmap with average political topic attention from September 6th to November 6th, 2019

By observing this heatmap (*cf.* Figure 6.15, p. 65) we see:

- Topics 10 and 51, with the highest increase on the elections day are the two most related to that specific topic;
- Topics 20, 44, and 61 have a minimal relative presence compared to the remaining topics;
- Topic 4 has an increase in the day after the elections, perhaps due to this party having elected a parliament deputy for the first time;
- Topics 10, 11, 29, 39, and 42 seem to have a simultaneous increase on September 16th. After some investigation, we concluded that this is most likely due to an election poll whose results were released on that day, announcing an unexpected prediction in the result. Appendix E (p. 116) contains a newspaper cover highlighting that.

Some of the above observations are further highlighted by plotting the ratio of variation between a day and the previous day (excluding the first) per topic (*cf.* Figure 6.16). Another relevant observation is that the election period stands out in the political topics; this is relevant to validate the assumption that topic-based approaches can indeed be a good starting point when investigating political events on Twitter. Finally, the election day seems to mark a threshold, after which the variations in the topics are much less frequent than before the elections.

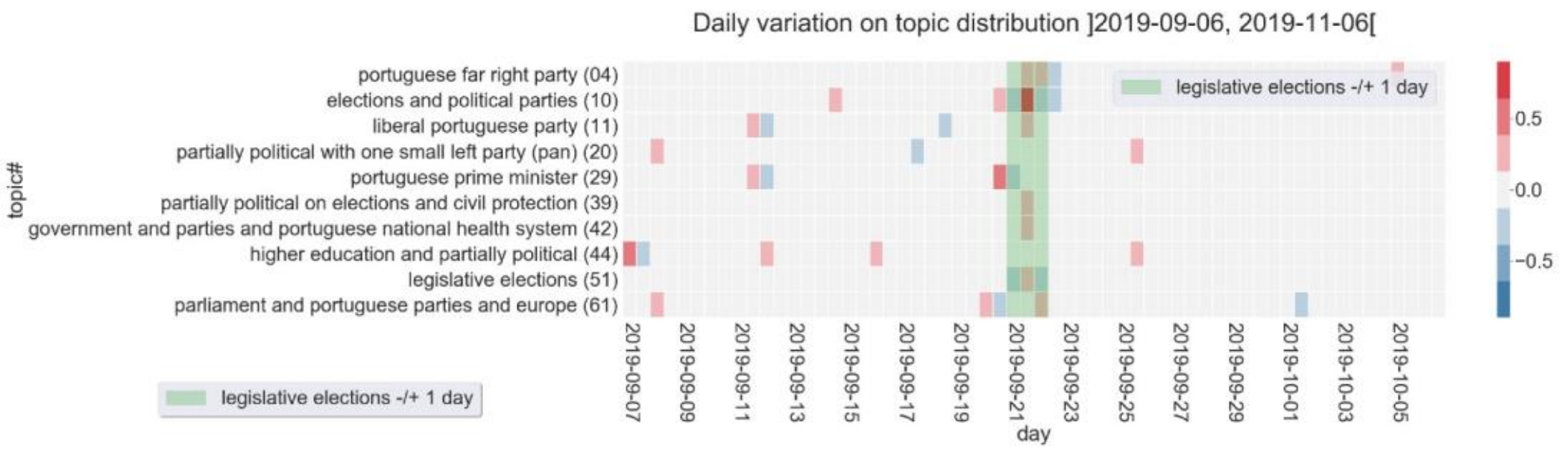

Figure 6.16: Heatmap with average political topic attention daily variation ratio from September 6th to November 6th, 2019

#### 6.3.1.5 Malicious Activity Detection

Now, a question:

*Is there a small number of accounts that is responsible for a significant part of a topic's discussion?*

To answer it, we calculated the cumulative contribution of each user for each topic, using our LDA model on the complete set of tweets each user posted – notice that the need to pool tweets into documents is only required when training the model. Since we are now using it and not training it, the length of the documents is not expected to be a limitation. Additionally, since we are focusing on the elections, we restricted our time window to start on October 1st (one week before the elections) and end two days afterward, on October 8th.

With this in mind, we sorted the users by decreasing total contribution to each topic and plotted the cumulative value of this list *vs.* the number of users required to achieve it. This is what Figure 6.17 shows. In this figure, we have separated political from non-political topics, and also the football topic – the one with much activity in Figure 6.14 (p. 65) – since it was the leftmost curve here. Indeed these curves will tend to be skewed leftwards when the number of accounts required to explain a given topic is smaller (for a given explained ratio). The actual shape of the curves, which looks logarithmic, indicated that overall a small number of users, around 10,000 are enough to explain approximately 80% and that the next 25,000 users are responsible for the remaining 20%. Notice that this behavior is generally the same across all topics. Political topics seem to approximately follow the same curves as non-political topics, although some are closer to the football topic than others.

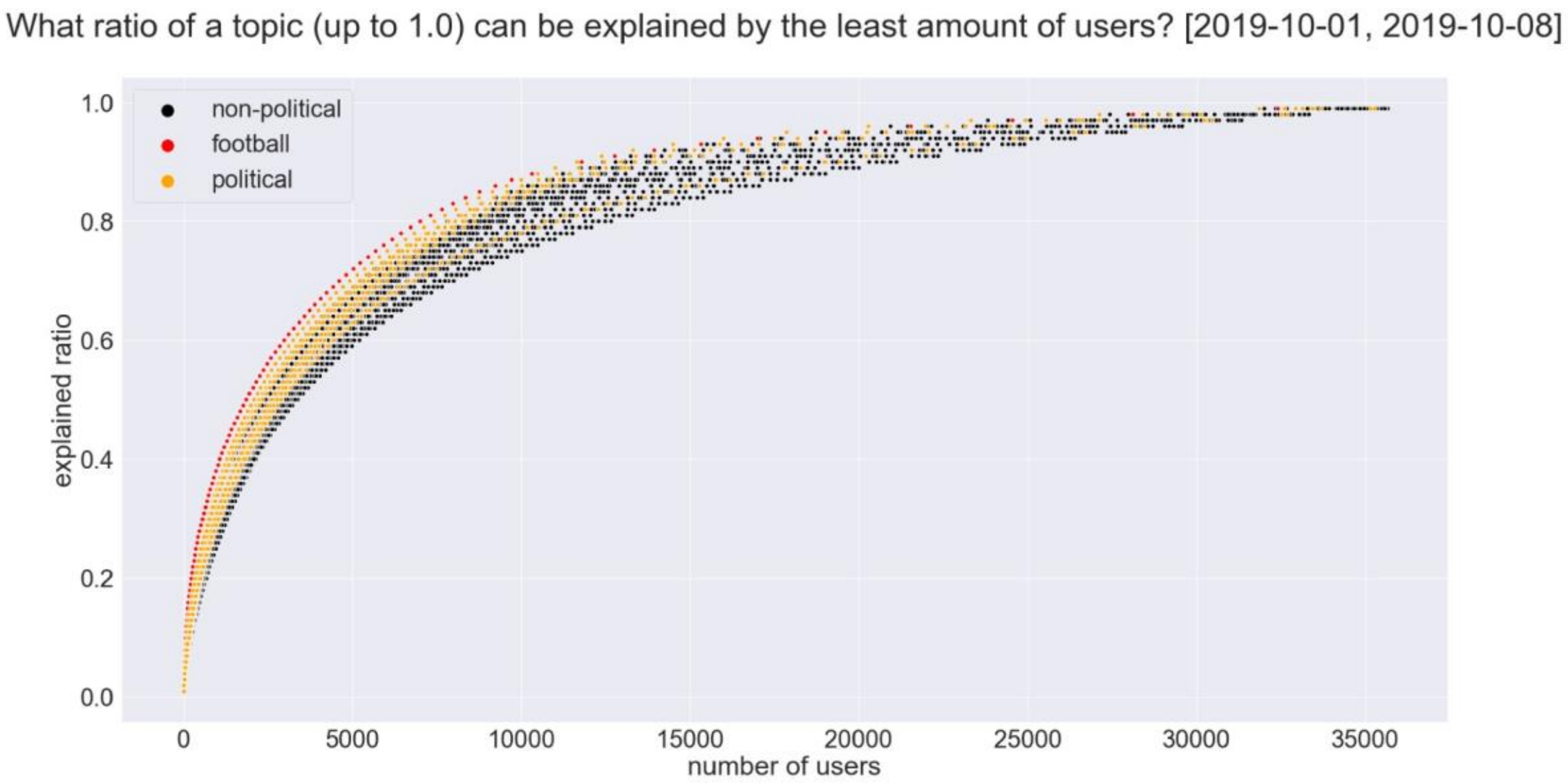

Figure 6.17: User accounts required to explain each topic's generated content from October 1st to October 8th, 2019

By isolating only the political topics and putting a threshold of 0.25 on the observed explained ratio, it becomes easier to see how some of these topics like *elections and political parties* are more skewed to the left, this means that overall, fewer accounts explain the same amount of content generated on a given topic. This is visible in Figure 6.18 (p. 68). So, indeed, *there is a small number of accounts responsible for a significant part of a topic's discussion*.

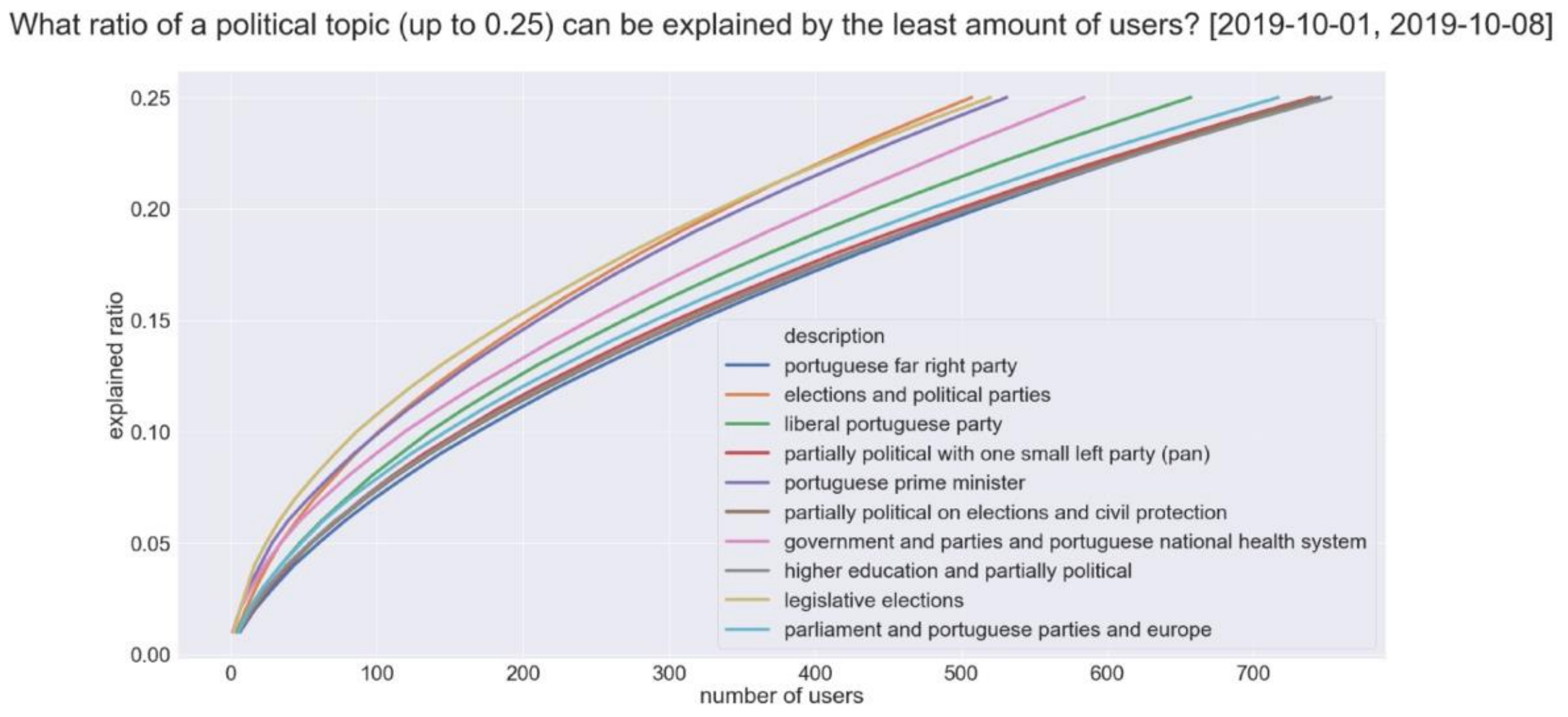

Figure 6.18: User accounts required to explain each political topic's attention from October 1st to October 8th, 2019

Another question rises:

*Are topics where a smaller amount of accounts explains the same amount of content more related to trolling activity?*

In order to answer this question, we decided to investigate the 75 top contributing accounts for each of the political topics in search of disproportionate distribution of troll activity. Instead of having the potentially 750 accounts in these 10 topics, there were only 270 accounts – 36% of the 750 maximum – meaning there are many accounts that contribute a lot to many political topics. In these accounts, we found two are in our peer's set, none has been suspended, and then we identified 17 new troll accounts and 17 clickbait accounts. Of the other accounts, we should highlight that some of them were actually news entities – that are expected to have a lot of activity, particularly in political topics.

Accounts were marked as trolls if we found at least two original tweets that fell into one of the categories described in [27], to which we add disinformation, fake news spreading, or hate speech. Appendix F (p. 117) contains the information we used to make these decisions, namely the malicious tweets identified for those accounts marked as trolls. Accounts were marked as clickbait when they had a systematic behavior of sharing links to websites without adding any relevant content – @opinadouro is one such example. In any case, we only use the accounts labeled as trolls and not the ones labeled as clickbait as a new set of malicious accounts that comes as an addition to those defined in Section 6.2 (p. 52).

Figure 6.19 (p. 69) and Figure 6.20 (p. 69) show how both troll and clickbait accounts are spread among the topics and the most contributing accounts. Although we expected the topics where a smaller number of accounts explained the same amount of content (in this case we focused 8% as a threshold since varying this value can change the order slightly) to have a disproportionately larger amount of either trolling or clickbait activity, this was not what the actual results show. However, we did find that this approach led us to easily uncover malicious accounts, a significantly higher number than expected.

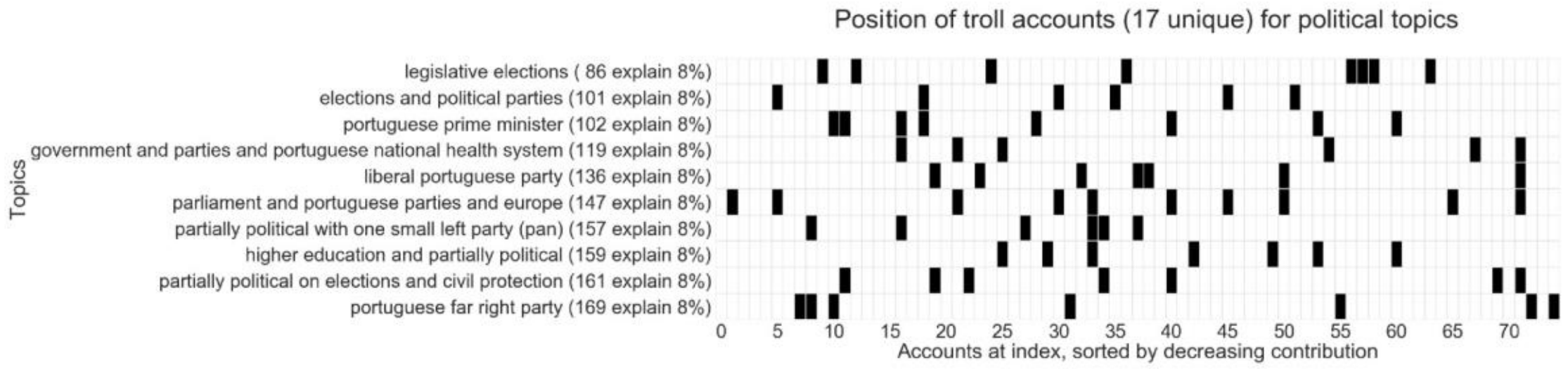

Figure 6.19: Punch card of troll accounts index among the accounts that contribute the most for each political topic

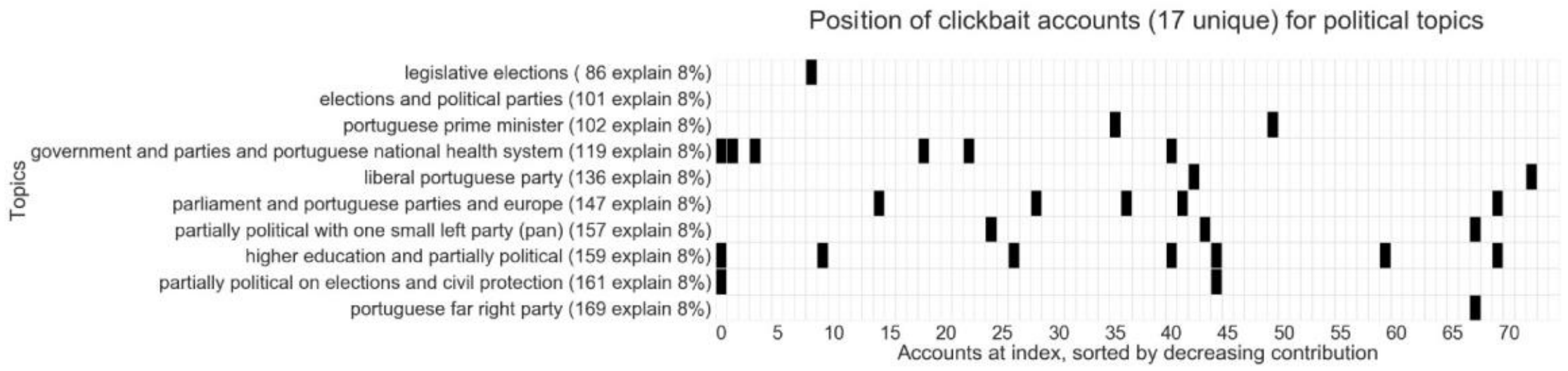

Figure 6.20: Punch card of clickbait accounts index among the accounts that contribute the most for each political topic

#### 6.3.1.6 Content Clustering

The final step we took using this topic model was to cluster accounts with a similar topic distribution. To achieve this we represented each watched user account as a vector averaging the topics distribution of all of their tweets from September 6th to November 6th, 2019. Of the 1,079,379 watched accounts, a total of 104,999 (9.7%) had at least one tweet to consider during this time interval. Table 6.3 (p. 70) shows how many accounts exist of each type, noting that the `malicious` type is the aggregation of `fakenews`, `peer`, `suspended`, and `lda_found_trolls` (*a.k.a.* `lda`). It should be noted that we made sure to avoid overlap in the malicious account types. So, when one account falls into two account types, we choose to remove it from the *weakest* type. The notion of weakness has to do with how informative each account type so we consider the decreasing order of information (from strongest to weakest) to be: `suspended`, `peer`, `lda`, `fakenews`. In practice, overlapping was uncommon but still existent, namely:

- 1 `fakenews` account was also in `suspended` – removed from `fakenews`
- 1 `peer` account was also in `suspended` – removed from `peer`
- 7 `lda` accounts were also in `fakenews` – removed from `fakenews`
- 55 `peer` were also in `fakenews` – removed from `fakenews`

Table 6.3: Number of accounts per account type – separating regular from malicious account types – for LDA dataset

| Account Type | #Accounts | %Total |
|---|---|---|
| regular | 102,665 | 97.78 |
| fakenews | 2,034 | 1.94 |
| peer | 149 | 0.14 |
| suspended | 134 | 0.13 |
| lda_found_trolls | 17 | 0.02 |
| malicious (*a.k.a.* non-regular) | 2,334 | 2.22 |

Considering the number of data-points, and the number of features in our vector – in this case, the number of topics (64), we chose K-means– actually K-means++ [5], a slightly modified K-means version – as the clustering algorithm. This was not our first choice, since we believed a hierarchical clustering approach could allow for an analysis at different clustering levels but the size of the dataset made many choices of algorithms – hierarchical clustering in particular – unfeasible. This infeasibility was first observed by the fact that experiments were taking exponential running times as the size of data-points used grew. This was also confirmed by an empiral benchmarking study [46] of runtime for different clustering algorithms implemented in Python. Since K-means ended up yielding interesting results (*cf.* Chapter 7, p. 89) there was no need to test other algorithms.

Having settled for K-means, we started by training models for a variable number of clusters in order to identify the best number of clusters to choose for the final model. Figure 6.21 (p. 71) shows the evolution of the final model inertia according to the number of clusters. In this case, inertia is the sum of distances of all the data-points to their closest cluster centroid. A greater number of clusters is expected to reflect in a smaller inertia value. In fact, when the number of clusters matches the number of data-points this values is, naturally, zero. The elbow method is a technique for identifying the ideal number of clusters by looking for the inflection in the inertia *vs.* number of clusters curve. Taking this into consideration we settled on using a K-means model with 64 clusters.

[46] https://hdbscan.readthedocs.io/en/latest/performance_and_scalability.html

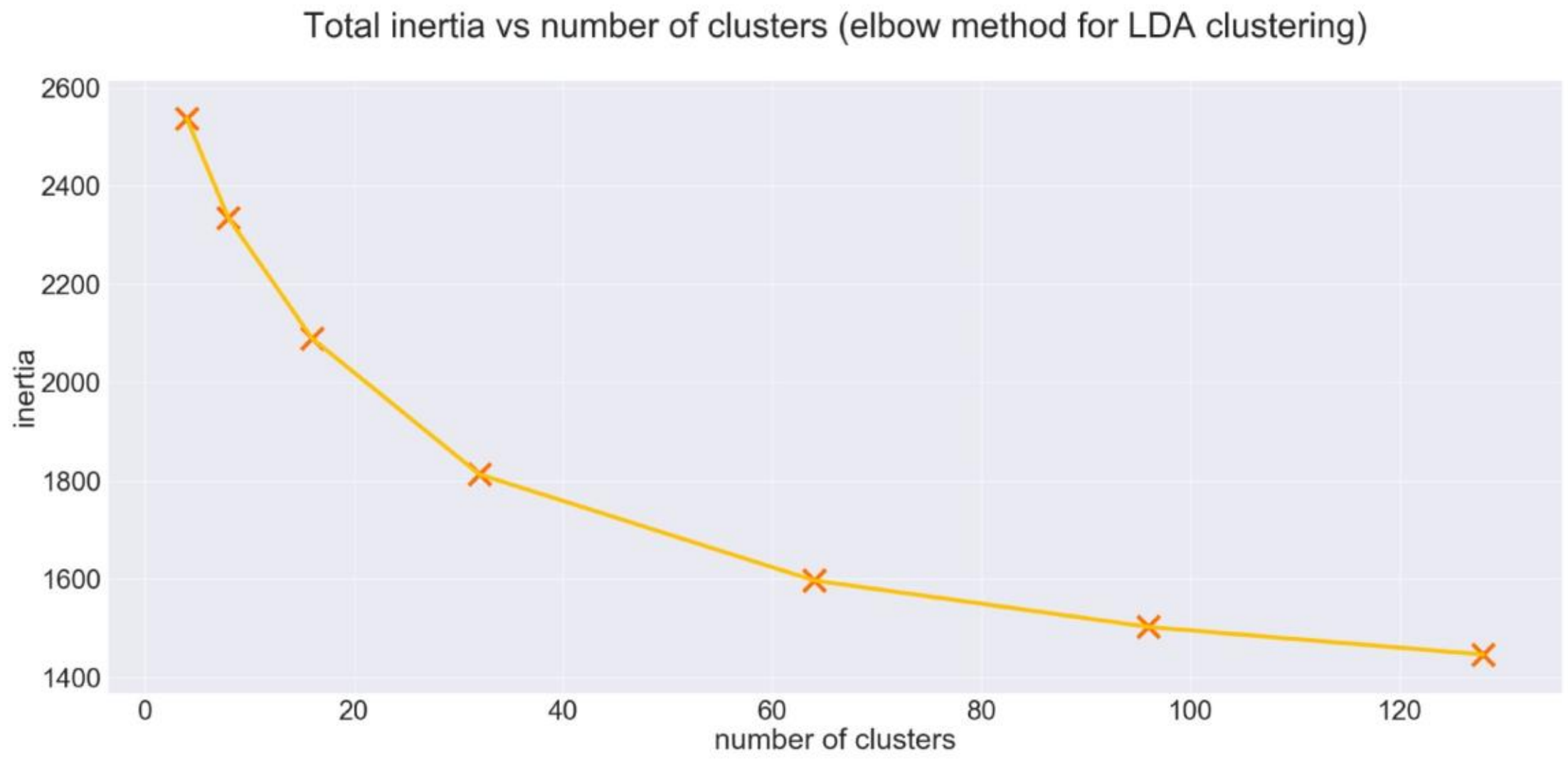

Figure 6.21: Inertia as a function of the number of clusters for elbow method - for LDA K-means clustering

The next step is inspecting how the different account types are distributed across the clusters. Figure G.1 (p. 122) (*cf.* Appendix G, p. 121) represents the percentage of each account type that is in each cluster, and also how this percentage is for all accounts. The 64 clusters and 5 account types make this figure harder to interpret, so we do not include it here although we consider it interesting as it aggregates the different account types and accentuates how different account types are distributed in the clusters. Figure 6.22 presents the same type of plot but only for the *peer*, *fakenews*, and *lda_found_trolls* account types.

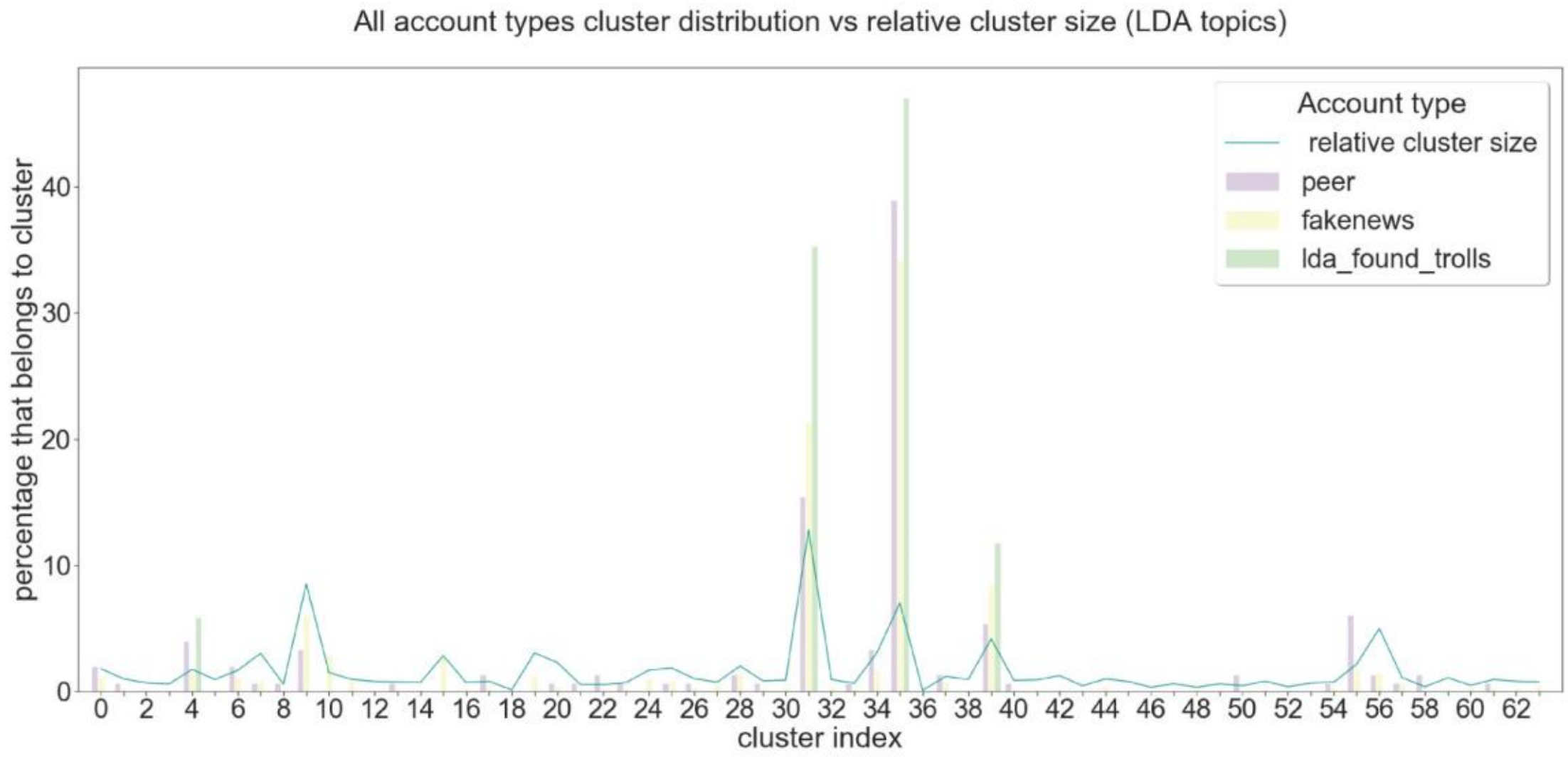

Figure 6.22: LDA cluster distribution for the *peer*, *fakenews*, and *lda* account types

Table 6.4 (p. 72) contains details on the top 3 largest clusters for each account type. These metrics include the id of the cluster ("Cluster"); the number of accounts of the given type in the given cluster ("Count"); the same value as a percentage ("Count (%)"); the cumulative sum of this percentage ("Cum. (%)"); the absolute percentage difference between the given cluster size ratio (for the given account type) and the total cluster size ratio ("% Delta"); and finally, the index at which this cluster is at when sorting a given account type's clusters by their delta ("Delta Index"). Since the *regular* account type represents 97.78% of the accounts, it is expected that the cluster distribution be very similar to its own distribution. Indeed, this results in minimal delta values for the *regular* class.

Table 6.4: Metrics for the top 3 clusters with the biggest amount (count) of each account type

| Type | Biggest | Cluster | Count | Count (%) | Cum. (%) | % Delta | Delta Index |
|---|---|---|---|---|---|---|---|
| regular | 1st | 31 | 12,962 | 12.63 | 12.63 | 0.18 | 2nd |
| | 2nd | 9 | 8,806 | 8.58 | 21.21 | 0.05 | 5th |
| | 3rd | 35 | 6,591 | 6.42 | 27.63 | 0.59 | 1st |
| suspended | 1st | 31 | 19 | 14.18 | 14.18 | 1.48 | 15th |
| | 2nd | 9 | 16 | 11.94 | 26.12 | 3.51 | 1st |
| | 3rd | 7 | 7 | 5.22 | 31.34 | 2.24 | 5th |
| peer | 1st | 35 | 59 | 39.33 | 39.33 | 32.33 | 1st |
| | 2nd | 31 | 23 | 15.33 | 54.66 | 2.53 | 7th |
| | 3rd | 55 | 9 | 6.00 | 60.66 | 3.85 | 3rd |
| fakenews | 1st | 35 | 733 | 34.97 | 34.97 | 27.16 | 1st |
| | 2nd | 31 | 442 | 21.09 | 56.06 | 8.49 | 2nd |
| | 3rd | 39 | 174 | 8.30 | 64.36 | 4.29 | 3rd |
| lda | 1st | 35 | 8 | 47.06 | 47.06 | 40.05 | 1st |
| | 2nd | 31 | 6 | 35.29 | 82.35 | 22.49 | 2nd |
| | 3rd | 39 | 2 | 11.76 | 94.11 | 7.60 | 4th |

This table further confirms what becomes apparent in the cluster visualization figure – that the account types *peer*, *fakenews*, *lda* are highly similar in terms of their cluster distribution, even when considering the overall distribution over the clusters. They share the top two biggest clusters (35, 31) with a large delta value at least for the first one (35) that are 32%, 27%, and 40%, respectively. Additionally, more than half of each of these account types are in those two largest clusters, 35 and 31.

The *suspended* accounts, in turn, have a distribution that is not too distinct from the *regular* accounts, although we can see that their presence has the largest delta for cluster 9, unlike the other types of malicious accounts.

Looking at some of these clusters allows us to understand the topical communities that those accounts form. Cluster 39, for instance, is mostly associated with accounts that discuss football.

This can be seen in the topical heatmap of Figure 6.23. Whereas, out of the top six topics in cluster 35, three are related to politics, one to international politics and the other two mixed topics including the United Nations and tourism in Portugal (*cf.* Figure 6.24).

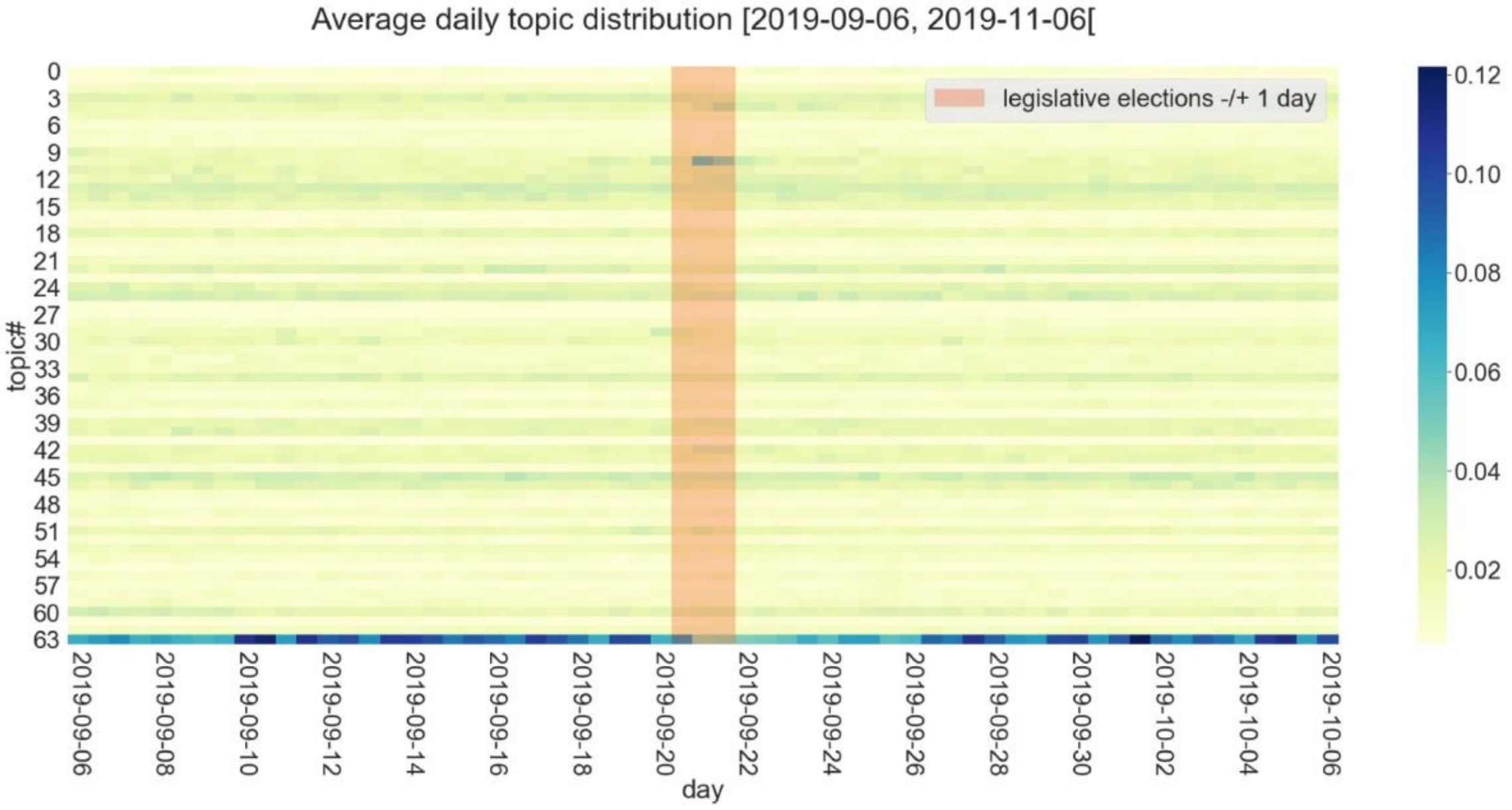

Figure 6.23: Heatmap with average topic attention from September 6th to November 6th, 2019 – cluster 39

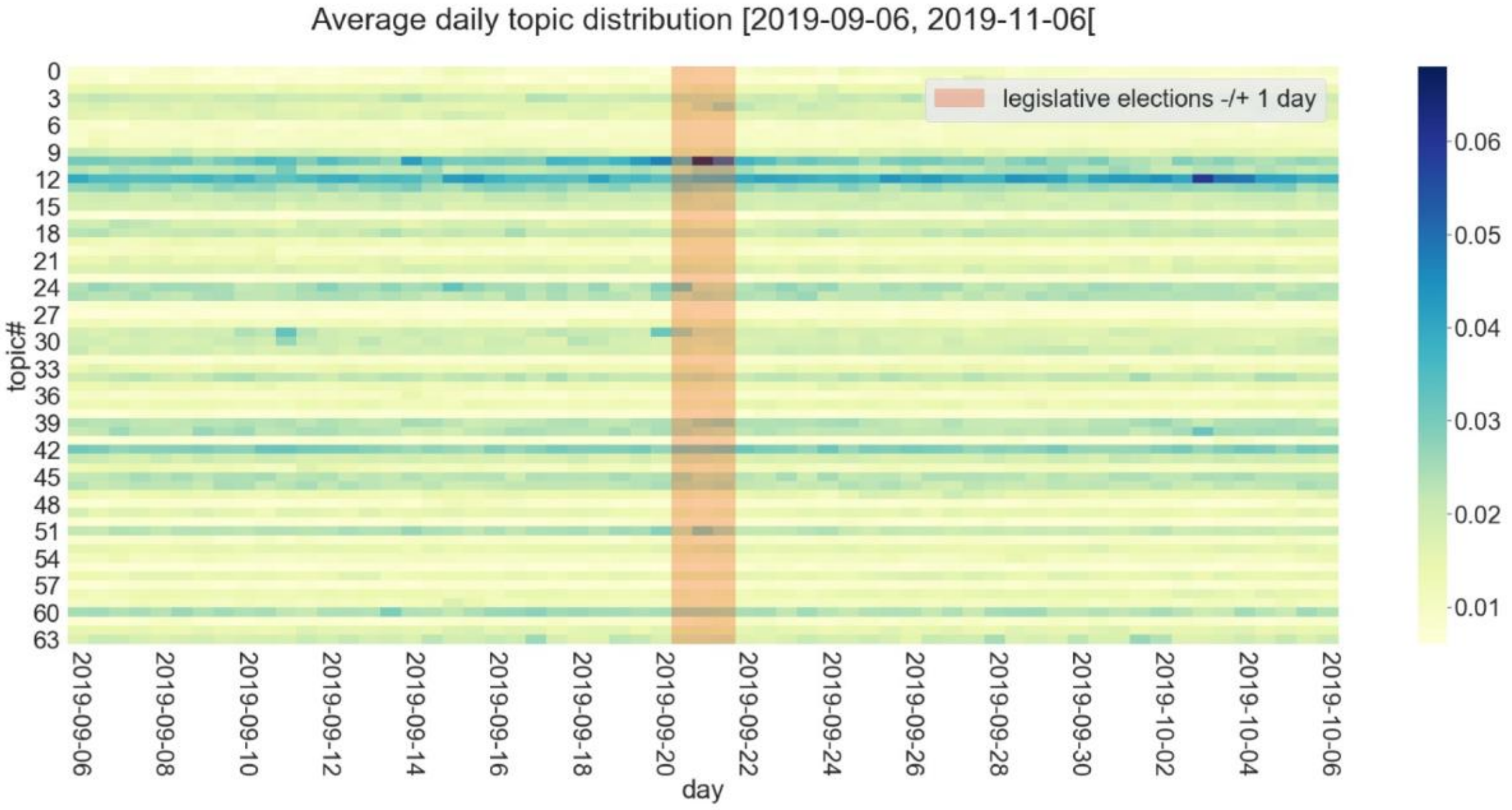

Figure 6.24: Heatmap with average topic attention from September 6th to November 6th, 2019 – cluster 35

In conclusion, we can observe that Twitter's suspensions patterns are not reflecting the type of accounts that fall into the three other malicious account types, *peer*, *fakenews*, and *lda*. Additionally, we believe topical activity analysis met our expectations in terms of both allowing to perform a high level exploration of how different topics are being discussed on Twitter, on identifying political topics and their variation through time, on identifying the accounts that contribute the most for a given topic, and on how groups of topical accounts can be identified and associated with our labeled data-points. All of this with the adjuvant that our LDA pooling strategy proved sound enough to lead to the above-reported results.

#### 6.3.1.7 Content-oriented Exploration Summary

In this section, we focused on using LDA for analyzing our dataset from a content perspective. To achieve this, we defined a strategy for performing document pooling that combined conversation pooling, hashtag pooling, and tweet pooling. We then tuned our topic model and checked how it performed during the period for which it was trained. We manually labeled each of the 64 topics in our model grouping them into politics, international politics, misc, and junk. Focusing on the political topics, we confirmed that the political topics had spikes in discussion focused on the election period, with the time before the election being more active in terms of political content than the period after. We identified 17 new troll accounts by looking at the 75 accounts that were producing the largest amounts of political content. We did not find a disproportionate distribution of those accounts in the topics with a larger amount of its content explained by those 75 users. We then clustered accounts based on their average topical distribution. We found that, within malicious accounts excluding suspended accounts, most were very similar to each other and disparate from regular accounts. Overall, we have found this analysis to produce meaningful results that reflect our expectations in terms of strongest political activity surrounding the election day, and also in terms of uncovered topics, as several were about national politics (10), some about international politics (8), and the remaining were either not discernable/junk (15) or miscellaneous (31).

### 6.3.2 Metadata-oriented

Following our content-oriented exploration, we focused on using metadata information and exploring it to gather a better knowledge of how different account types are associated with different political ideologies. Metadata stands between semantic content and interaction. In our case, we focused on the hashtag feature of Twitter. As explained in Section 2.2 (p. 8), users can associate a set of terms to a tweet in the form #hashtag. Hashtags are used by Twitter to measure trending topics and by users to search for information and tweets where a given hashtag appears. Our initial goal in this section was to assert hashtag usage as a measure of the political orientation of the previously identified malicious accounts, and potentially regular users as well. This task would be both a research challenge as well as an advantage for our other exploration efforts, as these could start exploring relations between accounts with different political affinities. We concluded that this idea was not feasible in the envisioned form for reasons that will be explained further along. However, we did manage to achieve very interesting results in terms of an actual hashtag embedding model, as well as using that model to analyze how our previously labeled account types relate to the main Portuguese political parties. This affinity is not comparable inter-parties, but rather intra-party. Hence it does not represent the political orientation but can be analyzed to measure the variations within each party.

In order to build a hashtag embeddings model we gathered all co-occurring the hashtags in a tweet. To train what is essentially a word-embedding model restricted to the set of hashtags used in the captured Twittersphere, we assumed each tweet to be a different document, and programmed the algorithm to ignore the proximity (order) of the hashtags used by enlarging the window size to be larger than the tweet with the most hashtags. Note that unlike the previous section (*cf.* Section 6.3.1, p. 59) we now use the complete tweets dataset, since we expect more data to be more beneficial than time-constrained data. Hashtags were also lowercased, and those that had non-ASCII characters – mostly related to non-Latin alphabet text – were excluded as they led to unexpected errors when training our models. The final dataset included 21,602,861 total hashtags, 1,610,106 of which are unique (7.45%) from a total of 6,861,372 tweets – after excluding tweets containing non-ASCII hashtags. The chosen library was Gensim's Word2Vec [47]. Table 6.5 contains the top 10 hashtags in our dataset, it is relevant to observe that many have to do with Covid-19 but also that the #portugal is the 4th top hashtag (#bbb20 is related to a Portuguese reality television show). This observation increases the trust in the collection process on the attempt to restrict data collection for the Portuguese Twittersphere.

Table 6.5: List of the top 10 hashtags found in our dataset (excluding tweets with single hashtags)

| Rank | Hashtag | Occurrences |
|---|---|---|
| 1st | #covid19 | 386,869 |
| 2nd | #coronavirus | 383,867 |
| 3rd | #covid_19 | 83,111 |
| 4th | #portugal | 71,662 |
| 5th | #covid-19 | 67,002 |
| 6th | #eu | 62,112 |
| 7th | #china | 47,295 |
| 8th | #brexit | 44,747 |
| 9th | #covid2019 | 39,791 |
| 10th | #bbb20 | 34,916 |

At a first moment, we manually tested the model to ensure that both the similarity between hashtags had been properly captured and that it was able to calculate analogies from those hashtags.

[47] https://radimrehurek.com/gensim/models/word2vec.html#gensim.models.word2vec.Word2Vec

These *ad-hoc* tests were a first attempt at testing our results because we had previously tried smaller samples of our dataset, for instance, including only tweets marked as being written in Portuguese – which yielded non-intuitive results at this stage mostly, we believe, due to being a much smaller dataset. Once we were confident that the current dataset and approach worked seemingly well in these *ad-hoc* tests, we then looked for a context on which we could ask the trained models – with different hyperparameters – to calculate analogies. The most prolific set of terms that we found to achieve this are country names and their capitals. So, we collected a list of 39 country-capital pairs (*cf.* Appendix H, p. 123) leading to a total of 741 different analogies to evaluate. Following this, we analyzed which hyperparameters needed to be tuned – for instance, not the window size, as it was fixed already – those were:

- `alpha` – learning rate;
- `size` – dimensionality of the resulting word vectors;
- `negative` – number of negative "noise word" samples to use.

Since there are a lot of infrequent hashtags in our dataset (89.7% have under 10 occurrences), we chose skip-gram as the underlying neural network architecture of Word2vec instead of Continuous Bag of Words (CBOW) (*cf.* documentation [21]). We tuned a model for several combinations of the above list and calculated the precision@1. This accuracy metric holds the ratio of analogies that were correct on a first guess basis. For instance, if the closest hashtag to the vector resulting from adding the vector $\overrightarrow{\#portugal} - \overrightarrow{\#lisbon}$ to the vector $\overrightarrow{\#madrid}$ is the vector $\overrightarrow{\#spain}$, then the precision value for this analogy is 1, otherwise it is 0. Figure 6.25 contains a heatmap of the average precision@1 for the different hyperparameter values tried. Although we tried higher values for the `negative` parameter, these yielded no better value. The final model chosen was therefore the one with the following parameters: `alpha=0.0275, size=75, negative=15`; with a final precision@1 of 73.4%.

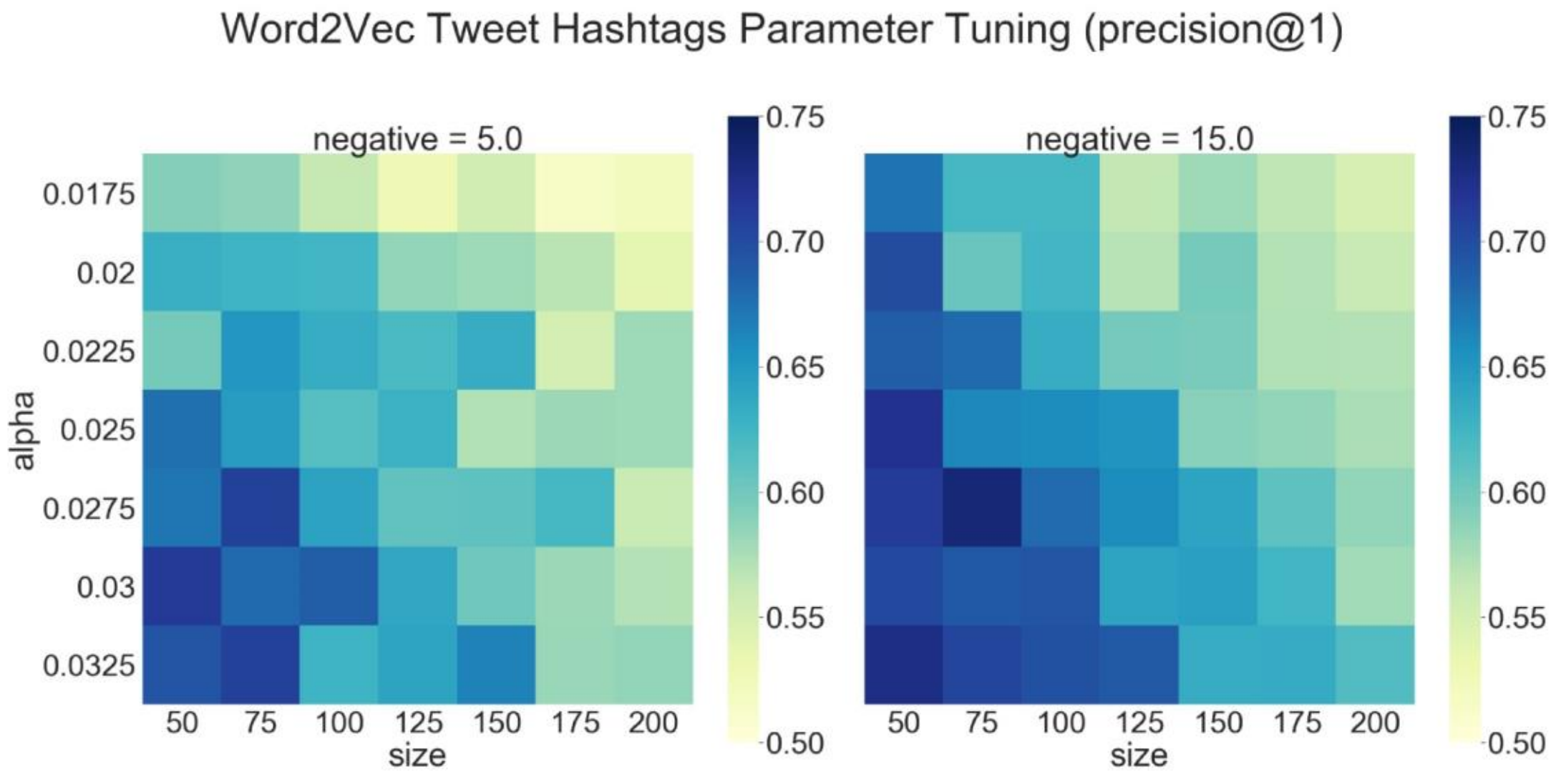

Figure 6.25: Heatmap of hyperparameter tuning for hashtag embeddings with precision@1

[21] https://code.google.com/archive/p/word2vec/

After tuning our model, we intended to use it to label each account's political affinity by calculating the average similarity between each Portuguese political party's top hashtags and the ones used by the account. However, we did not find a way of ensuring that the chosen hashtags were equally representative of each party. Although we could not achieve this initial goal without first solving the above problem, we still studied this similarity between accounts and political parties for the sole purpose of comparing how the different account types we had previously labeled (*cf.* Section 6.2, p. 52) behave within each political party. To achieve that, we manually selected the top six hashtags representative of each political party. This task was performed by looking at their most used hashtags and excluding the ones that were not specific to the party like #covid19. The hashtags chosen to represent each account were the top 15 most used hashtags by that account. We considered only Portuguese political parties that placed at least one deputy in Parliament after the October 6th legislative elections. These parties are listed below, approximately sorted downwards from left to right ideology:

pcp Partido Comunista Português; verdes Partido os Verdes; be Bloco de Esquerda; livre Partido Livre; ps Partido Socialista (*most voted in the elections*); psd Partido Social Democrata; pan Partido das Pessoas, dos Animais e da Natureza; il Iniciativa Liberal; cds Partido do Centro Democrático Social – Partido Popular; chega Partido Chega.

After plotting the average account similarity distribution by political party and by account type, in Figure 6.26 (p. 78), we can observe that there is an interesting phenomenon – the account types *peer*, *fakenews*, and *lda* are distinctly separated from the *regular* account type (and also the *suspended* account type). This should be read as a clear sign that these types of accounts are overall interacting much more with political content than *regular* and *suspended* accounts. Indeed the *suspended* accounts seem to have a similar distribution to the *regular* ones. In most political parties, we see that the median of *suspended* is below *regular* accounts, except for pan, il, and chega. Also, the *fakenews* accounts, although significantly skewed upwards comparing to *regular* accounts and closer to *peer* and *lda*, still have a distribution consistently under those two last types.

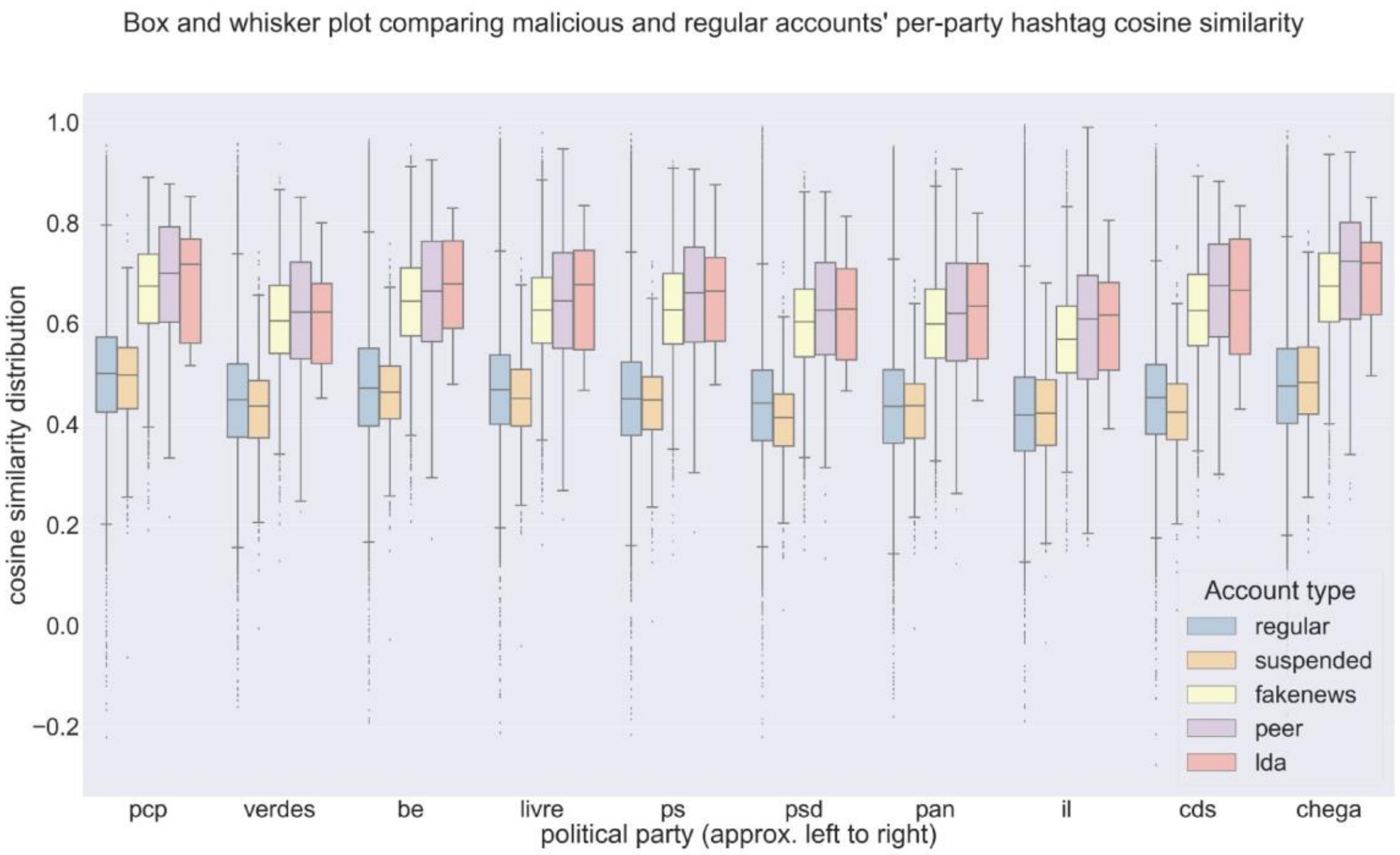

Figure 6.26: Boxplot of hashtag similarity between the different account types and Portuguese political parties

Again, we cannot look at the previous figure in terms of inter-party affiliation distribution for the reasons explained above. However, we can look at each party's distribution – which captures the similarity between the same hashtags – on its own. Table 6.6 (p. 79) contains four sub-tables, one for each account type, with a decreasingly ordered list of delta values. These delta values represent the difference between the average point of the *regular* accounts similarity to a given party and the account type specified on the leftmost column of each sub-table.

Table 6.6: Portuguese political parties sorted by decreasing mean delta for each account type

| Account Type | Party | Delta | Account Type | Party | Delta |
|---|---|---|---|---|---|
| suspended | chega | 0.004 | peer | chega | 0.21 |
| suspended | il | -0.000 | peer | cds | 0.20 |
| suspended | pcp | -0.011 | peer | ps | 0.20 |
| suspended | pan | -0.012 | peer | pcp | 0.19 |
| suspended | ps | -0.013 | peer | psd | 0.18 |
| suspended | be | -0.014 | peer | be | 0.18 |
| suspended | livre | -0.018 | peer | il | 0.18 |
| suspended | verdes | -0.019 | peer | pan | 0.18 |
| suspended | cds | -0.027 | peer | livre | 0.17 |
| suspended | psd | -0.031 | peer | verdes | 0.17 |

| Account Type | Party | Delta | Account Type | Party | Delta |
|---|---|---|---|---|---|
| fakenews | chega | 0.19 | lda | chega | 0.22 |
| fakenews | ps | 0.17 | lda | ps | 0.21 |
| fakenews | cds | 0.17 | lda | cds | 0.21 |
| fakenews | pcp | 0.17 | lda | be | 0.20 |
| fakenews | be | 0.16 | lda | pan | 0.19 |
| fakenews | pan | 0.16 | lda | pcp | 0.19 |
| fakenews | psd | 0.16 | lda | livre | 0.19 |
| fakenews | livre | 0.16 | lda | psd | 0.18 |
| fakenews | verdes | 0.16 | lda | il | 0.18 |
| fakenews | il | 0.15 | lda | verdes | 0.17 |

These sub-tables reveal a few things. First, the suspended account type does not differ much from the regular account types, revealing once again (*cf.* Section 6.3.1.6, p. 69) that the Twitter's suspensions are not representative of the types of accounts that are considered outright malicious, such as the trolls in *peer* and *lda*, neither of the less explicitly malicious *fakenews* account type. Second, we see that the political party chega has a unanimous top 1 rank in the average delta difference to the regular accounts for every account type, with significant differences from the top 2 party for every account type. This observation is curious as this party is known especially for its populist views which, in theory, are expected to drive more attention and separation from the norm. Excluding the small differences observed in *suspended* account types, we also see that the top 3 delta values belong, besides chega, to ps and cds. chega is a far-right party, cds is a democratic right party, and ps is a socialist party and actually the party in power. These results could be questioned if we assume that the hashtag choice process, although performed impartially and similarly for all parties, actually led to hashtags that better capture the difference between *regular*  and malicious account behavior. However, should that be the case, it means that the hashtags are themselves relatively the more "malicious" as the delta values we observe increase, which would lead to the same type of conclusions, simply under distinct assumptions.

#### 6.3.2.1 Metadata-oriented Exploration Summary

In this section, we created an embeddings model of Twitter hashtags for the entirety of our dataset. The final model was tuned against a total of 741 capital-country analogies leading to the best precision@1 of 73.4%. This model was initially intended to measure users' political orientation based on the hashtags they used. This goal was not possible since the choice of hashtags representing each party needed a level of equality that could not be measured or guessed. However, we manually gathered the six hashtags found to best represent each party and check how, for each party, the different types of accounts were distributed. We observed that suspended accounts are very similar to regular accounts for every party. The remaining malicious types of accounts are, like in Section 6.3.1 (p. 59), both very distinct from the other types and very similar in terms of proximity to political hashtags. For each of these types of accounts, we also observed that three parties – chega, ps, cds – showed the greatest distinction from regular users, with the populist farright party chega displaying a noticeably larger distance at the top of all differences. The fact that we were able to validate our model against capital-country pairs with a high precision@1, as well as the fact that the most common hashtags show the data collection process yielded data relevant to the Portuguese context gives some reassurance as to the meaningfulness of the embeddings model we built as well as the results stemming from it.

### 6.3.3 Structure-oriented

This final exploration effort complements our analysis of content (*cf.* Section 6.3.1, p. 59) and metadata (*cf.* Section 6.3.2, p. 74) by introducing an analysis on structure information – in an attempt to capture the interactions between different accounts and see how these form interaction groups or communities.

#### 6.3.3.1 Structure Embedding

Due to the limitations explained in Section 5.2 (p. 29), our dataset does not include the follower/followee relationship for the collected accounts. Although that could have been a good starting point, we believe that it is not the best alternative when the goal is to map the interaction between accounts, mainly because a follow relationship is, first of all, unidirectional – leading to a small number of accounts having a large number of followers, thus to an unbalanced network – and, second of all, this relationship does not necessarily map the strength of a relationship since there is no weight describing how strong the bond is. To overcome these limitations we first built a dataset including all the *retweet*, *reply*, *quote*, and *mention-by* interactions. This choice relates to the belief that the amount of information used would make up for the discrepancies that some of these relationships – *retweet*, *reply*, and *quote* – share with the follow relationship, especially the unilateral direction and consequent unbalanced network. Using this dataset we isolated all the pairs of accounts $A,B$ if:

- $A$ retweets a tweet from $B$;
- $A$ replies to a tweet from $B$;
- $A$ quotes a tweet from $B$ (a retweet with a comment);
- $A$ mentions $B$ in one of their tweets.

In line with the work done in Section 6.3.2 (p. 74), we embedded these relationships with a Word2vec approach, considering each document as a single $A,B$ pair. The result yielded a vector space that had some very dense regions. In practice, this means that when we try to find the most similar accounts to a given query account, the result consisted of a large number of accounts that had a cosine similarity of around 0.999 – a clear indicator that the embedding process was executed on very noisy data. Indeed the manual observation for some accounts known to the first author made no perceptible sense. We believe this was due to our embeddings model being fed directed relationships but being unable to look at them in this way, for one; and also because the data was noisy – for instance, many accounts are expected to retweet, reply or quote tweets from famous people or organizations.

Taking a step back, we realized that in order to capture accounts that formed interaction communities, our dataset should not depend on inherently unbalanced information. The solution lay in using the *mention* phenomenon. From intuition, mentions only occur when someone believes that the mentioned accounts are interested in interacting with a given tweet, as it represents an implicit request for interaction. We assume, from experience, that mentions are not common when the mentioned accounts are not expected to look at the tweet and interact with it, with a similar probability of doing so. Hence this phenomenon is expected to diminish the unbalanced relationships from above. So, we built a dataset for the Word2vec model by using tweets as documents with words corresponding to the accounts mentioned in a given tweet. Not using the mentioning account, reduces the risk of our assumption from above, if wrong, affecting our results. The final tweak introduced was that of only considering tweets that include at least two mentions, since this then allows us to capture accounts that, in the eye of the account mentioning the others, are somehow related, and are related enough to receive a simultaneous mention. Although these are mostly intuition-based reasons, we do believe that they are significant to avoid including the amount of noise the previous dataset used, and to capture an implicit network of where accounts with similar interests and interactions are highly connected. In the end, this new dataset included a total of 64,215,170 mentions for a total of 3,901,678 unique accounts (6.1%).

Indeed, we got better results when using this dataset. This notion stems from manual validation since, unlike in Section 6.3.2 (p. 74), we cannot easily create a set of analogies between accounts to validate our model. Because of this, the validation step for this model was based on manually observing the most similar accounts to a set of accounts known to the first author. The training hyperparameters used are `alpha=0.025, size=128, negative=5`. Additionally, we forced to model to ignore accounts that had less than 25 occurrences to improve the results. From 82 Exploration experience, this did not significantly reduce the number of labeled accounts captured by the model. Lastly, this model did not suffer from the density problem observed in the initial approach. The key idea here is that since we cannot perform a very good validation at this stage, we will have to use this model as-is and check if the outcome of that usage results in meaningful results – which it did, and that is why we did not need to go back and try other parameters. Table 6.7 contains the account occurrences by account type embedded in our final embeddings model.

Table 6.7: Number of accounts per account type – separating regular from malicious account types – for mentions dataset after removing low occurrence accounts

| Account Type | #Accounts | %Total |
|---:|---|---:|
| regular | 264,951 | 97.56 |
| suspended | 5,088 | 1.87 |
| fakenews | 1,306 | 0.48 |
| peer | 228 | 0.084 |
| lda_found_trolls | 13 | 0.005 |
| malicious (*a.k.a.* non-regular) | 6,635 | 2.44 |

Figure 6.27 (p. 83) contains a print-screen of a Uniform Manifold Approximation and Projection (UMAP) visualization of the embeddings by using TensorFlow's embedding projector [48]. First impressions indicate that the embeddings do contain different groups of accounts – based on the embedded co-occurring mention relationship.

[48] https://projector.tensorflow.org/

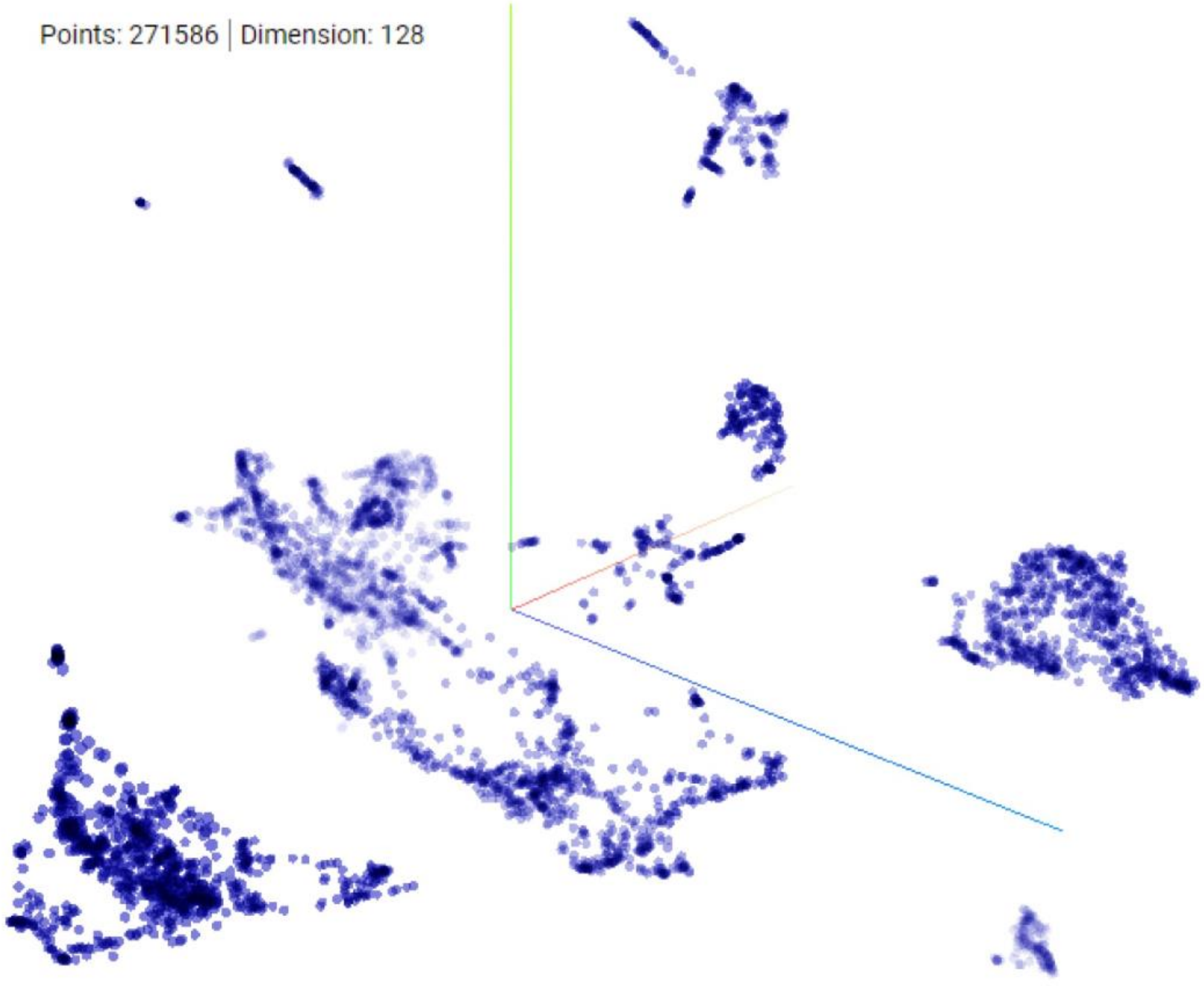

Figure 6.27: UMAP visualization of the mention relationship embeddings

#### 6.3.3.2 Structure Clustering

In the spirit of the approach developed in Section 6.3.1.6 (p. 69), to obtain non-overlapping groups of labeled accounts. We performed a similar study, in this case, on the embedding of tweet co-occurrence of mentions. In practice, what we achieved in the previous section was embedding a network of accounts where the edge between two accounts contains a weight equal to the number of co-occurrences between those two accounts. So, and for the same performance reasons as specified in Section 6.3.1.6 (p. 69), we chose K-means++ to embed the accounts according to their numerical 128-dimensional vector. Figure 6.28 (p. 84) shows a plot of the inertia – sum of distances of all the data-points to their closest cluster centroid – evolution for different cluster sizes in order to use the elbow method to choose the model with an appropriate number of clusters. In this case, we chose the model that uses 64 clusters, coincidentally matching the same number of clusters in Section 6.3.1.6 (p. 69).

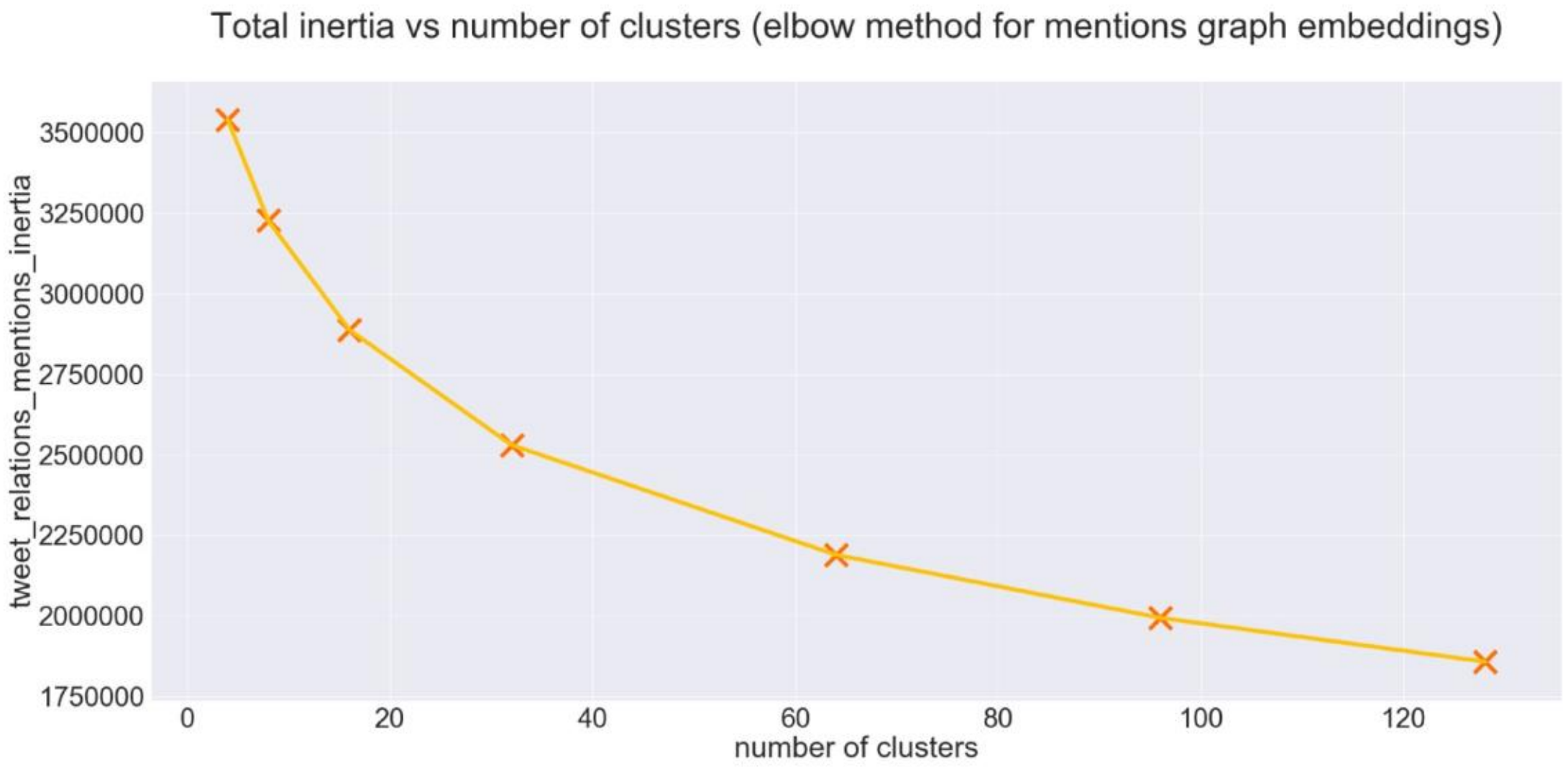

Figure 6.28: Inertia as a function of the number of clusters for elbow method - for mentions embeddings K-means clustering

Appendix I (p. 125) contains a rotated figure with the percentage distribution of accounts per clusters, for all account types, and was not included here since it becomes hard to read horizontally. However, we include Figure 6.29, which contains the same information restricted only to the *peer*, *fakenews*, and *lda_found_trolls* (*a.k.a. lda*) account types. This visualization is considered relevant because it shows how these three account types, unlike the remaining two –*regular* and *suspended* – are very similarly distributed in a small amount of clusters and with a percentage distribution that is much different from the actual percentage distribution of those clusters – this difference is called "Delta" in Table 6.8 (p. 85).

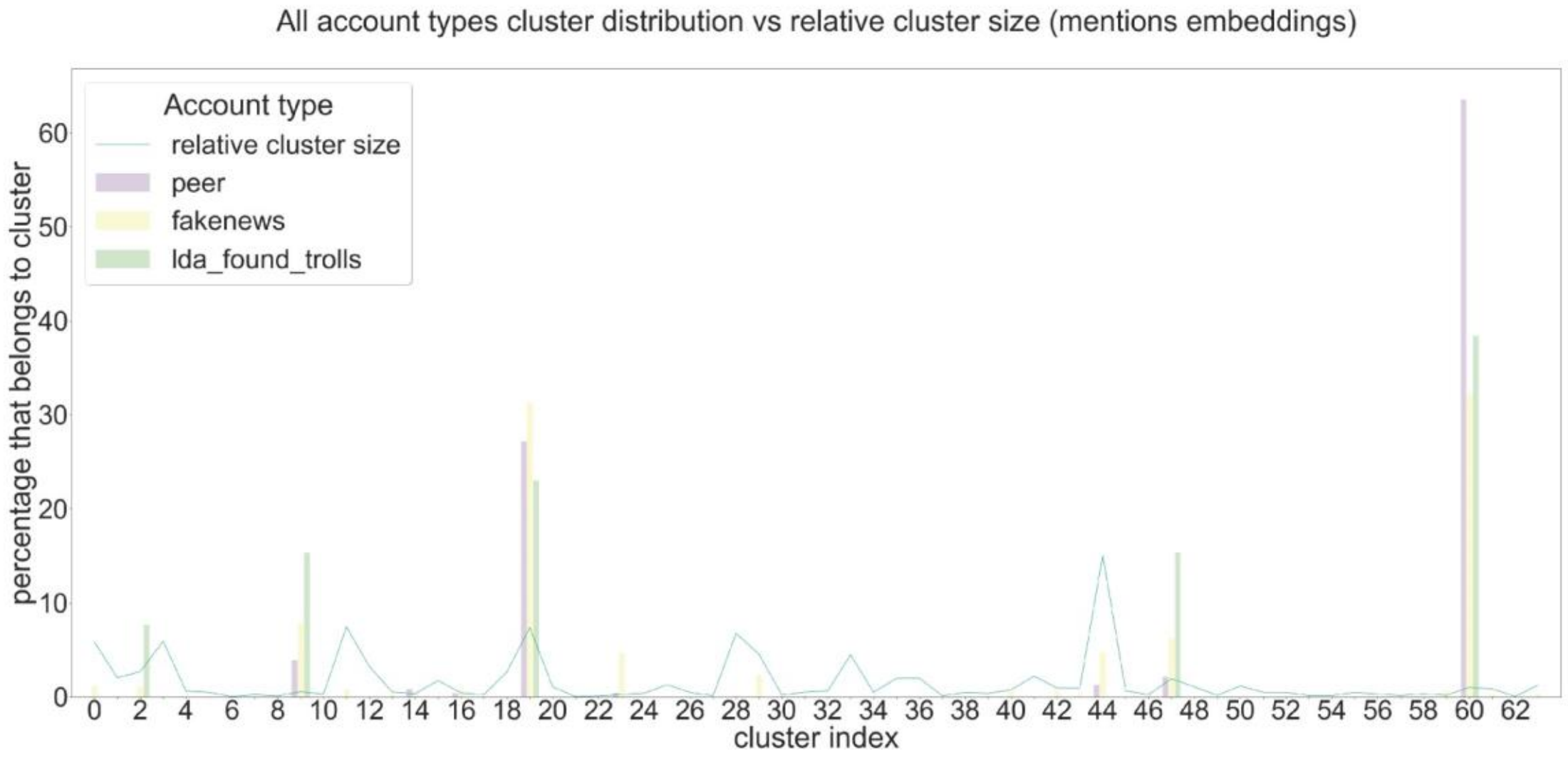

Figure 6.29: Mentions embeddings cluster distribution for the *peer*, *fakenews*, and *lda* account types

Indeed, Table 6.8 further shows that cluster 60 is the one where the greatest percentage of these three account types are contained. Namely, we see that *peer* accounts have 63.30% of its total size in cluster 60, *fakenews* accounts have 32.16% of its total size in cluster 60, and *lda* accounts have 38.46% of its total size in cluster 60. This would not be very meaningful if the relative cluster size was within the same range of values, but actually the "Delta" values are 62.58%, 31.14%, and 37.45%, respectively. This means that they are disproportionately distributed in cluster 60. We can also claim the exact same thing about clusters 19, and 9. These are the top 3 clusters which contain a cumulative value of 94.74%, 71.37%, and 76.92% of all the accounts in the three types specified before, respectively *peer*, *fakenews*, and *lda*, showing disproportionate distributions that are not as large but equally significant.

Besides this observation that these three different types of malicious accounts are disproportionally distributed and clustered together in terms of co-occurring mention interactions, we see that the *suspended* accounts seem to have a distribution that is not as extremely different from that of regular users but does however show few clusters with a disproportionate presence, more noticeably in cluster 0, with a "Delta" difference of 10.96%.

Table 6.8: Metrics for the top 3 clusters with the biggest amount (count) of each account type – mentions embeddings

| Type | Biggest | Cluster | Count | Count (%) | Cum. (%) | % Delta | Delta Index |
|---|---|---|---|---|---|---|---|
| regular | 1st | 44 | 39981 | 15.09 | 15.09 | 0.02 | 17th |
| | 2nd | 11 | 20129 | 7.60 | 22.69 | 0.16 | 3rd |
| | 3rd | 19 | 19257 | 7.27 | 29.96 | 0.09 | 5th |
| suspended | 1st | 44 | 881 | 17.32 | 17.32 | 2.25 | 6th |
| | 2nd | 0 | 854 | 16.78 | 34.10 | 10.96 | 1st |
| | 3rd | 28 | 356 | 7.00 | 41.10 | 0.24 | 45th |
| peer | 1st | 60 | 145 | 63.60 | 63.60 | 62.58 | 1st |
| | 2nd | 19 | 62 | 27.19 | 90.79 | 19.83 | 2nd |
| | 3rd | 9 | 9 | 3.95 | 94.74 | 3.40 | 10th |
| fakenews | 1st | 60 | 420 | 32.16 | 32.16 | 31.14 | 1st |
| | 2nd | 19 | 409 | 31.32 | 63.48 | 23.96 | 2nd |
| | 3rd | 9 | 103 | 7.89 | 71.37 | 7.34 | 4th |
| lda | 1st | 60 | 5 | 38.46 | 38.46 | 37.45 | 1st |
| | 2nd | 19 | 3 | 23.08 | 61.54 | 15.72 | 2nd |
| | 3rd | 9 | 2 | 15.38 | 76.92 | 14.84 | 4th |

What is most curious in these results – that we believe is a good indication that the embeddings model is working – is that it confirms the results obtained in Section 6.3.1.6 (p. 69):

- Suspension patterns differ from the trolling behavior identified in *peer* and *lda* account types, but also from the fake news posting behavior identified in the *fakenews* account type;
- There is an obvious similarity between fake news posting accounts and accounts that display trolling behavior;
- Besides content, structure also reveals strong similarities between *peer*, *fakenews*, and *lda* account types that is, at the same time, quite distinct from *regular* account activity.

#### 6.3.3.3 Structure-oriented Exploration Summary

In this section, we tested several approaches at building an embeddings model that would map the interaction structure found in our Twitter dataset. We found that mapping the co-occurrence of hashtags led to a better model, although this was only subject to manual validation since an analogy validation was not suitable. With this dataset we were able to embed the relationships between accounts in terms of how they are mentioned in the same parts of the Twittersphere. Even with our validation limitations, we found that clustering the accounts represented in this embeddings model to yield very interesting results: although not coming from a prior solid model validation, they are in agreement with the results of Section 6.3.1 (p. 59) and Section 6.3.1 (p. 59) increasing our belief in their meaningfulness by propagation.

## 6.4 Threats to Validity

In this section, we forward a few potential validity threats that are inherently present in our analysis. These have to do with both the collection process as well as the exploration. However, they belong in the exploration section since that is where they can be seen as threats to the outcome.

In terms of internal validity threats, that is, endogenous conditions to the way we collected and explored data, we highlight:

- Although we used the same dataset (*cf.* Section 6.1, p. 47) to perform the three independent exploration efforts described in Section 6.3.1 (p. 59), Section 6.3.2 (p. 74), and Section 6.3.3 (p. 80), we did not use the exact same parts of that dataset. For instance, in Section 6.3.1 (p. 59) we fixed a time window to analyze, whereas in the remaining two sections, we used the entire timeline, restricted according to the requirements of that section. This choice could be a problem since we are comparing the results of those different sections. However, these comparisons focus on results per the account types defined in Section 6.2 (p. 52) and complemented in Section 6.3.1.5 (p. 66) and, for these account types, there is a majority of overlap in the different explorations. Additionally, we only included larger or smaller time windows when the methodology under use is expected to benefit from that action. Using more data for embeddings typically leads to better results and, since we did not use time as a variable that could not be a threat. Nevertheless, this comparison between results on accounts that is not completely overlapping can be seen as a validation threat;

- When selecting suspended accounts in Section 6.2.1 (p. 52) we used the most recent data from Twitter. However, it is possible that in a few months, with the benefit of hindsight, there might be more suspended accounts that, at this point, have not yet been included in the set. This situation, although not under our control, could have some impact on the results if the study was repeated at a later time;
- The manual labeling process performed in Section 6.3.1 (p. 59) depended on being able to infer topics from a set of relevant words. Again, this was performed by the first author in the quality of an expert. This expertise derives from being a Portuguese citizen aware of a significant amount of both internal and external events, topics, personalities, and companies. However, there is always a margin for error as this effort was not done by independent expert subjects;
- Most of the conclusions drawn from data exploration are somehow related to political topics (*cf.* Section 6.3.1, p. 59), and even political parties and ideologies (*cf.* Section 6.3.2, p. 74). These were mostly drawn by the first author of this work who has his own political views – as is common in such scenarios and analysis. Even by making every effort to avoid political biases, we believe this aspect is worthy of mention;
- When identifying troll activity and malicious tweets, in Section 6.3.1.5 (p. 66), we defined a set of rules for considering a tweet as troll activity, and also a minimum of two malicious tweets in order to consider an account as a troll. In a transparency effort, we also included the identified tweets (sometimes more than two when doubt could arise), in Appendix F (p. 117). Even so, this analysis could fall under the same problem as the previous point: an implicit and unnoticed political bias;
- Once again, the first-author was the only subject responsible for classifying the content of the newly discovered fake news websites (*cf.* Section 6.2.3, p. 54), which is based on his knowledge and ability to fact check news.
- Finally, we need to highlight that our data collection process is very byzantine by design. This collection approach means it can collect relevant data without specifically being told to collect it but also that it can escape the design constraints of focusing on a given Twittersphere. We tried to measure this throughout the collection and analysis process and also reduce its impact by either excluding data (*cf.* Section 60, p. 44) or using approaches that are not expected to be affected by this diversity like embeddings (*cf.* Section 6.3.2, p. 74). In any case, this is a risk that we took, and one that we stand by.

From an external validity point of view – relating to the generalizability of the results – a couple of other threats arise, namely:

- The conclusions we drew stem from looking at data that is particular to a time window and a Portuguese context – we recognize that, although an effort exists to introduce factors and variables that could destabilize the outcomes – there is no guarantee that our conclusions will not age;
- Even assuming the previous point is a non-issue, a bigger one came about during the Exploration

development of this work: Covid-19. Although we tried to measure the impact of this phenomenon on our data, namely in Section 6.1 (p. 47) and also in Table 6.5 (p. 75), it is hard to predict what the lasting impacts of the pandemic will be on the results of our work.

# Chapter 7

# Conclusions

"Democracy could never be imposed at the point of a gun, but perhaps it could be sown by the spread of silicon and fiber."

Edward Snowden *in* Permanent Record

The current thesis touches a multitude of topics and approaches, in this section we go over each of these. We shall present our conclusions, revisit our hypothesis, enumerate contributions and future work ideas that we believe will make sense given all that has been learned.

Initially, we identified the inherent problem of the massive adoption of Social Networks (SNs) as sources of public information. Online Social Networks (OSNs) being at the center of a conflict between opinion manipulation and democratic values. The risk of compromised freedom of thought exists, it is recognized by the public, by governments, and by the very companies that built these OSNs (*cf.* Chapter 2, p. 5).

Following the contextualization of our work, we investigated what the state-of-the-art approaches focus on. This was achieved via a Systematic Literature Review (SLR) process that stands by itself as a structured survey. Under the flagship of understanding current approaches on malicious political content detection on Twitter, we learned the answers to three main questions. We learned from SRQ1 that current studies focus on extracting account, content, network, and activity data. We learned from SRQ2 that data is used as input for Machine Learning (ML) classifiers, for data representation approaches (like Latent Dirichlet Allocation (LDA) and embeddings), for community detection, and for other tools and intermediary processing stages. We learned from SRQ3 that these analyses tend to focus on performing *ad-hoc* data exploration, developing bot classifiers, or better mapping the taxonomy of the malicious activity in OSNs. The last lesson from this effort was understanding there are several systematic dangerous assumptions in the literature. One is assuming malicious behavior is not adaptable and to classification methods that rely on easily manipulated features. The outcome is that these efforts have no real-world application and thus fail to meet their proposed motivations. Another is assuming that really dangerous malicious behavior can be seen with a magnifier at the account-level (*cf.* Chapter 3, p. 15).

## 7.1 Hypothesis Revisited

We then proposed our hypothesis to defy these assumptions in an attempt to understand if high-level approaches can be used as an alternative, not forgetting that focus should be given to features that are not as easy to manipulate when manipulation is what we want to avoid:

> Do different high-level approaches to Twitter data analysis, when applied to distinct types of malicious activity, a) lead to meaningful results and b) agree among themselves?

To answer it, we analyzed Twitter data within the Portuguese Twittersphere. To analyze this data, we had to collect it first. To collect it, we designed a new collection method – Twitter Watch. Twitter Watch stands as a configurable, dynamic, easily-deployable, parallel and robust framework for data collection under a set of conditions. The most relevant conditions are its usage of a set of seed accounts and its mechanism to restrict content by languages. The resulting dataset was massive and, except for some limitations – like other Portuguese-speaking countries' content and indeed the Covid-19 pandemic impact –, adequate for our study. Following the release of this work, both Twitter Watch and the dehydrated dataset version will be open-sourced. Our dataset includes suspended account information, as per Twitter Watch's design. It was enriched by leveraging known fake news websites to uncover even more, and then all of those were used to identify accounts who shared them. It was also enriched through a collaboration with a peer working on a parallel project and through a later analysis stemming from an LDA model trained on this dataset. The union of these four labeled account types is considered the malicious set of accounts. This enriched dataset was then described and explored in three main approaches: content, metadata, and structure (*a.k.a.* network).

The content-oriented approach revolved around using LDA topic modeling and subsequent analysis and exploration to understand how politics work in the Portuguese Twitter. We analyzed the 75 accounts that explained the greatest cumulative portion of each political topic and uncovered 17 new trolls. We also grouped accounts by topical similarity through clustering. These clusters revealed that all the labeled malicious types, except the suspended accounts, are disproportionally distributed when compared to regular accounts, and that this disparity is extremely similar regardless of which of these three account types we are looking at – fake news posters, peer-shared trolls, LDA found trolls.

The metadata-oriented effort was initially envisioned as a means of finding political allegiance between the accounts in our dataset and the Portuguese political parties. We created an embeddings model of hashtags. Our initial goal was not feasible without solving an issue we failed to solve – how to find the hashtags that describe each political party equivalently – perhaps no perfect solution exists. However, we did manage to prove that building and tuning a hashtag embeddings model is possible, and that contribution may sparkle further research. Even with the apparent failure, we were able to perform an analysis of each party individually to find out how differently do malicious and regular users relate to it. We also found that the malicious account types, except the suspended ones, are much more politically separated than regular accounts.

The structure-oriented exploration delved into embedding interaction data in order to identify groups or communities within the same “sub-twitterspheres”. A few datasets were built with retweet, reply, quote, and mention interactions that failed to produce any decent embeddings model. The tables were turned when considering only co-occurring mentions. This new dataset revealed that mapping mentions could produce very appealing results in terms of accounts with similar interaction behavior. This was especially observable when we clustered accounts by their embedding values and reached the exact same conclusions as in the totally independent content-oriented approach. Namely, that within malicious accounts, suspended accounts are different, and the remaining three types are both very distinct from regular accounts as they are significantly similar between each other.

From all of these, we answer our hypothesis, as we have now seen that, for our preferential types of data analysis, these types of approaches yielded meaningful models which revealed a large agreement in terms of results, especially in understanding that Twitter’s suspension patterns are very distinct from the characterization of the remaining types of malicious accounts, whether in terms of content, metadata, or structure/interaction.

These different exploration endeavors focused on avoiding the pitfalls of the approaches we had previously identified in the SLR process. If someone intends to manipulate political opinions, either through individual behavior or orchestrated campaigns, there are things that are mainly hard to escape: the nature of what you want to manipulate – its content; and the targets of that manipulation – the interaction communities.

## 7.2 Contributions

All in all, we can highlight the following contributions:

Systematic Literature Review

Conducted a SLR process to understand the current state of research on the types of data, types of processing, and end-goals of malicious political activity detection;

Twitter Watch

Designed and implemented a configurable, dynamic, easily-deployable, and robust framework for Twitter data collection. This tool was used to collect data of the Portuguese Twittersphere from August 1st 2019 until June 20th 2020, with over 6,890,000 accounts and over 163,280,000 tweets collected in that period. Both Twitter Watch and the dataset will be open-sourced;

Fake News

Using a simple technique that uses a seed of 21 known fake news websites, we were able to isolate all the accounts who shared them on Twitter (715 unique). With the top 50 accounts that share the most significant amount of fake news, we isolated all their 200 most shared websites and found 18 new fake news websites;

Meaningfulness

We showed that using the collected dataset led us to build meaningful models for different types of data used. The LDA model revealed how the election period of October 6th 2019 influenced the discussion of political topics around that time. The hashtag embeddings model, after tuning, yielded a high precision@1 (73.4%) using capital-country analogies; The co-occurring mention embeddings model was found to be meaningful under the assumption that its subsequent results were in agreement with both previous exploration approaches, as well as a less reassuring manual validation;

Agreement

The three exploration approaches pursued – content, metadata, and structure – showed a significant agreement in terms of how the distinct types of malicious activity – suspended, fake news posters, peer shared trolls, trolls found with our LDA model – are related to each other and regular accounts. Emphasis on the suspended accounts that have a strong resemblance to regular accounts and the remaining malicious types having a high resemblance among themselves and very little resemblance with the suspended and regular accounts;

Hypothesis

As briefly explained in the above two points, and more detailed in Section 7.1 (p. 90), we answered our hypothesis after having established our preliminary study (*cf.* Section 4.1.1, p. 24) based on understanding the types of information we can get from each of the three exploration approaches undertaken.

## 7.3 Future Work

Our results are interesting but are only a first step into understanding what kind of high-level approaches work to analyze and detect malicious political content on Twitter. We believe many things can be considered for future work, namely:

Extrapolation in time

Pursue further research efforts that focus on validating our results through time, namely in future elections in Portugal, even though it is impossible to predict the existence of everlasting effects of the Covid-19 pandemic.

Extrapolation in context

Ensure that these results that were mostly focused on the Portuguese Twittersphere can be observed in other contexts. However, we expect countries where social media manipulation is more prolific to reveal different types of outcomes. For instance, the accounts most responsible for producing content on political topics might be a very sound way of finding highly widespread attempts at political manipulation.

Twitter Watch

In order to scale in a way that can be both more manageable in terms of computing requirements and the amount of data collected, we believe Twitter Watch would benefit from having a way to grow horizontally through the use of a distributed database. Additionally, we envision Twitter Watch as a platform not only for data collection but also for monitoring a given Twittersphere by being able to execute previously built models in real-time. More, we believe that expanding this framework to other social networks like Facebook could potentially result in the capitalization of the abstraction level we enforced on the tool, eventually leading to broader adoption;

Open-source

A final note on the future of this work pertains to the ability to make Twitter Watch, in the abovementioned capacity, open to any citizen, researcher, journalist, organization, or government as a means of political transparency and manipulation prevention.

# Epilogue

A word from the first author,

I saw the worst and the best of this SN, it was enlightening. I did not find any orchestrated manipulation effort, or maybe I did, and did not notice it. Perhaps that was actually the most startling observation stemming from this thesis. Let me explain. Orchestration was absent. Widespread extreme opinions were not. Racism, hate, and disinformation are prolific. I saw these proliferate in what can only be described as gullible *users*. Do take a moment to notice how we have extensively tried to use the term “account”, and not user, throughout this document. This is because users are expected to be made of flesh and bones, to be humans, people. Every person’s opinions and ideals influence the lives of many others, but the corrupted goals of a few are getting more attention, by marching on the back of a populism beast that can haze people into believing extreme nationalist messages the likes of which have proved to be catastrophic in the past. Twitter is the medium, we write the message. I believe technological efforts, like this one, are essential, but not enough to prevent intolerance from spreading. We need education, facts, and open-mindedness. Online Social Networks are a recent phenomenon. Most people did not grow up being taught about the dangers they pose. Not just to our privacy, but to our view on the world, to our beliefs, ideologies and, ultimately, our actions. This needs to change. The best way to fight the disinformation virus is not by using homeopathic machine learning, it is by inoculating people against it. Until then, I know efforts like this will not stop, nor should they, and neither should we tolerate the spread of false information and its proven power to lead people into intolerance. A simple action like reporting malicious accounts on Twitter can have a massive impact on how we help each other. We should tolerate everyone and everything, the only thing we cannot tolerate as accounts, as users, and as Humans is intolerance [51].

# Appendices

# Appendix A

# Example of a Twitter API User Object

```
{
"_id": NumberLong(718009445863788544),
"collected_at": ISODate("2020-04-05T17:40:29.333Z"),
"created_at": ISODate("2016-04-07T09:36:06.000Z"),
"depth": 0,
"description": "Primeiro-Ministro de Portugal e @antoniocostaps.
Sigam também o XXII Governo em @govpt.",
"favourites_count": 140,
"first_collected_at": ISODate("2020-04-05T17:29:47.216Z"),
"followers_count": 122171,
"friends_count": 308,
"location": "Portugal",
"name": "António Costa",
"profile_banner_url": "https://pbs.twimg.com/profile_banners/
718009445863788544/1465197909",
"profile_image_url": "http://pbs.twimg.com/profile_images/
960858468491190273/JHGIiOnB_normal.jpg",
"screen_name": "antoniocostapm",
"statuses_count": 2926,
"url": "https://t.co/m7nAuRvyly",
"verified": true
}
```

Listing 3: Example of a Twitter API user object for the current Portuguese prime minister

# Appendix B

# Fake News Websites List from Manual Investigation

Source 1 – DN.pt [49]

- https://noticiasdem3rda.com/
- https://www.bombeiros24.com/
- https://www.bombeiros24.pt/
- http://www.semanarioextra.com/
- https://jornaldiario.net/
- http://noticiario.com/
- http://www.magazinelusa.com/
- https://www.lusopt.com/
- https://www.altamente.org/
- https://www.lusonoticias.com/
- https://www.vamoslaportugal.com/
- https://www.lusopt.eu/
- https://www.curanatural.pt/
- https://www.muitofixe.pt/
- https://tafeio.com.pt/

Source 2 – Reddit's r/portugal [50]

- https://verdade.com.pt/

Source 3 – JN.pt [51]

- https://www.facebook.com/pg/RodrigoMoreno19/
- https://www.facebook.com/carregabenfica.pt
- https://www.facebook.com/SouBenfica1904/

---

49 https://www.dn.pt/edicao-do-dia/11-nov-2018/fake-news-sites-portuguesescom-mais-de-dois-milho html

50 https://www.reddit.com/r/portugal/comments/a6yg93/sites_de_fake_news_em_ portugal/

51 https://www.jn.pt/inovacao/rede-espalha-noticias-falsas-sobre-futebol-9376241. html

- https://www.facebook.com/benficalovers
- https://www.facebook.com/ViverBenfica1904
- https://www.facebook.com/Slb2015
- https://www.facebook.com/pg/obenfiquista.pt/
- https://www.facebook.com/estaincrivel/
- https://www.facebook.com/sogolo.pt/
- https://www.facebook.com/Levanta-te-e-Joga-885007101566580/
- http://noticias24.com.pt/
- http://noticiario.com.pt/
- http://carregabenfica.pt/
- https://www.carrega-benfica.pt/
- https://asnoticias.pt/
- https://noticias.com.pt/
- https://ligaportuguesa.pt/
- https://www.facebook.com/EnzoPerez.JF
- https://sonoticias.pt/
- https://livredireto.pt/
- https://remate.pt/
- https://futebol11.com/
- https://portoatemorrer.com/
- https://info24h.pt/

Source 3 – Author's knowledge

- https://www.noticiasviriato.pt/

# Appendix C

# Uncovered Fake News Websites from Inspection of Known Fake News Posters

Fake News

1. tuga.press
2. flash.pt
3. vortexmag.net
4. inimigo.publico.pt
5. tuasaude.com
6. partilhado.pt
7. soutodaboa.com

8. apost.com
9. sabiaspalavras.com
10. seuamigoguru.com
11. elucubrativo.blogspot.com
12. palavrasoltas.com
13. noticiasdevizela.pt
14. direitapolitica.com
15. lusojornal2015.blogspot.com
16. pensarcontemporaneo.com
17. postal.pt
18. magazinept.com

# Appendix D

# Portuguese Newspaper Covers from October 19th 2019

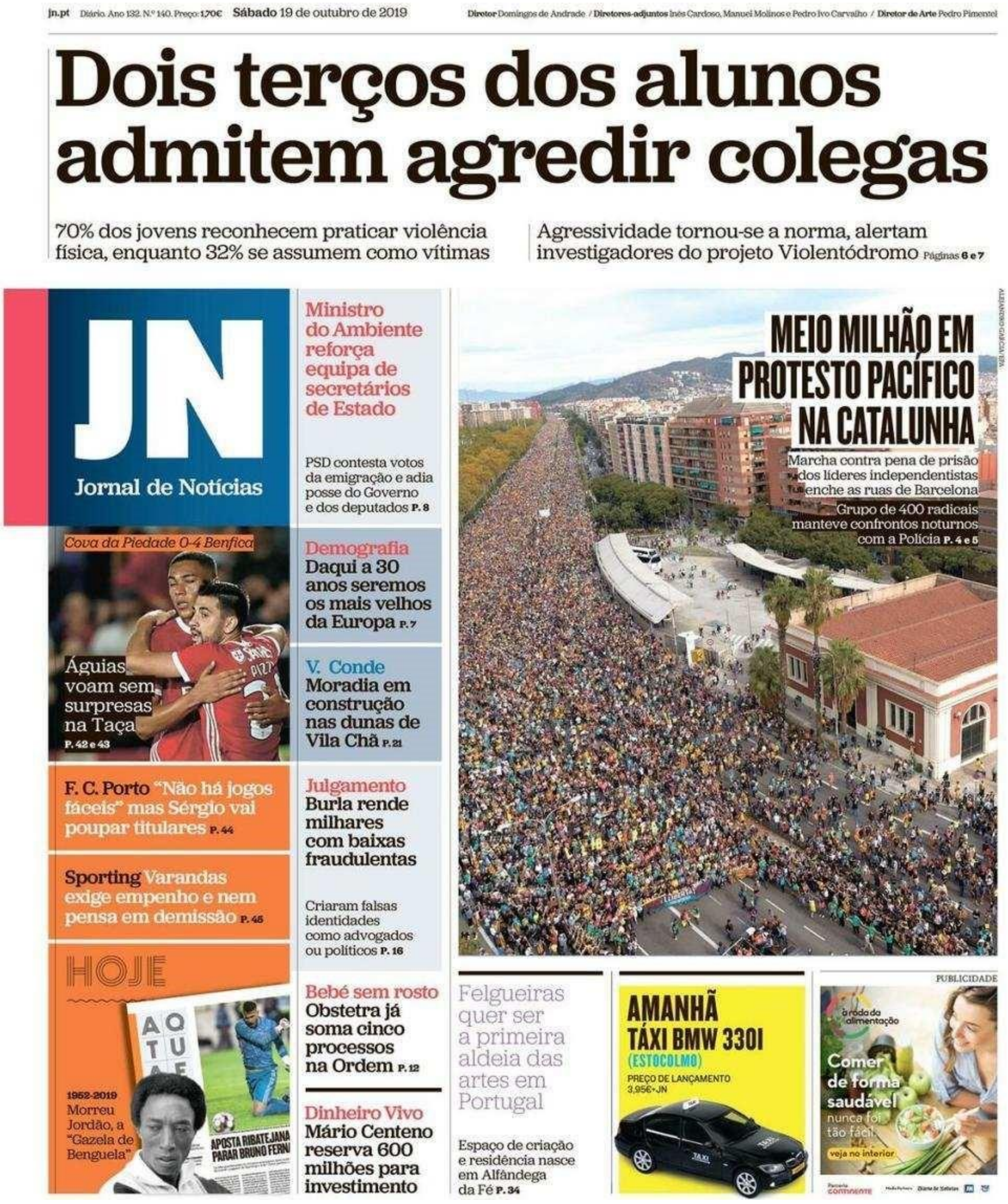

jn.pt Diário. Ano 132. N.º 140. Preço: 1,70€ Sábado 19 de outubro de 2019 Diretor Domingos de Andrade / Diretores-adjuntos Inês Cardoso, Manuel Molinos e Pedro Ivo Carvalho / Diretor de Arte Pedro Pimentel

**Dois terços dos alunos admitem agredir colegas**

70% dos jovens reconhecem praticar violência física, enquanto 32% se assumem como vítimas

Agressividade tornou-se a norma, alertam investigadores do projeto Violentódromo Páginas 6 e 7

JN Jornal de Notícias

**Ministro do Ambiente reforça equipa de secretários de Estado**

PSD contesta votos da emigração e adia posse do Governo e dos deputados P. 8

Cova da Piedade 0-4 Benfica

Águias voam sem surpresas na Taça P. 42 e 43

Demografia Daqui a 30 anos seremos os mais velhos da Europa P. 7

V. Conde Moradia em construção nas dunas de Vila Chã P. 21

F. C. Porto "Não há jogos fáceis" mas Sérgio vai poupar titulares P. 44

Julgamento Burla rende milhares com baixas fraudulentas

Criaram falsas identidades como advogados ou políticos P. 16

Sporting Varandas exige empenho e nem pensa em demissão P. 45

**MEIO MILHÃO EM PROTESTO PACÍFICO NA CATALUNHA**

Marcha contra pena de prisão dos líderes independentistas enche as ruas de Barcelona

Grupo de 400 radicais manteve confrontos noturnos com a Polícia P. 4 e 5

HOJE

1952-2019 Morreu Jordão, a "Gazela de Benguela"

Bebé sem rosto Obstetra já soma cinco processos na Ordem P. 12

Dinheiro Vivo Mário Centeno reserva 600 milhões para investimento

Felgueiras quer ser a primeira aldeia das artes em Portugal

Espaço de criação e residência nasce em Alfândega da Fé P. 34

Figure D.1: October 19th 2019 cover for Jornal de Notícias

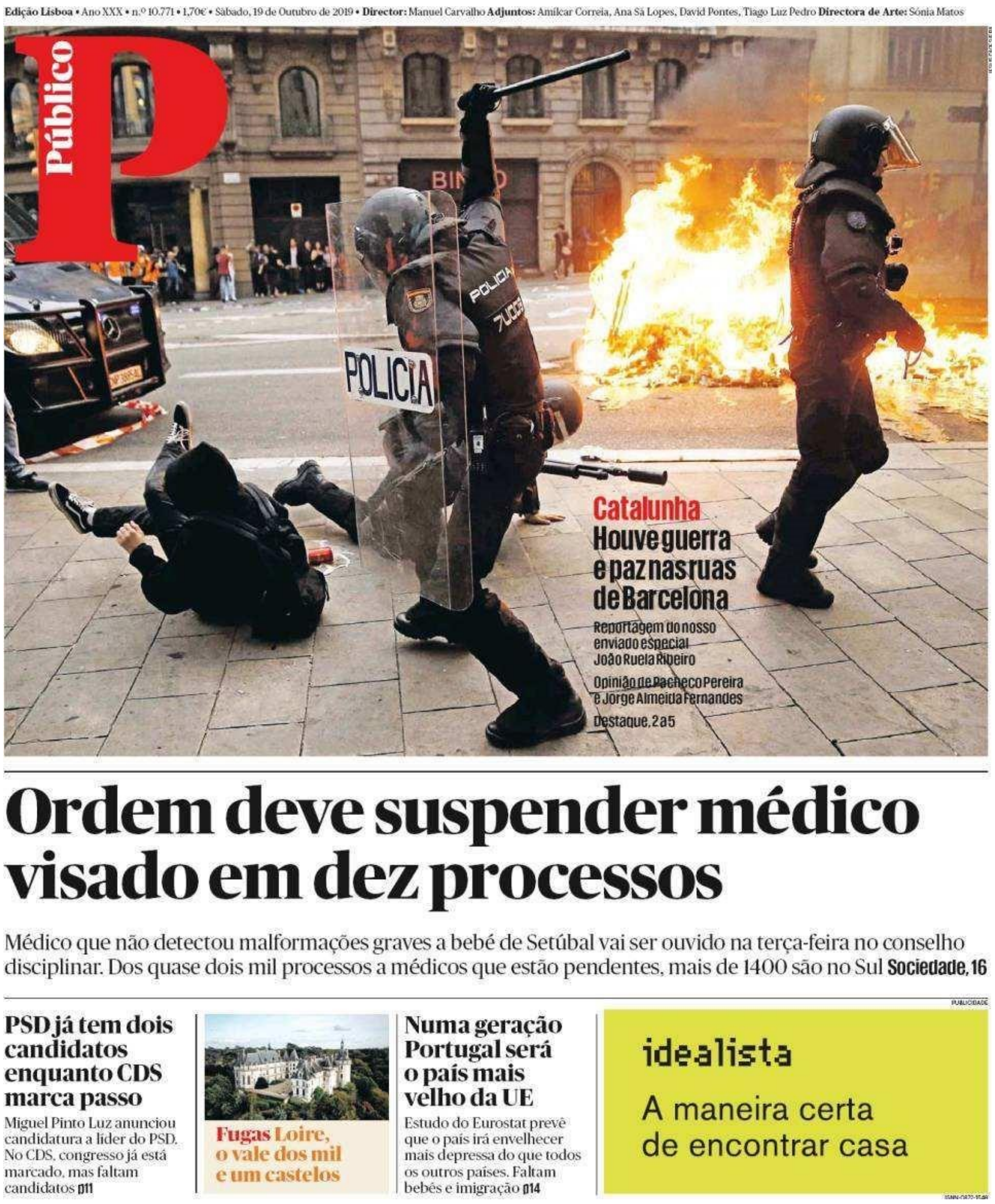

**Edição Lisboa** • Ano XXX • n.º 10.771 • 1,70€ • Sábado, 19 de Outubro de 2019 • **Director:** Manuel Carvalho **Adjuntos:** Amílcar Correia, Ana Sá Lopes, David Pontes, Tiago Luz Pedro **Directora de Arte:** Sónia Matos

Público

**Catalunha**
**Houve guerra e paz nas ruas de Barcelona**
Reportagem do nosso enviado especial João Ruela Ribeiro
Opinião de Pacheco Pereira e Jorge Almeida Fernandes
Destaque, 2 a 5

**Ordem deve suspender médico visado em dez processos**

Médico que não detectou malformações graves a bebé de Setúbal vai ser ouvido na terça-feira no conselho disciplinar. Dos quase dois mil processos a médicos que estão pendentes, mais de 1400 são no Sul **Sociedade, 16**

**PSD já tem dois candidatos enquanto CDS marca passo**
Miguel Pinto Luz anunciou candidatura a líder do PSD. No CDS, congresso já está marcado, mas faltam candidatos p11

**Fugas Loire, o vale dos mil e um castelos**

**Numa geração Portugal será o país mais velho da UE**
Estudo do Eurostat prevê que o país irá envelhecer mais depressa do que todos os outros países. Faltam bebés e imigração p14

PUBLICIDADE
idealista
A maneira certa de encontrar casa

Figure D.2: October 19th 2019 cover for Jornal Público

# Appendix E

# Portuguese Newspaper Covers from September 16th 2019

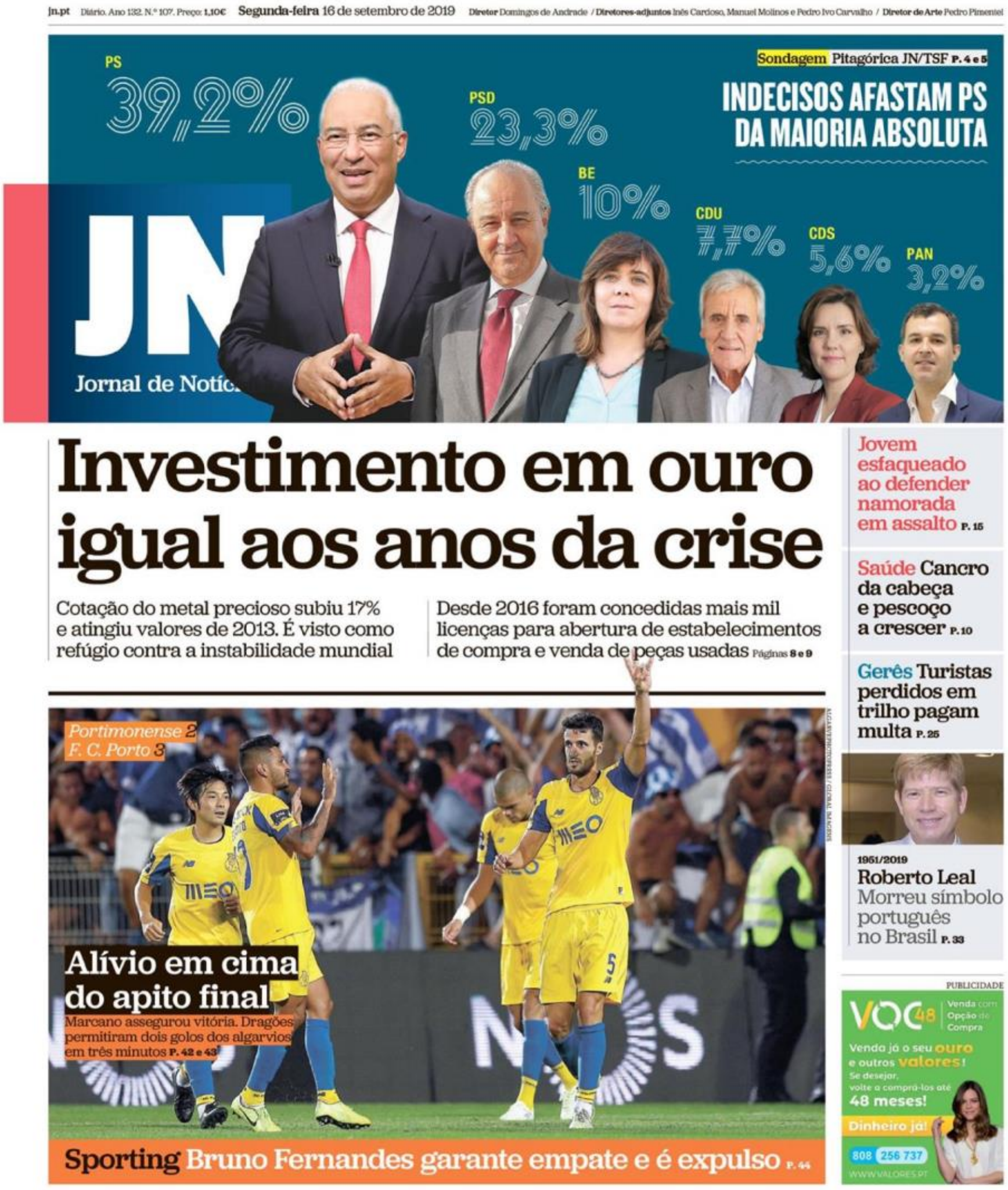
jn.pt Diário. Ano 132. N.º 107. Preço: 1,10€ Segunda-feira 16 de setembro de 2019 Diretor Domingos de Andrade / Diretores-adjuntos Inês Cardoso, Manuel Molinos e Pedro Ivo Carvalho / Diretor de Arte Pedro Pimentel

Sondagem Pitagórica JN/TSF P. 4 e 5

INDECISOS AFASTAM PS DA MAIORIA ABSOLUTA

PS 39,2%
PSD 23,3%
BE 10%
CDU 7,7%
CDS 5,6%
PAN 3,2%

JN
Jornal de Notíc

Investimento em ouro igual aos anos da crise

Cotação do metal precioso subiu 17% e atingiu valores de 2013. É visto como refúgio contra a instabilidade mundial

Desde 2016 foram concedidas mais mil licenças para abertura de estabelecimentos de compra e venda de peças usadas Páginas 8 e 9

Jovem esfaqueado ao defender namorada em assalto P. 15

Saúde Cancro da cabeça e pescoço a crescer P. 10

Gerês Turistas perdidos em trilho pagam multa P. 25

Portimonense 2
F. C. Porto 3

Alívio em cima do apito final

Marcano assegurou vitória. Dragões permitiram dois golos dos algarvios em três minutos P. 42 e 43

1951/2019
Roberto Leal
Morreu símbolo português no Brasil P. 33

Sporting Bruno Fernandes garante empate e é expulso P. 44

PUBLICIDADE

VOC48 Venda com Opção de Compra
Venda já o seu ouro e outros valores!
Se desejar, volte a comprá-los até 48 meses!
Dinheiro já!
808 256 737
WWW.VALORES.PT

Figure E.1: September 16th 2019 cover for Jornal de Notícias

# Appendix F

# Evidence of Malicious Activity in Manually Uncovered Trolls (LDA)

CSV list of *user_id; screen_name; tweet_id; tweet_text; malicious_reason; link_to_twitter*

- 2504570929; barba__rija; 1266449790386176002; Trump a tratar os senhores da OMS como eles merecem, mas daqui a uns tempos vamos saber toda a verdade do que se passou na China e como a OMS foi cumplice.; conspiracy seeding/hate speech; link
- 2504570929; barba__rija; 1278367116819935232; Portugal um destino turistico e com muitos emigrantes, sem a TAP essas mesmas pessoas não conseguiam chegar ao nosso país. Agora é repetir isto até à exaustão para se tornar verdade.; hate speech; link
- 4895874375; TioFCosta; 1269339229856903169; Tem o #Soros por trás...; conspiracy seeding; link
- 4895874375; TioFCosta; 1269356241169637378; Parte da teia de #Soros...; conspiracy seeding; link
- 4895874375; TioFCosta; 1267546691319840780; Sabem que #Soros e os seus correligionários judeus não financiam apenas a #Antifa mas também, direta e indiretamente, toda a grande indústria da música, cinema e comunicação social... É um poder imenso, ativado sempre que conveniente...; conspiracy seeding; link
- 4895874375; TioFCosta; 1267210216044707848; #Antifa is funded by Jews...; conspiracy seeding; link
- 4895874375; TioFCosta; 1267077658460016642; #Jewish; hate speech (retweet); link
- 976972802799603712; pedrocr75444218; 1269686566760431617; Os espanhóis a foderemnos o turismo. Já acreditam em mim?!?; hate speech; link
- 976972802799603712; pedrocr75444218; 1268167513466122240; "-Acho muito bem que em Loures é só comunas e ciganagem."; hate speech; link
- 1173313381488627712; Augusto58770700; 1268942980460806147; Obama só faz merda. Organizou as revoltas dos EUA e vai acabar condenado a pena de morte. Graças a Deus Trump existe e é o presidente da nação mais livre do planeta. Mais aqui...; hate speech; link
- 1173313381488627712; Augusto58770700; 1268942980460806147; Ex-diretor da CIA,

Steve Brannon, despedido por Trump, decidiu vingar-se e organizou as revoltas nos EUA. As revoltas são uma operação militar, não popular.; conspiracy seeding/disinformation; link

- 1173313381488627712; Augusto58770700; 1267894268045205509; O Covid foi uma operação política para retirar direitos. A operação acaba mal porque medicos, funcionários e o povo, descobriram o engano. Números forjados, alarme falso e chantagens.; conspiracy seeding/disinformation; link
- 749226888502050817; ordemespontanea; 1266322987818647553; Marcelo passou a noite num hotel para verificar se está tudo conforme indicações da (pide)DGS. Diz que sim, inclusivé o serviço de acompanhante de luxo.; disinformation; link
- 749226888502050817; ordemespontanea; 1261026872709853186; Marcelo escrutinado https://blasfemias.net/2020/05/14/marcelo-escrutinado/; fake-news; link
- 749226888502050817; ordemespontanea; 1256517861352394753; Factos: 1)a 8/Março realizou-se em Lisboa a manifestação feminista ilustrada na foto anexa. 2) o vírus tem um período de incubação até 14 dias. 3) a 18/Março é declarado o estado de Emergência pelo PR.; conspiracy seeding; link
- 749226888502050817; ordemespontanea; 1260546623383404544; Está um par de ineptos e indignos a dizer coisas na Autoeuropa.; hate speech; link
- 749226888502050817; ordemespontanea; 1247462593977073664; Marcelo goza à grande e sem qualquer pudor com mais de 90.000 pessoas suspeitas de ter #covid19, cerca de 11.800 confirmados e 140 recuperados. Os 311 mortos já não ouviram o PR.; conspiracy seeding; link
- 21890926; __tomaz; 1265287462559105025; Para o parasita Mamadou, o cigano não é racista... é um exemplar gestor de negócios.; hate speech; link
- 21890926; __tomaz; 1264964279012405256; As "pessoas" que, por falta de argumentos razoáveis, chamam, por tudo e por nada, fachos aos outros, deveriam ficar sujeitos a quarentena vitalícia e hereditária. A estupidez é muito mais perigosa que o #covid19!; hate speech; link
- 2542348724; fatos_ex; 1270343639827243009; Esses são os 13 segredos que a indústria da beleza não quer que você saiba.. https://t.co/BpSmR8RvKL?amp=1; fake-news; link
- 2542348724; fatos_ex; 1270426687092617218; 8 receitas naturais eficazes para clarear os dentes em casa https://t.co/YZeZuO9yJp?amp=1; fake-news; link
- 746140256; oiddtc; 1179079958402125827; Chega para lá, Cristina Ferreira. O André Ventura é a transferência televisiva do ano. Da CMTV para a ARTV. O meu obrigado, desde já, ao fenomenal distrito de Lisboa por ter eleito tão asquerosa personalidade. Filhos da puta. #Legislativas19; hate speech; link
- 746140256; oiddtc; 1180212803912777729; Óptima manobra de campanha por parte do PS. Parabéns, Costa. És um burro e também to diria na cara.; hate speech; link
- 746140256; oiddtc; 1181097634381094913; Obrigado Setúbal por elegeres esta besta. #Legislativas19 https://t.co/Pmcay7lSg5; hate speech; link
- 746140256; oiddtc; 1181090361340567553; A gaja do Livre estava tão contente que nem entalava o discurso. Vai ser bonito com os tempos-limite na AR. #Legislativas19; hate speech;

link

- 746140256; oiddtc; 1240372029804761091; Portugal, caralho! - Marcelo Rebelo de Sousa #covid19pt; disinformation; link

- 298195731; theumilk99; 1179717261273292802; Quero que a marinhs se foda a gaja é nojenta portanto enfiem lá o Expose no cu ya ya tenho MT pena coitadinha , whatever mas por favor calem se já não posso ouvir essa novela de criança; hate speech; link
- 298195731; theumilk99; 1179717261273292802; Quem lê isto É Gay, Apanhei-te De Novo Chupa Boi LOL; hate speech; link
- 298195731; theumilk99; 1181005045170606080; @HeroiAmarelo @gmgr20 Masninguém quer saber o que tu achas mete te no caralho; hate speech; link
- 1129097944605569024; PedrodeCastroS1; 1189511731384606720; [image]https://pbs.twimg.com/media/EIH-mf7XUAAo8bn?format=jpg&name=900x900; hate speech/images; link
- 1129097944605569024; PedrodeCastroS1; 1183691470211039232; Uma das 13 estátuas de Jesus mais famosas do mundo é portuguesa | ncultura https://t.co/JbzWD3OXPm?amp=1; fake news; link
- 990210416176451589; MariaCordRosa; 1180460544127623168; Uma mulher preta, gaga e racista e o André Ventura entram num bar, oh wait! É a Assembleia da República; hate speech; link
- 990210416176451589; MariaCordRosa; 1180982500660781057; @EAdlEssEAd Já foste a Estremoz? Beja? Elvas? Aquilo parece o Ciganistão; hate speech; link
- 990210416176451589; MariaCordRosa; 1181209266579034114; O truque da Black Gaga é o célebre "Vou chamar-lhes putas antes que elas me chamem a mim"; hate speech; link
- 60743301; ilspeciale_; 1179493894138060801; RT @joaoduarte97: Dá para apertar o pescoço a quem achou que este plantel era suficiente?; hate speech; link
- 60743301; ilspeciale_; 1179493894138060801; Benfica europeu o caralho que te foda oh Orelhas!!!; hate speech; link
- 51691926; casadoscaes; 1220701291023826945; Se havia dúvidas de que o estado não é pessoa de bem... https://blasfemias.net/2020/01/24/ nao-vivemos-num-estado-de-direito-mas-antes-num-estado-mafioso/ ; fakenews; link
- 51691926; casadoscaes; 1220605309724565505; Trump é o 1º Presidente Americano a Participar na Marcha Pela Vida em Washington https://www.noticiasviriato .pt/post/trump-e-o-primeiro-presidente-americano-a-participar-na-marchapela-vida-em-washington; fakenews; link
- 269361642; SAIDLE; 1266842420932677632; Um independente? Amigo íntimo de Sócrates e das negociatas; conspiracy seeding; link
- 269361642; SAIDLE; 1266842420932677632; Este é o mesmo primeiro ministro que foi de férias quando os portugueses morriam queimados em virtude da ineficiência do SIRESP que ele comprou e que não funciona https://t.co/ePc2HzPZY1; disinformation; link

- 563093420; indecisor; 1180241856183046146; RT @hipersurf: Expresso descobre que homem que tentou agredir idoso é do PS; disinformation; link
- 563093420; indecisor; 1179569212681068545; Ele não se importa, pode sempre comprometer o futuro dos portugueses a troco do poder.; conspiracy seeding; link

- 825622897; ReporterSombra; 1270355191943086082; Não são mentiras. São medos reais. Resultam da história da própria humanidade e assolaram em pequena e grande escala os nossos antepassados. Está-nos gravado no ADN. https://t.co/JZZmKMKFXo?amp=1; fakenews; link
- 825622897; ReporterSombra; 1270355191943086082; Somos o que comemos?Comemos o que somos? - Por Célia Meira https://t.co/pJYtb0PCUt?amp=1; fakenews; link

## Appendix G

# LDA clustering – All Account Types Distributed per Cluster

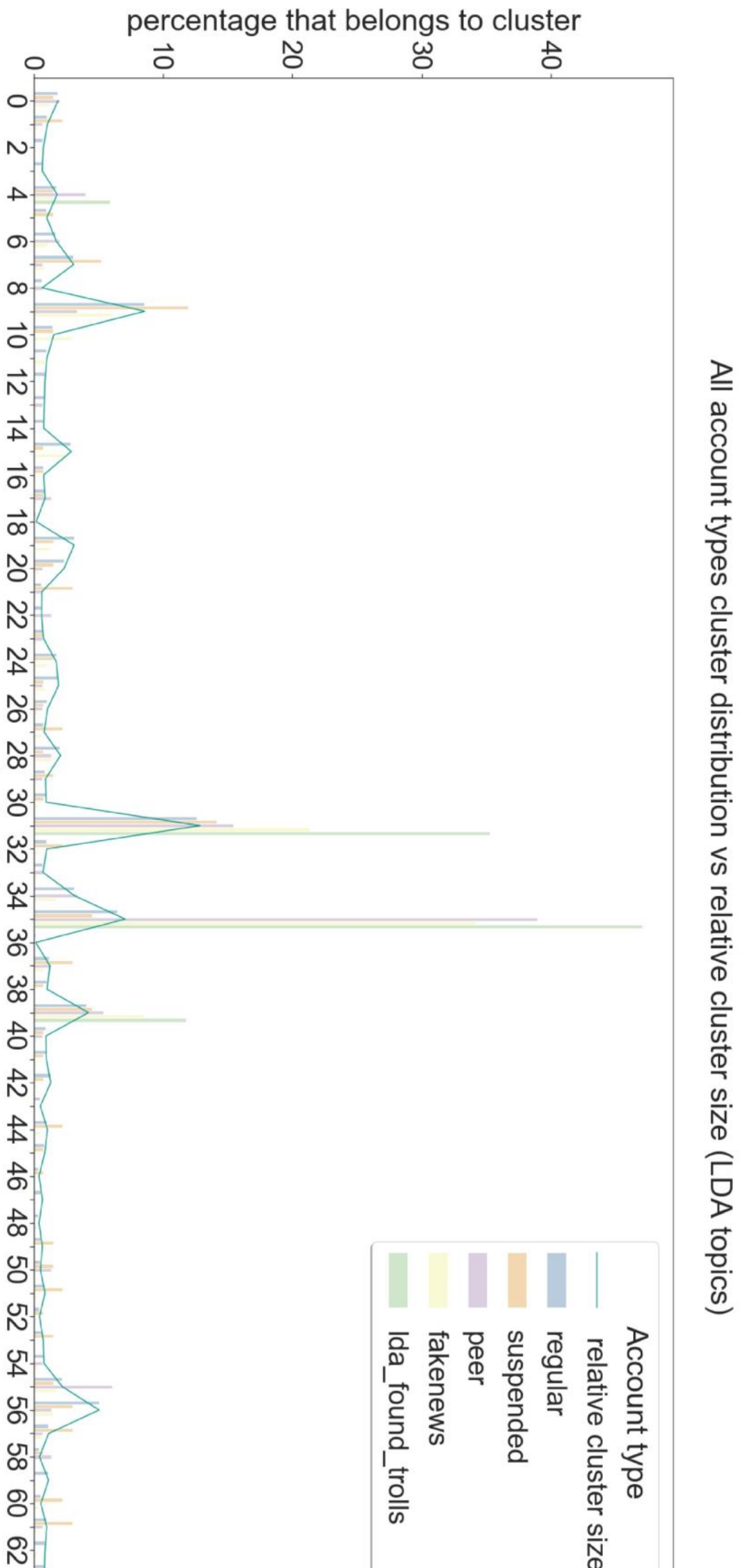

Figure G.1: LDA cluster distribution for all account types

# Appendix H

# List of Capital to Country Analogy Pairs

Table H.1: Capital to country hashtag pairs

| capital | country | capital | country |
|---|---|---|---|
| madrid | spain | sofia | bulgaria |
| paris | france | tokyo | japan |
| ottawa | canada | tunis | tunisia |
| berlin | germany | washington | unitedstates |
| london | england | stockholm | sweden |
| athens | greece | prague | czechia |
| amsterdam | holland | montevideo | uruguay |
| budapest | hungary | bratislava | slovakia |
| vienna | austria | moscovo | russia |
| zagreb | croatia | algiers | algeria |
| bern | switzerland | luanda | angola |
| belgrade | serbia | amman | jordan |
| brussels | belgium | riga | latvia |
| bucharest | romania | vilnius | lithuania |
| cairo | egypt | dakar | senegal |
| dublin | ireland | copenhagen | denmark |
| seoul | southkorea | quito | ecuador |
| oslo | norway | lima | peru |
| kiev | ukraine | warsaw | poland |
| maputo | mozambique | - | - |

Appendix I

# Mentions Embeddings Clustering – All Account Types Distributed per Cluster

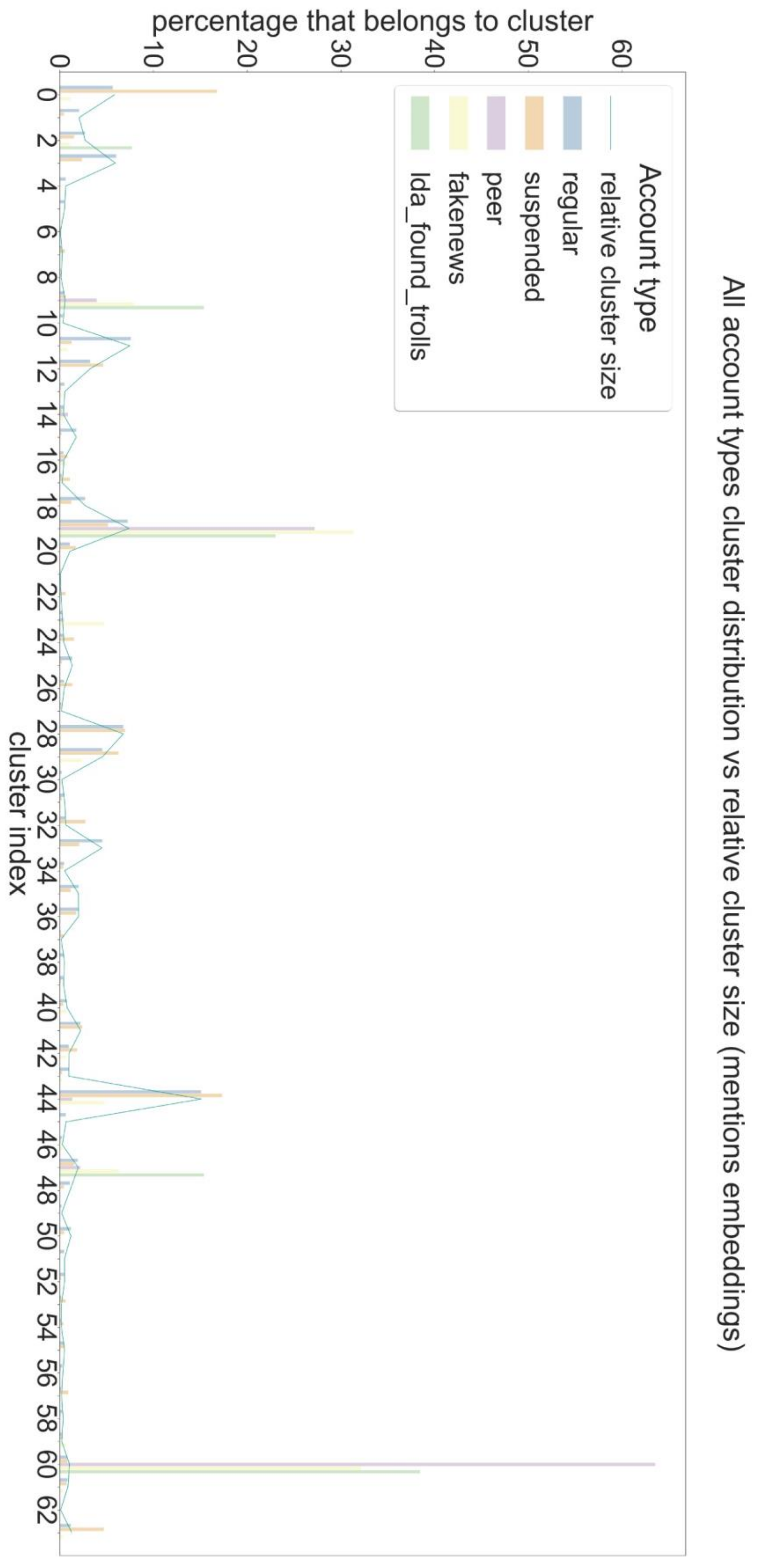

Figure I.1: Mentions embeddings cluster distribution for all account types